\documentclass[a4paper,onecolumn,accepted=2023-05-18]{quantumarticle}
\pdfoutput=1
\usepackage[utf8]{inputenc}
\usepackage[english]{babel}
\usepackage[T1]{fontenc}
\usepackage{multirow}
\usepackage{tikz}
\usepackage{lipsum}
\usepackage{graphicx}
\usepackage{amsmath,stackengine}
\usepackage{latexsym}
\usepackage[style=numeric,backend=bibtex,sorting=none]{biblatex}
\usepackage{amsfonts}
\usepackage{amssymb}
\usepackage{array}
\usepackage{color}
\usepackage{subfigure}
\usepackage{verbatim}
\usepackage{mathtools,leftidx}
\usepackage{enumitem}
\usepackage{hyperref}
\usepackage[normalem]{ulem}

\stackMath
\newcommand{\ket}[1]{\left\vert#1\right\rangle}
\newcommand{\bra}[1]{\left\langle#1\right\vert}

\newcommand{\ps}[2]{\prescript{}{#1}{#2}}

\newcommand{\A}{\alpha}

\newcommand{\Fi}{\iffalse}
\newcommand{\cmmntemp}[1]{\ignorespaces}
\newcommand{\cmmnt}[1]{\ignorespaces}

\definecolor{darkblue}{rgb}{0.4, 0.4, 0.9}
\newcommand{\old}[1]{{\color{black}#1}}

\newcommand{\new}[1]{{\color{black}#1}}

\DeclareMathOperator{\arcsec}{arcsec}

\DeclareMathOperator{\tr}{tr} 
 
\DeclareMathOperator{\ii}{i}

\begin{document}

\title{Quantum Rabi interferometry of motion and radiation}

\author{
Kimin Park}
\email{park@optics.upol.cz}
\affiliation{Department of Optics, Palacky University, 77146 Olomouc, Czech Republic}
\affiliation{Center for Macroscopic Quantum States (bigQ), Department of Physics, Technical University of Denmark, Building 307, Fysikvej, 2800 Kgs. Lyngby, Denmark}
\author{
Petr Marek}
\affiliation{Department of Optics, Palacky University, 77146 Olomouc, Czech Republic}
\author{Ulrik L. Andersen}
\affiliation{Center for Macroscopic Quantum States (bigQ), Department of Physics, Technical University of Denmark, Building 307, Fysikvej, 2800 Kgs. Lyngby, Denmark}
\author{Radim Filip}
\affiliation{Department of Optics, Palacky University, 77146 Olomouc, Czech Republic}

\begin{abstract}
The precise determination of a displacement of a mechanical oscillator or a microwave field in a predetermined direction in phase space can be carried out with trapped ions or superconducting circuits, respectively, by coupling the oscillator with ancilla qubits.
Through that coupling, the displacement information is transferred to the qubits which are then subsequently read out. 
\old{However, unambiguous estimation of displacement in an unknown direction in the phase space has  not been attempted in such oscillator-qubit systems.}
Here, we propose a hybrid oscillator-qubit interferometric setup for the unambiguous estimation of  phase space displacements in an arbitrary direction, based on feasible Rabi interactions beyond the rotating-wave approximation. 
Using such a hybrid Rabi interferometer for quantum sensing, we show that the performance is superior to the ones attained by single-mode estimation schemes and a conventional interferometer based on Jaynes-Cummings interactions. 
Moreover, we find that the sensitivity of the Rabi interferometer is independent of the thermal occupation of the oscillator mode, and thus cooling it to the ground state before sensing is not required.
We also perform a thorough investigation of the effect of qubit dephasing and oscillator thermalization.
We find the interferometer to be fairly robust, outperforming different benchmark estimation schemes even for large dephasing and thermalization.

\end{abstract}

\maketitle

\section{Introduction}

Quantum sensing is about estimating unknown   processes of interest  using a finite ensemble of probes and detectors with a sensitivity that goes beyond the reach of classical sensing~\cite{DegenRMP2017,GiovannettiNatPho2011Metrology, Sidhu2020, sensinghighenergyphysics2018,DAlessandroBook2021control}.
Historically,  optical probes have been at the center of this field due to the experimental accessibility of lasers and non-classical light resources, high-efficiency detectors, and  the strong robustness of light to external noise sources~\cite{PirandolaNatPhot2018PhotSens, pop00003,pop00009,Polino2020}. 
Quantum optical interferometers exploit the interference between a probe and a reference beam to detect weak signals \cite{DEMKOWICZDOBRZANSKI2015345}, a prominent example being the detection of gravitational waves from black hole mergers~\cite{LigoPRL2016,LIGO2020}. 
A similar approach can be adopted for microwave traveling waves~\cite{Lang2013microwaveHOM}, microwave cavity fields~\cite{Gao2018MwMemInterf}, matter waves~\cite{Bongs2019atominterferometer,Cronin2009atominterferometer},  between light and atomic ensembles~\cite{PezzeRMP2018AtomEnsemble,ChenPRL2015}, and  optomechanics~\cite{tsangPRL2010,motazedifard2020,Bemani2021}.


Numerous sensing proposals use quantum non-Gaussian states as probes for improved estimation of phase and displacement \cite{dalvit_filho_toscano_2006,TerhalPRA2017GKP,BraunRMP2018Sensing}. For example, \cmmnt{a} quantum displacement sensing has been performed with Fock states and non-Gaussian superpositions of Fock states of the phononic modes of trapped ions \cite{2020FockDispersive,McCormick2019}.  
Similarly, for superconducting circuits, the preparation of microwave Fock states and their superpositions has been recently mastered~\cite{Premaratne2017,Wang2017,Pfaff2017NatureConversion,Gely2019} as well as the preparation of non-Gaussian acoustic modes of a quartz crystal \cite{Chu2018}, which can be also used for measuring displacements. 
Superconducting transmon has been recently used for sensing magnons~\cite{Lachance-Quirion2017, Wolski2020, Lachance-Quirion2020}, search for dark matter~\cite{Dixit2020}, microwave radiometry~\cite{Wang2019, Kristen2019,Wang2021} including sensing with a non-Gaussian probe in a Fock-state-superposition~\cite{Wang2019HeisenSupercond}. 
All these non-Gaussian states are however non-optimal for the detection of a phase space displacement. 
The optimal single-mode state for unambiguous estimation of displacement is the  Gottesman-Kitaev-Preskill  state~\cite{TerhalPRA2017GKP} which however are experimentally challenging to generate.
This challenge  can alternatively be  mitigated by using a correlated squeezed probe and continuous-variable measurement~\cite{Park2022}.

Recently, significant progress has been made in exploiting the advanced hybrid qubit-continuous-variable (CV) interface for quantum sensing, e.g. \old{as it is available for}  quantum dots~\cite{pop00001}, color centers~\cite{Schirhagl2014}, trapped ions~\cite{KienzlerPRL2016,Bruzewicz2019APRtrappedion,FluhmannHomeNature2019}  \old{or} superconducting circuits~\cite{Wendin2017superconducting,GuPR2017Superconducting,Touzard2019,BlaisNatPhys2020superconducting, SchoelkopfNature2020,ClerkNatPhy2020SuperCon,Kwon2021Superconduting,BlaisRMP2021CircuitQED}.
Amplification of a displacement through squeezing  was recently tested on trapped ions~\cite{BurdScience2018SqzAmpDisp}.
For the phase estimation, Ramsey interferometry~\cite{Ramsey1949, RiehlePRL1991Ramsey,CadoretPRL2008Ramsey,AriasPRL2018Ramsey} has been used specifically for metrology with trapped ions~\cite{Leibfried2004HeisenbergMultipartite} and superconducting circuits~\cite{Wang2019HeisenSupercond}.  
In such hybrid systems, a \old{short burst of} mechanical force and electromagnetic radiation exerted by classical fields~\cite{DegenRMP2017,Brownnutt2015} generate a small displacement in phase space of the weakly coupled mechanical or microwave oscillator~\cite{TerhalPRA2017GKP}, which can be estimated indirectly by coupling it to a qubit system.

%

The \old{strong} coupling between the oscillator and the qubit can be \old{accurately described by the Rabi Hamiltonian which does not invoke} the rotating-wave approximation (RWA) \old{as is done for the} \old{conventional} Jaynes-Cummings \old{(JC)} Hamiltonian. 
Such a \old{Rabi} Hamiltonian has been shown to enable the preparation of different CV quantum states~\cite{Hastrup2019GKP, Hastrup2020squeezing}, the construction of CV quantum gates~\cite{ParkNJP2018}, \old{slowing down the decoherence of CV states~\cite{Park2021}, conversion of the states between oscillators and qubits~\cite{HastrupPRL2022}}, and the realization of certain phase transitions~\cite{hwang15,Cai2021}. 
These Rabi interactions also appear to be favorable to quantum interferometry \old{of photon scattering events, termed cat-state spectroscopy} as demonstrated on single qubits on a trapped ion crystal~\cite{HempelNatPho2013catspectroscopy,GilmoreScience2021Displacement}.
\old{This idea has been extended to a driven interferometry of a spin-dependance force~\cite{Martinez-Garaot2018}, and to a motion echo \cite{McCormick2018}.}
\new{Recently, a quantum-sensing protocol was proposed that leverages the phase transition of the Rabi model focused on frequency estimation~\cite{GarbePRL2020CriticalRabiSensing}.
These critical quantum sensors also exhibit resilience against thermal noise~\cite{DiCandia2021KerrCritialSensor,ChuPRL2021}.
}
\old{However, a} Rabi interferometer \old{simultaneously estimating} both of the conjugate variables of the oscillator  \old{has} not \old{been considered}  yet and is \old{indispensible} for unambiguous sensing of a displacement in an arbitrary direction  \old{ in phase space caused by an unknown mechanism}~\cite{TerhalPRA2017GKP,IvanovPhysRevA2020TwoDisplac}.

In this work, we \new{introduce a technique}\cmmnt{present a method} for unambiguously estimating the displacement of an oscillator \new{by utilizing its}  coupling to  discrete level systems. 
This method is based on a hybrid interferometric setup \new{that employs} Rabi couplings~\cite{NoriRabiRMP2019,SolanoRabiRMP2019} \new{of the oscillator} only to a pair of auxiliary two-level systems. 
\new{This hybrid interferometer} can be used to measure mechanical force,  \old{electromagnetic} radiation, \new{and other physical phenomena} by \new{determining} the  displacement \old{in an arbitrary direction in  phase space} of a  single-atom oscillator and microwave cavity mode.
This interferometric scheme \new{offers improved}\cmmnt{has} sensitivity \old{quantified} by classical Fisher information (CFI) (for the definition, see Appendix \ref{sec:Fisher}), \old{that is} quadratically increasing with \old{the} coupling strength, superior to the JC interferometer  which show\old{s} a \new{limited} sensitivity.
We compare \old{the hybrid interferometric} scheme to the non-interferometric protocols using equivalent resources where the Rabi gates \old{are} used \old{only} to  prepare \old{single-mode} quantum  non-Gaussian states  \cite{IvanovPRAp2015ForceSensor, Ivanov2016, IvanovPRA2018} \old{and to measure \old{the signal} together with qubit detectors}.  

We also \new{demonstrate} that the unwanted interference between the non-commuting Rabi interactions can be avoided by \cmmnt{choosing proper}\new{selecting appropriate} Rabi coupling strengths, or \cmmnt{be} actively cancelled \cmmnt{through}\new{with} an engineered two-qubit interaction.
For \old{a proof-of-principle experiment,} we \cmmnt{consider}\new{evaluate} \old{realistic} experimental imperfections and decoherence effects on the estimation \old{precision}, under which the superiority \old{of the interferometer} is maintained \old{over all levels of the considered noises}. 
Interestingly, the dynamical range of the interferometer \cmmnt{witnessed by the full width \old{at} the half maximum CFI} is wider than the non-interferometric setups \old{as well}.
\old{Notably,} we find that even \old{a} naturally existing  thermal state probe is as useful as a pure state probe, which alleviates the \old{necessity} of \new{pre-}cooling the mechanical mode \old{and preparing highly non-classical states,} and \old{thus} enhances the experimental feasibility.
\old{This} protocol can be readily implemented both in  superconducting circuits~\cite{Touzard2019,BlaisNatPhys2020superconducting,SchoelkopfNature2020,Kwon2021Superconduting,ClerkNatPhy2020SuperCon,BlaisRMP2021CircuitQED} and trapped ion systems~\cite{KienzlerPRL2016,Bruzewicz2019APRtrappedion,FluhmannHomeNature2019}, and be applied to radiometry and magnon sensing~\cite{Wang2019,Wang2021, Wolski2020,Lachance-Quirion2020,Kristen2019}. 

\section{Displacement estimation using Rabi interactions}



An important physical quantity \old{that needs to be estimated} for precise control is a short-time   external force,  which  weakly displaces the mechanical motion of trapped ion~\cite{LeibfriedRMP2003,Biercuk2010,Gilmore2017,Affolter2020},  or  microwave field in a cavity \old{\cite{ritschRMP2012,Wang2019}} at a time scale below the decoherence time.    
\old{In trapped ions and superconducting circuits, qubits   coupled to an internal cavity  field can  sense such a \old{small} displacement (See Table I of ~\cite{DegenRMP2017} and references therein). 
}
\old{Here}, we focus on the  estimation of the effect of  force component\old{s} on a single oscillator \cmmnt{spatial degree of freedom} regardless of the \old{physical} mechanisms,  described by \old{a complex} parameter $\alpha=\alpha_\mathrm{r}+\mathrm{i}\alpha_\mathrm{i}$ (with $\alpha_\mathrm{r,i}\in \mathbb{R}$) of an unknown displacement of
$\hat{D}[\alpha]=\exp[\alpha \hat{a}^\dagger-\alpha^* \hat{a}]$ (with field  operators $\hat{a}$ and $\hat{a}^\dagger$).
Without \cmmnt{the} force or radiation, we assume that the probe \cmmnt{can also be  affected by}\new{may also experience} a \old{known  phase shift, represented by a unitary operator} $\hat{\mathcal{R}}[\theta]=\exp[\mathrm{i} \theta \hat{n}]$ where $\hat{n}=\hat{a}^\dagger\hat{a}$.
We note that  \old{an unknown} \old{target} \old{mechanical} force generates a \old{change} in \old{the} momentum  as the primary effect, while the \old{known} phase rotation \old{angle $\theta$}  is a secondary effect \old{caused by} \old{the} simultaneous\old{ly occurring}  oscillator free evolution.
\old{Together, they create an unknown displacement} in phase space.
The rotation effect $\hat{\mathcal{R}}[\theta]$ can be estimated independently in advance to calibrate the interferometer for the displacement measurement. 
On the other hand, for microwave radiation, the \old{displacement} can \cmmnt{happen}\new{occur} simultaneously in both the position  and momentum quadratures of \old{the} electric field during free evolution.
Note that the qubit \old{may not be}  directly impacted by the external force, but only coupled to the disturbed field for \cmmnt{the} read-out, \cmmnt{which may add}\new{providing} flexibility \old{to} the setup design \old{in the choice of the qubit}.
In the main text, we focus on the estimation of \old{the} \old{arbitrary} displacement, \old{assuming that the \old{extra} pure phase rotation has been a priori estimated.} 
This \old{rotation estimation} is summarized in Appendix~\ref{sec:rotation}.

In  hybrid \old{quantum} systems~\old{\cite{xiang12,kotler16,MonroeRMP2021,Kurizki2015, Lachance-Quirion2020}},  e.g. a mechanical or microwave oscillator \old{coupled to} two-level atomic or transmonic system\old{s}, the probe and ancilla   of different nature \old{and/or dimensions} are \old{interacting and being controlled} for \cmmnt{the} estimation.
Hybrid interferometry \old{ of such systems} (as described in Appendix \ref{sec:signal}) can be studied within \old{the} RWA   at resonance, i.e. \old{exploiting} JC interaction with \old{Hamiltonian} $H^\mathrm{JC}_\mathrm{int}=\hat{\sigma}_+\hat{a}+\hat{\sigma}_-\hat{a}^\dagger$ \cite{ShoreKnight1993JC,FinkNature2008,schindler2013quantum,Casanova2010JC}, where  Pauli matrices acting on \old{the} qubit space are denoted as $\hat{\sigma}_j$ for $j=\mathrm{x},\mathrm{y},\mathrm{z}$ and $\hat{\sigma}_\pm\old{=\frac{\hat{\sigma}_\mathrm{x}\pm i \hat{\sigma}_\mathrm{y} }{2}}$. 
Recently, for both trapped ion\old{s}~\cite{KienzlerPRL2016,Bruzewicz2019APRtrappedion,FluhmannHomeNature2019} and superconducting microwave circuits~\cite{Touzard2019,SchoelkopfNature2020,Kwon2021Superconduting} among many other\old{s},  the Rabi coupling  $H_\mathrm{int}^\mathrm{Rabi}=\hat{\sigma}_\mathrm{x} \hat{X}_\theta$ \old{containing} anti-JC term  $H_\mathrm{int}^\mathrm{AJC}=\hat{\sigma}_+\hat{a}^\dagger+\hat{\sigma}_-\hat{a}$ \cmmnt{became}\old{has become} experimentally \old{accessible and} precisely controllable, exhibiting promising application\old{s} beyond RWA.
Here $\hat{X}_\theta=(\hat{a}\mathrm{e}^{\mathrm{i} \theta}+\hat{a}^\dagger \mathrm{e}^{-\mathrm{i} \theta})/\sqrt{2}$ is a generalized quadrature operator of   the oscillator with phase $\theta$, \old{for example the} position $\hat{X}=\hat{X}_{0}$ and \old{the} momentum $\hat{P}=\hat{X}_{\mathrm{\mathrm{\pi}}/2}$.
For optical experiments, \old{such a} coupling \cite{SpillerNJP2006}  \old{is} \old{currently} under development \cite{ParkNJP2020Rabi}.

\old{The} quantum  probe and ancilla states \old{as resources for sensing}  are assumed \old{here} to be initially    separable  for a fair assessment of the estimation power of the setups, \old{while correlated probe-ancilla states can be considered for future extensions}.
The \old{states }\old{of} \old{the} qubit ancilla \old{are assumed to} be  \old{
easily} prepared, manipulated, and measured with high precision as \cmmnt{is most often the case} in real experimental systems. 
\old{In contrast,} in the experimental platforms of trapped ions and superconducting cavity circuits, the  oscillator mode is not directly measurable, \old{but} only \old{indirectly} through the detection \old{of} the coupled qubit system.
Non-classical states of \old{the} probes can be considered as a realistic resource to further enhance the estimation \old{protocol}, as \old{such states have been} prepared for both trapped ions \cite{KienzlerPRL2016,Bruzewicz2019APRtrappedion,FluhmannHomeNature2019,HackerCatNatPho2019} and superconducting circuits \cite{Touzard2019,SchoelkopfNature2020,Kwon2021Superconduting}. 
However, for many \old{future schemes with solid-state optomechanical systems~\cite{yin13}},  the probe oscillator is \old{typically} in \old{non-squeezed} \old{Gaussian states at} thermal equilibrium, and preparation of pure states \old{requires} cooling \old{of the oscillator} to extremely low temperature.
In addition, the non-classical states do not bring beneficial effect\old{s} to \old{the} \old{estimation by the} interferometer, as will be explained later. 
\cmmnt{Therefore}\new{In this study}, we focus \cmmnt{our study} on readily accessible thermal probes and leave non-classical probes for future \cmmnt{investigation}\new{research}. 


\begin{figure*}[thp]
\includegraphics[width=\textwidth]{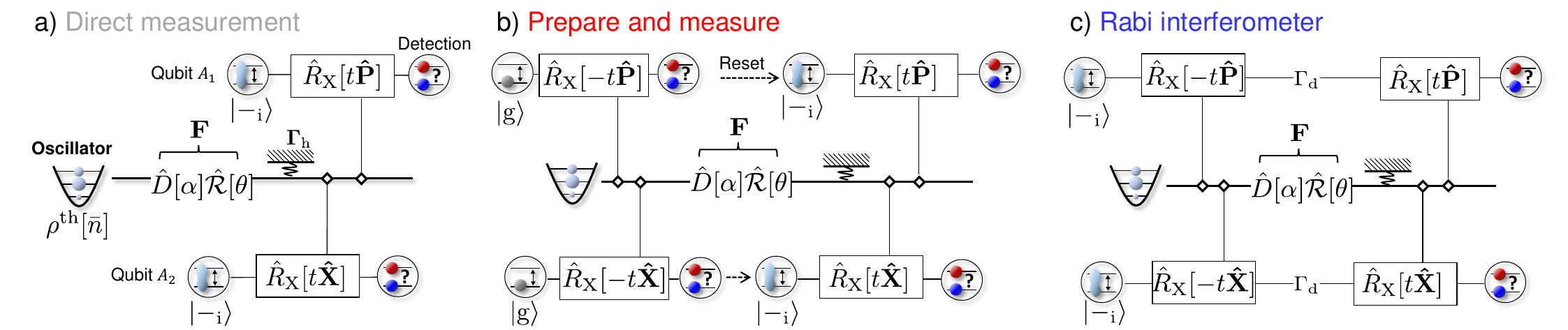}  
\caption{
Setups to \old{unambiguously} estimate a short-time external classical force $\mathbf{F}$ \old{with} an \old{unknown} direction \old{in phase space}  by simultaneous  estimation of  parameters $\alpha_\mathrm{r}=\operatorname{Re}(\alpha)$ and $\alpha_\mathrm{i}=\operatorname{Im}(\alpha)$ of displacement $\hat{D}[\alpha]$ using two sets of \old{Rabi} interactions $\hat{R}_\mathrm{x}[t \hat{X}]=\exp[\mathrm{i} t \hat{\sigma}_\mathrm{x} \hat{X}]$
 and $\hat{R}_\mathrm{x}[t \hat{P}]=\exp[\mathrm{i} t \hat{\sigma}_\mathrm{x} \hat{P}]$  and their inverses\cmmnt{ for   each displacement component}.
 The same setup can be used to independently estimate  the free-evolution phase $\theta$ of rotation in phase space $\hat{\mathcal{R}}[\theta]$ \old{as in Appendix \ref{sec:rotation}}.
  The oscillator is initially in a thermal state $\rho^\mathrm{th}[\bar{n}]$ with average \old{excitation} number $\bar{n}$, including the oscillator ground state at vanishing excitation limit $\bar{n}\to 0$.
 Oscillator heating $\Gamma_h$ \old{from coupling to a thermal environment} is considered as a threat to \old{all of the} estimation strategies\cmmnt{as a practical consideration}.
 a)  A  direct-measurement setup   using  Rabi interactions  
 and  qubit detectors only after the target displacement  \cmmnt{as a measurement device} (Rabi detector).
b) \old{A non-interferometric} setup for \old{a} prepare-and-measure strategy where \cmmnt{still only} \old{a} local state of the oscillator \old{(a compass state)} is \cmmnt{experiencing the effect of}\new{affected by} the force.
\cmmnt{Reset prepares a new state of the same two-level system.}\new{The qubit is reset before the detection.}
c) \old{A} Rabi interferometer where the ancillary qubits initially prepared in  the eigenstate of $\hat{\sigma}_\mathrm{y}$ become entangled to the oscillator \cmmnt{by}\new{through} Rabi interactions and remains \old{so}  throughout  \cmmnt{signal} encoding of the displacement and rotation,  to be measured by \old{a} Rabi detector made of \old{the inverse Rabi interactions and} \old{a} qubit detector.
\cmmnt{Here}\new{In this last setup,} the qubit dephasing $\Gamma_d$ \old{can} \old{additionally} deteriorate the estimation precision \old{due to the non-local nature of the protocol} as the qubit ancillary mode is \old{now} \cmmnt{participating}\new{involved} in the estimation. 
\new{The CFIs of these setups are listed in Table \ref{tab:CFItab}.}
The titles of the sub-figures \new{are color-coded to} \cmmnt{in colors} match \cmmnt{those of} the \new{corresponding} curves in Fig.~\ref{fig:measure-prepare}.
}
 \label{fig:RabiD}
\end{figure*}

  We consider \old{three}  \old{simple and comparable} \old{protocols} \old{exploiting} Rabi interactions  \old{as} depicted in Fig.~\ref{fig:RabiD}: \cmmnt{the} direct measurement,  \cmmnt{the} non-interferometric (prepare-and-measure), and \cmmnt{the} interferometric sensing setups.
   \old{For an unambiguous estimation of the displacement,} we assume that the rotation \old{$\hat{\mathcal{R}}[\theta]$} can be pre-estimated \old{separately} before the displacement\old{, as it requires a pre-displaced probe in the oscillator as \old{discussed} in Appendix \ref{sec:rotation}}.
\cmmnt{For the analysis, we assume} The unknown target displacement \new{is assumed} to be \cmmnt{weak and thus consider the}\new{in the} weak-displacement limit (WDL),  $|\A|\ll 1$.
These  protocols can equivalently \old{be tweaked to} estimate  the amplitude $|\alpha|$ and the phase $\arg(\alpha)$ simultaneously \old{by a different set of Rabi interactions}. 
The direct measurement scheme \old{in} Fig.~\ref{fig:RabiD}a uses a single-mode  probe  in a thermal state at various temperature\old{s} (including \old{oscillator} ground state) and \cmmnt{the} \old{Rabi} interactions  with the \old{qubit} ancillas \old{only} to measure the oscillator \old{after the applied force}.
 \cmmnt{Here} The \new{optimal} qubit state $\ket{\psi}_{\mathrm{A}_{1,2}}$ \cmmnt{can be optimized for the best performance, which} is found  to be  the eigenstate of $\hat{\sigma}_\mathrm{y}$ in the WDL.
 The non-interferometric setup in Fig.~\ref{fig:RabiD}b, \old{on the other hand,} uses \old{Rabi} interactions \new{for} both \cmmnt{as a part of the} preparation of the oscillator probe state as well as a part of the detection process.
The first set of Rabi interactions  prepares a superposition of coherent state\old{s}  in the oscillator, \old{while} the second set of interactions with the optimal qubit states prepared as the eigenstate of $\hat{\sigma}_\mathrm{y}$ and qubit detection in the energy eigenbasis comprises a Rabi detector.
The reset prepares two-level systems for the \cmmnt{second set}\new{Rabi detector}.
In contrast, \old{for} the interferometer  in Fig.~\ref{fig:RabiD}c the qubit ancillas  participate in \old{all stages of} the \old{setup} \old{from the preparation}, during the signal \old{accumulation, and}   up to the measurement.
\cmmnt{We stress that} The Rabi interferometer  is the main setup of interest, \old{while} the   \old{other}  schemes  are  considered mainly for comparison, to clarify \new{the role of interferometer and its non-locality in performance enhancement}\cmmnt{if and to what degree the interferometer and its non-locality are   crucial for the enhancement in the performance}.
\new{In Table \ref{tab:CFItab}, we summarized the CFIs and QFIs for the setups considered.  }

The degrees of freedom  in these setups \new{are}\cmmnt{comprise} the oscillator probe \cmmnt{in the}   $\mathrm{C}$, and the ancillary qubits  $\mathrm{A}_{j=1,2}$.
The initial ancillary qubit  can be  set \old{in any state} as  $\ket{\psi}_{\mathrm{A}_j}=c_\mathrm{e}\ket{\mathrm{e}}_{\mathrm{A}_j}+c_\mathrm{g}\ket{\mathrm{g}}_{\mathrm{A}_j}$ with  qubit  eigenbasis $\{\ket{\mathrm{e}}_{\mathrm{A}_j}, \ket{\mathrm{g}}_{\mathrm{A}_j}\}$, where the coefficients $c_{\mathrm{e},\mathrm{g}}\in \mathbb{C}$ can be chosen for the optimal performance \old{of estimation at all stages, and thus are known}. 
The initial state of the oscillator is  assumed \old{universally}  to be \old{in} a \old{naturally existing} thermal state probe $\rho^\mathrm{th}[\bar{n}]=\sum_{n=0}^\infty \frac{\bar{n}^n}{(\bar{n}+1)^{n+1}}\ket{n}_\mathrm{C}\bra{n}$,  with Fock \old{basis} $\ket{n}_\mathrm{C}$ and average \old{excitation} number $\bar{n}$ representing the various levels of the initial cooling for realistic experiments.
The \old{lowest energy} state \old{$\rho^\mathrm{th}[0]=\ket{0}_\mathrm{C}\bra{0}$}  for \old{the} microwave cavity field and \old{the phononic oscillator} is considered as an \new{ideal} limit when the cooling \old{in the preparation stage} is \cmmnt{ideal}\new{perfect}.

The CFI  with the  hypothetical  \cmmnt{$\hat{X}_\theta$-} quadrature  detector (denoted as $F$, defined in (\ref{eq:CFIdef})),  and the quantum Fisher information (QFI, \old{denoted as} $Q$, defined as (\ref{eq:QFIdef})), \old{of}  Gaussian state\old{s} (e.g. vacuum/ground state) \old{and  thermal state $\rho^\mathrm{th}[\bar{n}]$ } set the  benchmark in the estimation of  \old{single displacement variables $\A_\mathrm{r}=\operatorname{Re}(\alpha)$ and $\A_\mathrm{i}=\operatorname{Im}(\alpha)$ for $k=\mathrm{r},\mathrm{i}$:} 
\begin{align}
Q_{k}[\rho^\mathrm{th}[\bar{n}];\alpha_{k}]=F^{x_\theta}_{k}[\rho^\mathrm{th}[\bar{n}];\alpha_{k}]=4/(1+2\bar{n})
\label{eq:sql}
\end{align}
 \old{(classical benchmark)~\cite{TerhalPRA2017GKP}}, also \old{related to the} standard quantum limit.
 \cmmnt{We  see that it has a lower value}\new{They are low} when cooling is not perfect (i.e. $\bar{n}>0$), even though the number of \old{particles in the probe} is larger than the vacuum/oscillator ground state. 
In most experimental realizations, the oscillator cannot be directly measured, and \cmmnt{under this condition, we are forced to use an indirect detection scheme for the estimation}\new{ and thus an indirect detection scheme is used}, which \cmmnt{inevitably} leads to a smaller \old{CFI} than the \cmmnt{above-stated } classical benchmark.

We note that utilizing an input squeezed state in the oscillator \cite{Hastrup2020squeezing} \cmmnt{in the setups to be introduced} can \cmmnt{have a QFI and CFI larger than the classical benchmark in}\new{improve} the estimation of one displacement component $\A_\mathrm{r}$, but \cmmnt{smaller in }\new{negatively affect the} estimation of \old{the} conjugate component $\A_\mathrm{i}$, and thus it does not \cmmnt{help}\new{aid} in the simultaneous estimation of both \new{components}.

\begin{widetext}
\begin{table}[ht] \centering \caption{Summary of the CFIs and QFIs.}
 \resizebox{\textwidth}{!}{
\begin{tabular}{|c|c|c|} \hline Setup & CFI & QFI  \\ \hline 
Direct measurement  (Fig. \ref{fig:RabiD}a) & $\frac{8 t^2 \cos ^2\left(2 \sqrt{2} t \alpha _k\right)}{ \mathrm{e}^{2 t^2 \left(2 \bar{n}+1\right)}-\sin^2 \left(2 \sqrt{2} t \alpha _k\right)}$  (\ref{eq:CFIthermalDirect}) 
& $\frac{4}{1+2\bar{n}}$ (\ref{eq:sql})
\\ \hline Prepare and measure  (Fig. \ref{fig:RabiD}b)& $\frac{8  t^2 \cos ^2\left[2 \sqrt{2} t \alpha _k\right]}{4 \left(1\pm \mathrm{e}^{-t^2 \left(1+2 \bar{n}\right)}-\mathrm{e}^{-2 t^2 \left(1+2 \bar{n}\right)}\pm \mathrm{e}^{-3 t^2 \left(1+2 \bar{n}\right)}\right)^{-2}-\sin ^2\left[2 \sqrt{2} t \alpha _k\right]}$ (\ref{CFIPnMnonzero})
& $\frac{ \left(8 t^2+4\right) \mathrm{e}^{ t^2 \left(2 \bar{n}+1\right)}+8 \bar{n} \left( \mathrm{e}^{ t^2 \left(2 \bar{n}+1\right)}-1\right)-4}{ \mathrm{e}^{ t^2 \left(2 \bar{n}+1\right)}-1}$ (\ref{eq:PnMQFI})
\\\hline Rabi interferometer  (Fig. \ref{fig:RabiD}c) & $\frac{8 t^2  \cos ^2\left(2 \sqrt{2} t \alpha _k\right)}{4\left(2-p(1+\mathrm{e}^{-4 t^2})\right)^{-2} - \sin ^2\left(2 \sqrt{2} t \alpha _k\right)}$ (\ref{eq:CFIDepy})
& $8 t^2+\frac{4}{1+2\bar{n}}$ (\ref{eq:QFIsim})
\\ \hline
\end{tabular}
}
\label{tab:CFItab}
\end{table}
\end{widetext}

\old{\subsection{Direct measurement}}

\cmmnt{We first consider the  setup in Fig.~\ref{fig:RabiD}a which we refer to as the direct measurement approach. 
In this case, Rabi couplings between the probe and the ancillas are only applied during the actual measurement. }
\new{The direct measurement approach in Fig.~\ref{fig:RabiD}a is used for displacement estimation in a system where Rabi couplings are only applied during measurement.}
At \cmmnt{the} resonance between the oscillator probe and \old{the} ancillary qubit \old{modes}, \old{two non-commuting} unitary Rabi interactions    $\hat{R}_\mathrm{x}[t\hat{X}]=\exp[\mathrm{i} t \hat{\sigma}_\mathrm{x} \hat{X}]\equiv \hat{R}_\mathrm{X}$ and $\hat{R}_\mathrm{x}[t\hat{P}]=\exp[\mathrm{i} t \hat{\sigma}'_\mathrm{x} \hat{P}]\equiv \hat{R}_\mathrm{P}$ \old{(prime denotes the second qubit)}  with \old{dimensionless} strength $t$ \old{(a product of the interaction time and the interaction strength)}  \old{are applied} \old{to} transfer  \cmmnt{the} information about the displacement to the qubit detector. 
\old{Using} this direct-measurement \old{approach}, we can \old{jointly estimate the  components ($\alpha_\mathrm{r}$ and $\alpha_\mathrm{i}$) of the complex displacement parameter}. 
Each \old{ancilla coupling} corresponds to a  \old{detection module} for the estimation of $\alpha_\mathrm{r}$ ($\alpha_\mathrm{i}$)  exploiting $\hat{R}_\mathrm{X}$ ($\hat{R}_\mathrm{P}$).
\new{This scheme is similar to the estimation setups in \cite{TerhalPRA2017GKP}, which described the displacement estimation protocol using so-called Gottesman-Kitaev-Preskill (GKP) states.
A preliminary numerical study suggests that estimation using GKP probes may have a slight enhancement over vacuum probes.
}

We assume here  no prior knowledge  about the target parameters $\A_{k=\mathrm{r},\mathrm{i}}$, and  \old{therefore,} we use a fixed \old{ancillary} qubit eigenstate  of $\hat{\sigma}_\mathrm{y}$ with eigenvalue $\pm1$: 
\begin{align}
\ket{\phi^\mathrm{opt}}_\mathrm{A}=2^{-1/2}(\ket{\mathrm{e}}_\mathrm{A}\pm \mathrm{i}\ket{\mathrm{g}}_\mathrm{A})=\ket{\pm_\mathrm{i}}_\mathrm{A},
\label{eq:optqub}
\end{align}
which gives the largest values of  \old{the}  CFI \old{among all qubit states} \old{in}  the WDL. 
\cmmnt{In a setup with a fixed qubit state,} The CFI of $\A_\mathrm{\new{k}}$ \old{by the qubit detection \old{in $\hat{\sigma}_\mathrm{z}$-eigenbasis} $\{\ket{\mathrm{e}}_{A_1}\bra{\mathrm{e}}, \ket{\mathrm{g}}_{A_1}\bra{\mathrm{g}}\}$} depends \old{generally} on the \old{actual} value of $\alpha_\mathrm{\new{k}}$,  \old{but} approaches its maximum in \old{the} WDL. 
This behavior is different from the constant \cmmnt{behavior of} QFI \old{using} a thermal probe for all $\A_k$ \old{as in (\ref{eq:sql})}, which can be approached only with the prior knowledge of the target value $\A_k$.  
\new{
In the presence of noise and beyond WDL, the qubit basis may need to be adjusted adaptively based on the accumulated data, as in Appendix \ref{sec:adaptive}. 
}
When a thermal probe \old{and the qubit state $\ket{\phi^\mathrm{opt}}$} are used \old{as the input and the ancilla}, the CFI \old{is reduced for a  non-zero $\bar{n}$ \old{for both $k=\mathrm{r},\mathrm{i}$}:}
\begin{align}
    F_{k}^\mathrm{direct}[\rho^\mathrm{th}[\bar{n}];\A_k,t]=\frac{8 t^2 \cos ^2\left(2 \sqrt{2} t \alpha _\old{k}\right)}{ \mathbb{A}_\mathrm{th}-\sin^2 \left(2 \sqrt{2} t \alpha _\old{k}\right)}    \stackrel{\alpha_\old{k}\ll 1}{\approx}8 t^2\left(\mathbb{A}_\mathrm{th}^{-1}-8 t^2 \A_k^2 \frac{\mathbb{A}_\mathrm{th}-1}{\mathbb{A}_\mathrm{th}^2}\right) , \label{eq:CFIthermalDirect}
\end{align}
\old{as} only the thermal effect $\mathbb{A}_\mathrm{th}=\mathrm{e}^{2 t^2 \left(2 \bar{n}+1\right)}$ is \old{monotonously} increasing  \old{with} $\bar{n}$.
The maximum value of this CFI is \new{found} at the optimal Rabi strength $t_\mathrm{opt}=(4 \bar{n}+2)^{-1/2}$ \new{giving the extremal point of CFI} as
\begin{align}
&F_{k}^\mathrm{direct}\left[\rho^\mathrm{th}[\bar{n}];\A_k,t_\mathrm{opt}\right]\stackrel{\alpha_k\ll 1}{\approx}\frac{4}{\mathrm{e} \left(1+2 \bar{n}\right)}+\frac{16 \alpha _k^2}{\mathrm{e}^2 \left(1+2 \bar{n}\right)^2}-\frac{16 \alpha _k^2}{\mathrm{e} \left(1+2 \bar{n}\right)^2}
\label{eq:CFIdirectmax}
\end{align}  
\old{in the} WDL. 
It \cmmnt{establishes}\new{sets} a practical benchmark for \old{the} following schemes.

The CFI beyond \old{the} WDL    is also an important measure of performance. 
\cmmnt{We remark that} The CFIs \cmmnt{for all examples} in this work do not have \cmmnt{the} odd-order term in $\A_k$  due to the \new{sign} symmetry\cmmnt{in the sign of the signal}, and thus the second-order term is the lowest order dependence on $\A_k$, \old{as is evidenced in (\ref{eq:CFIthermalDirect})}.  
The rate \cmmnt{by}\new{at} which a CFI changes with $\alpha_k$  can be found \cmmnt{by}\new{from} the normalized curvature in the WDL as 
\begin{align}
\mathbb{C}=-\left.\dfrac{\partial^2 F_{k}[\A_k]}{\partial \A_k^2 }\right|_{\A_k=0}  (2 F_k[\A_k=0])^{-1}.
\end{align}
For the CFI in (\ref{eq:CFIthermalDirect}), \old{the rate is found to be} $\mathbb{C}=8t^2$ for a large Rabi strength $t\gg 1$ \old{exhibiting} a narrower dynamic range for a large $t$ regardless of $\bar{n}$. 
 The half-width-at-half-maximum \old{(HWHM)} of CFI  \old{is a measure of} the dynamic range, given by
 \begin{align}
 \alpha_k^\mathrm{(HWHM)}=\frac{\arcsec\left[2 \mathbb{A}_\mathrm{th}-1\right]}{4 \sqrt{2} t}\stackrel{t\gg 1}{\to}\frac{\mathrm{\mathrm{\pi}}}{8\sqrt{2}t}.
 \label{eq:HWHM1}
 \end{align}
 The CFI in (\ref{eq:CFIthermalDirect}) is zero at multiple values of $\A_k$, and (\ref{eq:HWHM1}) is located at  the half of the one closest to $0$ at $\A_k=\frac{\mathrm{\mathrm{\pi}}}{4\sqrt{2}t}$.
 \new{We note that the deviation of HWHM from that predicted by the curvature is relatively small ($\lesssim 5\%$). }
At the optimal strength $t=t_\old{\mathrm{opt}}$, it \old{becomes} $\A_k^\mathrm{\old{(HWHM)}}\approx 0.45\sqrt{1+2\bar{n}}$, \new{showing a broader dynamic range when $\bar{n}$ is larger}.
The product of the maximum CFI \old{at WDL} in (\ref{eq:CFIthermalDirect}) and the square of the HWHM in (\ref{eq:HWHM1}) \cmmnt{to} \new{that can} be compared with other schemes as a way to show the trade-off relation \new{between maximum CFI and dynamic range} is given as $ \frac{\mathbb{A}_\mathrm{th}^{-1}}{4} \arcsec^2 \left[2 \mathbb{A}_\mathrm{th}-1\right] \stackrel{t\gg 1}{\approx} \frac{\mathrm{\pi}^2}{16}\mathrm{e}^{-2t^2(2\bar{n}+1)}$, asymptotically approaching $0$.
It shows that an increase in maximum CFI is always \cmmnt{compensated}\new{offset} by a smaller dynamic range.
At optimal strength $t=t_\mathrm{opt}$ this product is given as a constant $\frac{\arcsec^2[2\mathrm{e}-1]}{4\mathrm{e}}$ regardless of $\bar{n}$.
 

 \cmmnt{An \old{encompassing figure}  of merit that reflects both the dynamical range and the maximum precision is the average CFI.}
 \new{The average CFI is a figure of merit that combines the dynamical range and maximum precision.}
As the CFI is periodic \old{in $\A_k$} due to the qubit nature of the detection, \cmmnt{we can average out the CFI in}\new{it can be averaged over}  a single period of $\A_k\in \left[-\frac{\mathrm{\pi}}{4\sqrt{2}t},\frac{\mathrm{\pi}}{4\sqrt{2}t}\right]$, which gives 
\begin{align}
F^\mathrm{(av)}_{k}[\rho^\mathrm{th}[\bar{n}];t]=\frac{8 t^2}{\mathbb{A}_\mathrm{th}+\sqrt{\mathbb{A}_\mathrm{th}^2-\mathbb{A}_\mathrm{th}}}.
\label{eq:CFIthermalDirectav}
\end{align}
\old{This is} reduced by \old{increasing} $\bar{n}$ \old{and $t$} \old{beyond the optimal strength $t_\mathrm{opt'}[\bar{n}]$ different from $t_\mathrm{opt}$}, asymptotically approaching $0$, as in  Fig.~\ref{fig:DirectThermal}a.
\old{These results indicate that} a non-ideal cooling significantly reduces the CFI of  the direct measurement strategy.
\new{In Bayesian quantum metrology, the problem of dynamical range is inherently incorporated into the model, unlike in the Fisher information approach, allowing operationally meaningful statements about the achievable limits in practice~\cite{GoreckiPRL2019}.}


\new{The problem of displacement estimation in an unknown direction has symmetry with respect to phase space rotation, but the proposed schemes break this symmetry by measuring $\alpha_r$ and then $\alpha_i$, leading to different formulas for each. 
}
When \cmmnt{the two parameters,}\new{both} $\alpha_\mathrm{r}$ and $\alpha_\mathrm{i}$, are simultaneously estimated (as in Fig. \ref{fig:RabiD} a), the CFI for $\alpha_\mathrm{r}$ is still given by (\ref{eq:CFIthermalDirect}), while for $\alpha_\mathrm{i}$ it acquires a modulation factor of $\cos^2\left[2 t^2\right]$: 
\begin{align}
F_\mathrm{i}^\mathrm{direct}[\rho^\mathrm{th}[\bar{n}];\A_\mathrm{i},t]\stackrel{t\gg 1}{\rightarrow}  \old{\cos ^2\left[2 t^2\right] \frac{8 t^2 \cos ^2\left(2 \sqrt{2} t \alpha _\mathrm{i}\right)}{ \mathbb{A}_\mathrm{th}-\sin^2 \left(2 \sqrt{2} t \alpha _\mathrm{i}\right)}.}
\end{align}
This modulation factor \cmmnt{is arising}\new{arises} from the interference of the two non-commuting Rabi interactions \old{yielding an} additional factor $\mathrm{e}^{\pm i t^2/2}$ in the coefficients of the hybrid entangled state \old{before the qubit detection}.
At $t^\mathrm{c}=\sqrt{c\mathrm{\pi}/2}$\old{, where} $c\in\mathbb{Z}$, \old{the modulation factor attains its maximum value} and the CFI \old{becomes} equal to that of the independent estimation.
\old{However, due to the non-matching $t^c$ and $t_\mathrm{opt}$, the direct measurement strategy will  suffer either from interference or sub-optimal precision.}
\new{If we adopt a symmetric scheme achieved by a trotterization of alternating two types of weak Rabi interactions such as $\lim_{N\to \infty}\left(\exp[\ii \frac{t}{N}\hat{\sigma}_x \hat{X} ]\exp[\ii \frac{t}{N}\hat{\sigma}'_x \hat{P} ]\right)^N=\exp[\ii t\hat{\sigma}_x \hat{X} +\ii t\hat{\sigma}'_x \hat{P} ]$,  the estimation of both displacement components could give the same performance asymptotically. }
Alternatively, we can actively counteract this redundant factor with a  two-qubit entangling interaction $\exp[-\mathrm{i} t^2 \hat{\sigma}_\mathrm{x} \hat{\sigma}_\mathrm{x}']$ \old{before or after the Rabi interaction, generating a proper qubit entanglement}.
This interaction can be engineered using \cmmnt{by} a geometric-phase effect of a set of Rabi interactions as
\begin{align}
    \exp[-\mathrm{i} t^2 \hat{\sigma}_\mathrm{x} \hat{\sigma}_\mathrm{x}']&=\exp[-\mathrm{i} \frac{t}{\sqrt{2}}\hat{\sigma}_\mathrm{x}  \hat{X}]\exp[-\mathrm{i} \frac{t}{\sqrt{2}}\hat{\sigma}'_\mathrm{x}  \hat{P}]\exp[\mathrm{i} \frac{t}{\sqrt{2}}\hat{\sigma}_\mathrm{x}  \hat{X}]\exp[\mathrm{i} \frac{t}{\sqrt{2}}\hat{\sigma}'_\mathrm{x}  \hat{P}]. \label{eq:twoqubit}
\end{align}
\cmmnt{for any qubit-oscillator state.  }




\subsection{Non-interferometric prepare-and-measure setup }
\label{subsec:PnM}

An advanced \old{non-interferometric} estimation protocol \old{is presented}  in Fig.~\ref{fig:RabiD}b. 
In this  \new{prepare-and-measure} scheme, Rabi interactions with ancillary qubits \old{$\mathrm{A}_1$ and $\mathrm{A}_2$} are used both for the \old{conditional} oscillator-probe state preparation and \old{detection}\cmmnt{, i.e. a prepare-and-measure strategy}.
 The individual estimation process of the \old{real component of the displacement parameter} $\alpha_\mathrm{r}$  described below  \old{is  identical to the estimation process of $\alpha_\mathrm{i}$  \old{when they are estimated separately}  \cmmnt{for which the Rabi interactions of}\new{using} $\hat{R}_\mathrm{P}$ and  $\hat{R}_\mathrm{P}^\dagger$ \cmmnt{are used} instead of $\hat{R}_\mathrm{X}$ and $\hat{R}_\mathrm{X}^\dagger$}.

The estimation process is made of three stages: preparation, encoding, and measurement.
In \cmmnt{the} preparation, Rabi interactions and postselection \new{using} \cmmnt{by the} subsequent qubit detectors \cmmnt{enable}\new{generate} the generation of coherent state superpositions \cmmnt{which}\new{that} are highly sensitive to \cmmnt{the} displacements~\cite{Zurek2001Compass,munroPRA2001cat}.
The optimal state is an odd balanced superposition of coherent states, as \cmmnt{a result of its large}\new{it has high} quantum coherence \new{as} measured by the total Wigner function negativity~\cite{AlbarelliPRA2018}.
\cmmnt{For the preparation of}\new{To prepare} such probe states, \cmmnt{we   take $\ket{\mathrm{g}}_{\mathrm{A}_{1,2}}$ as the input ancillary\old{} states of both  qubits}\new{input ancillary states $\ket{\mathrm{g}}_{\mathrm{A}_{1,2}}$ are used}. 
\cmmnt{After the \old{application of the} Rabi interactions  and the \old{probabilistic} projection onto the qubit excited state $\ket{\mathrm{e}}_{\mathrm{A}_{1,2}}\bra{\mathrm{e}}$\old{, we get}}\new{The application of Rabi interactions and probabilistic projection onto the qubit excited state $\ket{\mathrm{e}}_{\mathrm{A}_{1,2}}\bra{\mathrm{e}}$ results in}  a  superposition of four coherent state\old{s} \old{(known as a compass state~\cite{ZurekNature2001}). }
\cmmnt{If a projection onto the qubit ground state $\ket{\mathrm{g}}_{\mathrm{A}_{1,2}}\bra{\mathrm{g}}$ occurs in any one of the two qubit detection, 
the estimation precision is  \old{slightly} \old{reduced}.}
\new{A projection onto the qubit ground state $\ket{\mathrm{g}}_{\mathrm{A}_{1,2}}\bra{\mathrm{g}}$ prepares a state with a slightly reduced estimation precision.}
\cmmnt{The remaining part of the setup - encoding and detection - is identical to the setup for the direct estimation discussed in the previous section.
Interestingly, the prepare-and-measure and direct measurement setups require only sequential interactions with the qubit ancilla achievable by a single ancilla and therefore give an advantage in the experimental implementation.}
\new{Encoding and detection are the same in both prepare-and-measure and direct measurement setups and only require sequential interactions with a single ancilla, making it easier to implement experimentally.}

The  CFI  \old{for the independent estimation} of the two parameters using a thermal state \old{probe} can be found as (see Appendix~\ref{sec:PnMFis} \new{for details}) 
\begin{align}
   & F_{k}^\mathrm{PnM}[\rho^\mathrm{th}[\bar{n}];\A_k,t]=\frac{8  t^2 \cos ^2\left[2 \sqrt{2} t \alpha _k\right]}{4 \mathbb{B}_\pm-\sin ^2\left[2 \sqrt{2} t \alpha _k\right]}\stackrel{\mathrm{WDL}}{\approx}\frac{2 t^2}{\mathbb{B}_\pm}-\frac{4 (4 \mathbb{B}_\pm-1) t^4 \alpha _k^2}{\mathbb{B}_\pm^2},  
 \label{CFIPnMnonzero}
\end{align}
where the non-monotonous thermal effect is \cmmnt{for the prepare-and-measure strategy} $\mathbb{B}_\pm=\left(1\pm \mathrm{e}^{-t^2 \left(1+2 \bar{n}\right)}-\mathrm{e}^{-2 t^2 \left(1+2 \bar{n}\right)}\pm \mathrm{e}^{-3 t^2 \left(1+2 \bar{n}\right)}\right)^{-2}\stackrel{t\gg 1}{\to}1$. 
Here the signs $\pm$ refer to \old{those} of the superposition of coherent states that depend on the outcome of the qubit detection at the preparation stage.
\old{This thermal effect can be saturating faster for a large $\bar{n}$ compared to the ground state as $\mathbb{B}_\pm\stackrel{\bar{n}\gg 1}{\to}1$.}
The CFI in the \old{WDL} \old{in (\ref{CFIPnMnonzero})} has a non-monotonous behavior against $t$ as shown in Fig.~\ref{fig:measure-prepare}a,b.
\old{An analysis of such a non-monotonous behavior  is summarized in Appendix \ref{sec:non-monotonocity}.}
\cmmnt{We note that} This CFI \old{of the prepare-and-measure strategy} \old{in (\ref{CFIPnMnonzero})} \cmmnt{has a larger value}\new{is larger} than the maximal CFI of the direct measurement strategy for \old{strengths} $t\gtrsim 1.21$.
 The average CFI \old{of the prepare-and-measure strategy} \cmmnt{is found to}  \old{$F^\mathrm{(av)}_{k}[\rho^\mathrm{th}[\bar{n}];t]=\old{8 t^2-\frac{4 \sqrt{4 \mathbb{B}_\pm-1} t^2}{\sqrt{\mathbb{B}_\pm}}}\stackrel{t\gg 1}{\to}4(2-\sqrt{3})t^2$} \cmmnt{, and} is larger than the maximum \old{average} CFI by the direct measurement strategy for $t\ge 0.879$.
 \old{The normalized curvature is 
$\old{\mathbb{C}=\frac{2 (4 \mathbb{B}_\pm-1) t^2 }{\mathbb{B}_\pm}\stackrel{t\gg 1}{\to} 6t^2}$, slightly lower than \cmmnt{what was obtained for}\new{that of} the direct-measurement approach, \old{implying a wider dynamic range}.}
\old{The} HWHM of the CFI is given \old{asymptotically} as $\A_k\old{^\mathrm{(HWHM)}}\stackrel{t\gg 1}{\approx}\old{\frac{\arctan \left[\frac{2}{\sqrt{3}}\right]}{2 \sqrt{2} t}}\cmmnt{\stackrel{t\gg 1}{\approx}\old{1.1\frac{\mathrm{\pi}}{8\sqrt{2}t}}}$,  slightly increased compared to the direct measurement case.
This implies that the dynamical range is slightly extended as well.
For small displacements, a product of the square of HWHM of CFI and Fisher information in (\ref{CFIPnMnonzero}) is a constant in the asymptotic $t$ limit given approximately as $\frac{1}{4} \arctan^2 \left[\frac{2}{\sqrt{3}}\right]\cmmnt{\approx 1.21 \frac{\pi^2}{64}}$; therefore, they form a trade-off between the \old{maximum CFI} and the dynamical range, \old{in contrast to monotonously decreasing product of the direct measurement strategy.}

\new{The estimation protocol remains robust even with a thermal input state,}
 as the asymptotic expression for the CFI for $t\gg 1$ 
 is given by 
 $2t^2$ regardless of \new{the mean photon number} $\bar{n}$ (see Fig.~\ref{fig:DirectThermal} b).
\cmmnt{This observation implies that the oscillator does not have to be pre-cooled \old{before} the estimation protocol to  access the \old{asymptotic} quadratic trend.}\new{The input state does not need to be pre-cooled to achieve this trend.} 
   The dependence of the modulation of CFI on $\A_k$ is the same as in (\ref{CFIPnMnonzero}), \old{linked to the robustness to the initial thermal noise}.
   \cmmnt{We can interpret this robustness as being born out of the robustness of the qubit detectors to excess noise.}\new{This robustness may be due to the qubit detectors being resistant to excess noise.}   
This \cmmnt{can be partially evidenced by considering}\new{ is further supported by the observation that the CFI for} a coherent state $\ket{\beta}$ \cmmnt{, whose CFI} is asymptotically given by $ F^\mathrm{PnM}_{\alpha_k}[\ket{\beta}\bra{\beta};t]\stackrel{t\gg 1}{\approx}2t^2$ regardless of $\beta$. 
\cmmnt{Therefore, all the superpositions and mixtures of them \new{may} have the same scaling \new{in the cases considered, even though such an inference is not true in general}.
We also numerically verified that the \old{same} scaling is obtained when states from the Fock basis are randomly mixed into the probe.}\new{Random mixtures of states from the Fock basis also exhibit the same scaling.}
\new{GKP states can be actually generated by several Rabi interactions \cite{Hastrup2019GKP}, and estimation by GKP probes with a Rabi detector can be seen as a generalization and improvement of the prepare-and-measure method. }

The QFI for  thermal states \new{after the first interaction and qubit detection} is given by (see Appendix \ref{sec:quaddet})
\begin{align}
&Q_{k}^\mathrm{PnM}[\rho^\mathrm{th}[\bar{n}];\A_k,t]&\nonumber\\&=\frac{ \left(8 t^2+4\right) \mathbb{A}_\mathrm{th}^{1/2}+8 \bar{n} \left( \mathbb{A}_\mathrm{th}^{1/2}-1\right)-4}{ \mathbb{A}_\mathrm{th}^{1/2}-1}\stackrel{\bar{n}\ll 1}{\approx} 4+8 t^2+\frac{8 t^2}{\mathrm{e}^{t^2}-1}.
\label{eq:PnMQFI}
\end{align}
\new{This QFI is different from (\ref{eq:sql}) due to the additional preparation step before the signal displacement.
}
This is independent of $\A_k$ in contrast to the CFI, as the optimal detector may implicitly use the knowledge \cmmnt{about}\new{of} the true value of the estimation target.
\cmmnt{Interestingly,} This QFI \cmmnt{is increasing}\new{increases} with the average boson number $\bar{n}$ without a bound \new{as $8\bar{n}$}\cmmnt{.
In addition, it has a larger value}\new{, and is higher} than the QFI of the Rabi interferometer in (\ref{eq:QFIsim}).
This result implies a prospect of a better detection scheme that can surpass the Rabi interferometer.

Now if we estimate $\A_{r,i}$ simultaneously, we arrive at a more complicated form of the CFI \old{(see (\ref{eq:CFIPnMSimul}) in Appendix \ref{sec:PnMFis}} than \old{those of} the individual \old{component} estimations in (\ref{CFIPnMnonzero}), as each displacement component interferes with the estimation of the other\cmmnt{ component}.
 The difference \old{between individual and simultaneous estimation} is \cmmnt{compared}\new{shown} in Fig.~\ref{fig:simultPnM} \old{of Appendix \ref{sec:PnMFis}}.
 \cmmnt{We note that} The CFI of $\A_\mathrm{r}$ has again an asymptotic modulation factor $\cos^2[2t^2]$  arising from the interference of two non-commuting Rabi interferometers, which is universal for all true values of $\A_\mathrm{r}$.  
This factor again disappears at $t^\mathrm{c}=\sqrt{c\mathrm{\pi}/2}$, \old{or can be actively cancelled by (\ref{eq:twoqubit})} \old{or a proper qubit entangled state preparation}.
\new{Again, by trotterizing two types of weak Rabi gates, a more symmetric displacement estimation can be achieved.}
 

  
 

\subsection{Rabi interferometer}


\cmmnt{We finally consider} \new{In} the interferometric estimation scheme illustrated in Fig.~\ref{fig:RabiD} c), \cmmnt{where} qubit  ancilla are interacting  \old{unitarily} with \old{the} oscillator  \old{by Rabi couplings} before and after  the unknown displacement (\cmmnt{and/}or phase rotation)  \old{before} the \old{final qubit} detection. 
\cmmnt{For a fixed and known phase-space variable, the Rabi interferometric scheme was already carried out experimentally and thus proves its feasibility}\new{The feasibility of the scheme for a fixed, known phase-space variable has been demonstrated experimentally }~\cite{HempelNatPho2013catspectroscopy}.
\cmmnt{In what follows,} We first describe the scheme for the estimation of the individual parameters and then move on to the scheme for the simultaneous estimation of the parameters.
\old{The} inverse Rabi interaction $\hat{R}_\mathrm{X}^{-1}$ \old{or $\hat{R}_\mathrm{P}^{-1}$} applied \old{before} the signals  can be  engineered by either $\mathrm{\pi}$-phase rotations \old{of the oscillator}, or by using additional Rabi interactions with \old{a} strong drive at \old{the} opposite phase as discussed in Appendix~\ref{sec:inverseRabi}.
The total unitary transformation of the interferometer  
up to the qubit detection is \old{described} as
\begin{align}
&\old{\hat{U}_\mathrm{interf}}=\old{\hat{R}_\mathrm{P}}\hat{R}_\mathrm{X} \hat{D}[\alpha]\hat{R}_\mathrm{X}^{-1}\old{\hat{R}_\mathrm{P}^{-1}}\nonumber\\
&=\hat{D}[\alpha]\hat{R}_\mathrm{x}[\sqrt{2}\alpha_\mathrm{r} t]\old{\hat{R}'_\mathrm{x}[\sqrt{2}\alpha_\mathrm{i} t]},
\label{eq:totalaction}
\end{align}
and is composed of a \old{signal} oscillator displacement  and  qubit rotations $\hat{R}_\mathrm{x}[\phi]=\exp[\mathrm{i} \phi \hat{\sigma}_\mathrm{x}]$ whose angles are proportional to the displacement components $\alpha_\mathrm{r}$ \old{and $\alpha_\mathrm{i}$}. 

The CFI associated with the estimation of $\alpha_k$ \cmmnt{is found to \old{be}} (see Appendix \ref{sec:CFIRabi}) \new{is}
\begin{align}
F^\mathrm{interf}_{k}[\rho^\mathrm{th}[\bar{n}];\A_k,t]
=8 t^2,
\label{eq:Freal}
\end{align}
surpassing the \cmmnt{prefactor of the} \old{asymptotic} scaling $2t^2$ of the non-interferometric setups \old{in (\ref{CFIPnMnonzero})}. 
\cmmnt{Note that the} \new{This} scaling \cmmnt{in Eq.~(\ref{eq:Freal})} holds \cmmnt{regardless of the true}\new{any} value of $\alpha_k$, \cmmnt{. 
This aspect overcomes all limiting} \new{and eliminates} trade-offs between CFI and measurement range for the direct and prepare-and-measure method in the absence of noise. 
Moreover, it \cmmnt{beats}\new{surpasses} the maximal CFI for the direct measurement strategy for $t=0.429$, the benchmark \old{of (\ref{eq:CFIdirectmax})} at $t=0.707$, and the prepare-and-measure strategy for all $t$.    
It approaches asymptotically for large $t$   the scaling of practically inaccessible CFI in  Eq.~(\ref{eq:PnMCFIquadrature}) \old{that uses the quadrature detection on the oscillator,} \old{or} QFI  \old{in (\ref{eq:QFIsim})} \old{that may require infeasible  detectors}.
\new{In addition, the interferometer works equally for all states in the oscillator in the absence of imperfections, and thus preparation of a complex non-Gaussian states such as GKP states is not needed.}
\old{Remarkably,} the CFI in eq.~(\ref{eq:Freal}) holds for any non-pure probe state, $\rho$, which is  \old{initially} separable \old{from the} input qubit e.g.  a thermal state $\rho^\mathrm{th}[\bar{n}]$.

\old{The CFI in (\ref{eq:Freal}) guarantees in principle no sensitivity bias within the dynamical range of the Rabi interferometer.}
When the data is finite, the dependence on $\A_k$ appears around the values $\A_k=\pm\frac{\pi}{4\sqrt{2}t}$ due to the statistical fluctuation or noise (\cmmnt{which will be more} visible on Fig. \ref{fig:measure-prepare}).
This independence of CFI to $\A_k$ arises due to the implied local estimation method.
\new{However,} there exist\old{s} one practical limitation resulting from the periodicity of the qubit rotations $\hat{R}_\mathrm{x}[\sqrt{2}\alpha_k t]$ \cmmnt{over $\alpha_k$ and thus the \old{probabilities} $P_g(\alpha_k)$ and $P_e(\alpha_k)$ \old{if a global estimation of arbitrary strength displacement $\A_k$ is tried}. 
This periodicity}\new{that} narrows the range   \old{$\alpha_k\in[-\frac{\mathrm{\pi}}{2\sqrt{2}t},\frac{\mathrm{\pi}}{2\sqrt{2}t}]$} in which unique estimation can be performed.
\cmmnt{The higher precision in estimation is therefore achieved at a cost of \old{a} smaller dynamical range. This type of trade-off between  precision and range of estimation is \old{a} rather common trait in many sensing protocols (see e.g. \cite{Bonato2016, DegenRMP2017,HerbschlebNatComm2021DynamicRange}). We note that  \old{estimation} of \old{the} target values beyond this range is possible by combining two setups with slightly different coupling strengths. For an example, see Appendix  \ref{sec:AssymmetricStrength}.}
\new{This leads to a trade-off between precision and range of estimation, a common trait in many sensing protocols~\cite{Bonato2016, DegenRMP2017,HerbschlebNatComm2021DynamicRange}. 
Estimation beyond this range is possible by combining two setups with different coupling strengths (see Appendix~\ref{sec:AssymmetricStrength}).}


 \cmmnt{Only} For comparison, \cmmnt{we also compute} the QFI of the interferometric scheme \cmmnt{which} is \cmmnt{found}\new{computed} by considering the state of the system right after the signal has been imposed and thus allowing for an optimal detection strategy. 
 \cmmnt{We find} 
  \new{The QFI approaches the CFI}
\begin{align}
Q_{k}^\mathrm{interf}[\rho^\mathrm{th}[\bar{n}];t]=8 t^2+\frac{4}{1+2\bar{n}}\old{\stackrel{t,\bar{n}\gg 1}{\to}8t^2}. 
\label{eq:QFIsim}
\end{align}
\cmmnt{which is seen to approach the CFI} for $t,\bar{n}\gg 1$. 
\new{Again, this is different from other QFI in (\ref{eq:PnMQFI}) in that there is now projective qubit detection before the signal displacement. }
The extra term in the expression for the QFI, in addition to the quadratic scaling in the CFI, stems from the residual information \cmmnt{contained} in the oscillator state \cmmnt{and which}\new{that} is not measured in the interferometric setup.
This contribution \cmmnt{, however,} becomes insignificant for large $t$ or large $n$. 
\cmmnt{We note that} An additional Rabi detector \cmmnt{can be added} after the interferometric setup \cmmnt{to}\new{can} fill the gap between QFI and CFI partially, as the oscillator is still \old{displaced}  after the interferometer.
For example, with a single additional \old{Rabi} detector \old{after the interferometer,} we get  the sum of the CFIs of the interferometer \old{(\ref{eq:Freal})} and \old{the optimal} direct measurement \old{(\ref{eq:CFIthermalDirect})} $F_{k}[\rho^\mathrm{th}[\bar{0}];\A_k,t]\approx 8t^2+\frac{4}{\mathrm{e}}-\frac{16 \alpha_k ^2}{\mathrm{e}}(1-\mathrm{e}^{-1})$.

  \cmmnt{We also note that} The ancillary state residing in the equator subspace \old{of}  $\hat{\sigma}_\mathrm{x}$-basis can be chosen freely as they all lead to the same CFIs and QFIs. 
 However, \cmmnt{for certain environmental noise sources, some choices are more beneficial than others}\new{certain ancillary states may provide better results for certain noise sources,} 
 as \cmmnt{will be} discussed \old{in Sec. \ref{sec:comparison}}.

The interferometric scheme can be \old{straightforwardly} extended to include \cmmnt{many qubit ancillas interacting independently and simultaneously through Rabi interactions with the oscillator for enhanced precision}\new{multiple, independently interacting qubit ancillas for improved CFI} \cmmnt{.
Using such a scheme, the CFI can be increased to}  $F^{\ket{\mathrm{g}}^{\otimes m}}_{k}[\rho^\mathrm{th}[0];\A_k,t]=8 t^2 m$, where $m$ is the number of qubit ancillas.
\cmmnt{It can be simply understood from}\new{This is due to} the independent probabilities for each ancilla (see (\ref{eq:RabiProb}) in Appendix \ref{sec:CFIRabi}), as each joint application of $\hat{R}_\mathrm{x}$ and $\hat{R}_\mathrm{x}^{-1}$ with each ancilla adds $8t^2$ to the \old{CFI}.
In Appendix~\ref{sec:entangledqubit}, \cmmnt{we \old{briefly} investigate a further extension to the scheme by including the generation of input}\new{ the scheme is further extended by creating a} qubit-oscillator entangled probe using Rabi interactions. 
This preliminary study shows \cmmnt{a promising further enhancement, although a more thorough investigation is needed for the full exploration of such a possibility}\new{potential for improvement, but further investigation is needed}.
\old{We also remark that exploiting higher-order Rabi interactions engineered from multiple applications of Rabi interactions can enhance the precision further as discussed in Appendix \ref{sec:non-linear}.}


\section{Robustness of Rabi interferometry}
\label{sec:comparison}



\textbf{Qubit dephasing.---} \new{The} dephasing error on qubit \old{ancilla is} described by a map $\Gamma_\mathrm{d}^{(p)}[\rho]= (1-\frac{p}{2})\rho+\frac{p}{2} \hat{\sigma}_\mathrm{z} \rho \hat{\sigma}_\mathrm{z}$ with \new{a} dephasing parameter \old{$p\in[0 (\text{no dephasing}),1 (\text{complete dephasing})]$}. 
\old{It} occurs simultaneously with the signal displacement
~\cite{Bruzewicz2019APRtrappedion,Kjaergaard2020SuperCondQub}, thereby introducing a degradation in the performance which can be only avoided by \cmmnt{using} error correction. 
As a result of qubit dephasing,  
 the CFI \old{of the Rabi interferometer} changes to
 \begin{align}
   &F_{k}^\mathrm{\Gamma_d^{(p)}}[\rho^\mathrm{th}[0];\A_k,t]=\frac{8 t^2  \cos ^2\left(2 \sqrt{2} t \alpha _k\right)}{4\mathbb{D}- \sin ^2\left(2 \sqrt{2} t \alpha _k\right)}\nonumber\\
   &\stackrel{\A_k\ll 1}{\approx}\frac{2 t^2}{\mathbb{D}}-\frac{4 (4 \mathbb{D}-1) t^4 }{\mathbb{D}^2}\alpha _k^2.
   \label{eq:CFIDepy}
 \end{align}
where we assume that the initial qubit state is $\ket{\pm_\mathrm{i}}$. The dephasing effect is given by the parameter $\mathbb{D}=\left(2-p(1+\mathrm{e}^{-4 t^2})\right)^{-2} $ attaining the value $1/4$ for no dephasing and $1$ for complete dephasing in the asymptotic limit $t\gg 1$.
  The \old{maximum} value of the CFI in (\old{\ref{eq:CFIDepy}}) has an \old{asymptotic} scaling, $2t^2 (2-p)^2$, \cmmnt{that}\new{which} is reduced to the scaling of prepare-measure protocols ($2t^2$) at complete dephasing.
 Here again, the CFI \cmmnt{is monotonous with}\new{depends monotonously on} $t$ regardless of $p$ and $\alpha_k$ in the asymptotic limit.
The normalized curvature is found to \old{be}
 $ \mathbb{C}=\frac{2 (4 \mathbb{D}-1) t^2 }{\mathbb{D}}$
 which is zero for \old{no dephasing} and attains its maximum asymptotic value of $6t^2$ for complete dephasing, thereby coinciding with \cmmnt{the curvature}\new{that} of the prepare-and-measure strategy. 
 For values of $\A_k$ \old{beyond the WDL}, the dependency \old{of the CFI} on $t$ is non-monotonous, having multiple local optimal values for $t$ that depend on the true value of $\A_\mathrm{r}$.
 This CFI has its first zero at $\A_k=\frac{\mathrm{\pi}}{4\sqrt{2}t}$ regardless of $p$, while the  HWHM is found to \old{be}
 \begin{align}
 &\A_k^\mathrm{(HWHM)}
 =\tan ^{-1}\left[ \sqrt{\frac{4\mathbb{D}}{4 \mathbb{D}-1}}\right]\nonumber\\
 &\stackrel{p\ll 1}{\approx}\frac{\mathrm{\pi} }{4 \sqrt{2} t}-\frac{\sqrt{p} \sqrt{\mathrm{e}^{-4 t^2}+1}}{2 \sqrt{2} t},
 \end{align}
 which becomes narrower for \old{larger} values of $t$ and $p$. 
  This dynamic range is approximately twice as large as that of the prepare-and-measure strategy.
The product of the CFI at WDL and the square of HWHM is given at asymptotic $t$ as  $\frac{1}{4} (2-p)^2 \arctan ^{2}\left[\frac{2}{\sqrt{4-p} \sqrt{p}}\right]$ independent of $t$, again showing a trade-off between the dynamic range and maximum CFI for a fixed $p$.
This product is larger than the prepare-and-measure strategy for all $p$.
\new{
The curvature at the origin is a local property, whereas the HWHM exhibits the property of a dynamic range that extends far from the origin. 
However, in our case, the deviation is small, with a difference of less than $5\%$.
}
 
The average CFI is given as 
 \begin{align}
&F_{k}^\mathrm{(av)}[\rho^\mathrm{th}[\bar{0}];\A_k,t]= \old{8 t^2-4 t^2\frac{ \sqrt{4 \mathbb{D}-1}}{\sqrt{\mathbb{D}}}}\nonumber\\
&\stackrel{t\gg 1}{\rightarrow}8 t^2-4 t^2 \sqrt{(4-p) p}.
\label{eq:avCFI}
 \end{align}
 Without dephasing, this expression reduces to $8t^2$ as previously found, while at complete dephasing, it becomes $\frac{4 t^2}{2+\sqrt{3}}$, which is \old{equal to} the asymptotic average value of CFI for the prepare-and-measure strategy.
 This equality to the prepare-and-measure strategy arises due to the mathematical equivalence of the complete qubit dephasing and the projective measurement onto the qubit energy eigenbasis.
 \new{
 In the regime where $t$ is small, utilizing information about the qubit detection outcome in the prepare-and-measure strategy for the preparation state can result in a slightly improved estimation precision, as shown in equation (\ref{CFIPnMnonzero}). 
 However, in the large t regime, this increase in precision can be negligible.
 }
 \cmmnt{We note that} The same equations (\ref{eq:CFIDepy}-\ref{eq:avCFI}) hold for \cmmnt{the} thermal state probes, as \cmmnt{the} qubit dephasing only linearly transforms the qubit detection probabilities from those in the ideal setup, \cmmnt{which are the same for all}\new{regardless of the} probe states of the oscillator.

The CFI for a single displacement component estimation \cmmnt{by}\new{using} a single qubit ancilla \cmmnt{in} (\ref{eq:CFIDepy}) is \old{different from the CFIs of}  joint estimation of the displacement parameters \cmmnt{when}\new{using} two-\old{qubit} ancillas \cmmnt{are used}  as \new{shown} in Fig.~\ref{fig:RabiD}c \cmmnt{under the existence}\new{in the presence} of dephasing and other noises.
The same modulation factor $\cos^2[2t^2]$ is multiplied \old{to the CFI} of  one of the displacement components,  \cmmnt{as in}\new{similar to} the unambiguous estimation by the prepare-and-measure strategy (see Appendix \ref{sec:PnMFis}).
\cmmnt{which}\new{This factor} can be negated at  \old{the same specific Rabi strengths} $t_c=\sqrt{c\frac{\mathrm{\pi}}{2}}$ for $c\in\mathbb{Z}$.
\cmmnt{Similarly as for}\new{Like} other estimation methods, an active correction by (\ref{eq:twoqubit}) is possible.

\begin{figure*}[thp]
\includegraphics[width=\textwidth]{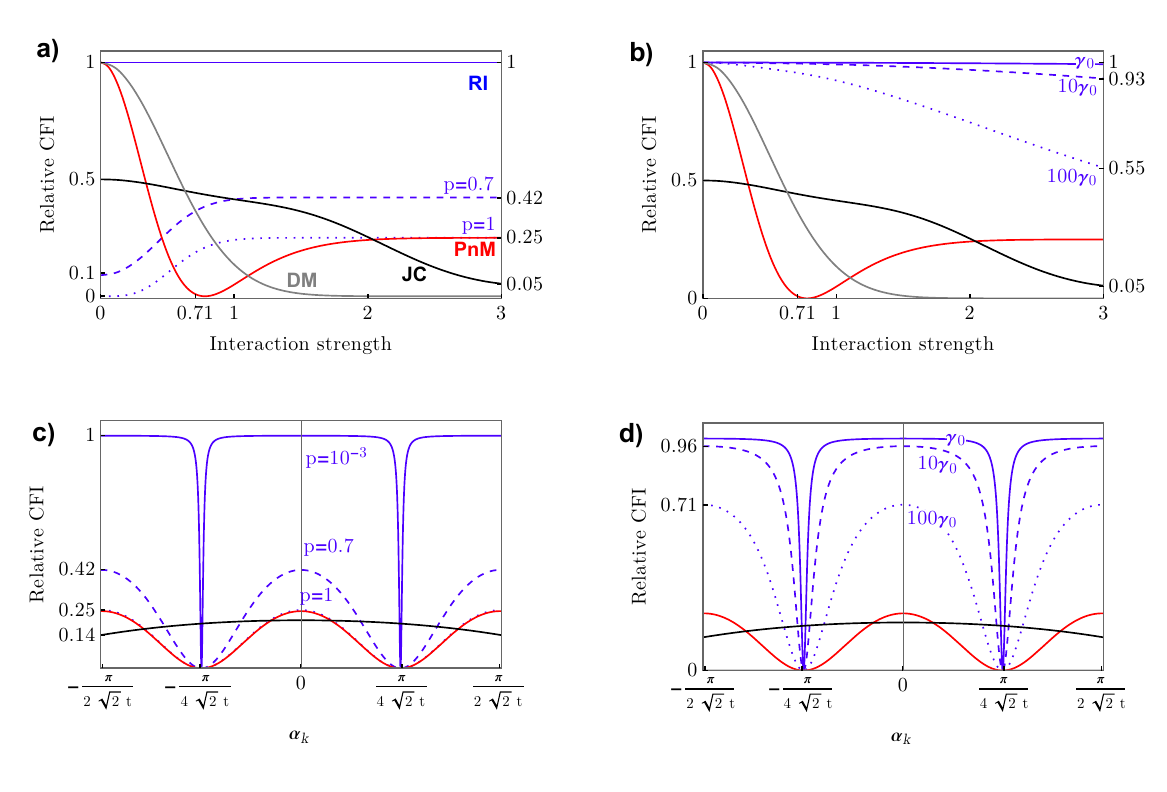}
\caption{\new{Comparison of CFIs summarized in Table \ref{tab:CFItab}.} 
a) Classical Fisher information  re-scaled by the quadratic scaling of an ideal Rabi interferometer \old{in (\ref{eq:Freal})} about \old{the} individual estimation of $\alpha_\mathrm{r}$ (or equally $\alpha_\mathrm{i}$) \new{at various interaction strength $t$} for Rabi interferometer \old{(RI, blue, \new{Fig. \ref{fig:RabiD}c)}}  \cmmnt{vs.} 
  JC interferometer \old{(JC, black, \new{SM Appendix \ref{sec:JC inter}})}, prepare-and-measure strategy (PnM, red, \new{Fig. \ref{fig:RabiD}b}), direct estimation (DM, gray, \new{Fig. \ref{fig:RabiD}a}) at WDL.
The oscillator \new{is initially} in a \old{ground} state,  \new{and the effect of}  \cmmnt{under} various qubit dephasing parameter $p$ \new{is shown}, where CFI of the interferometer is reduced to \old{that of the} prepare-and-measure \old{strategy} at a large $t$ for a complete dephasing $p=1$. 
 b) \old{The relative CFI} under  oscillator heating stronger than known experimental parameters \cmmnt{equivalent to}\new{in} the setup from \cite{Ballance2016}.
\cmmnt{We can see that}  The Rabi interferometer surpasses the JC interferometer, direct Rabi measurement, and prepare-and-measure \old{Rabi} strategies more than twice even at a vanishing strength, \old{even}   under an infinite level of \old{heating} noises. 
\old{c, d) The CFI \old{beyond WDL} under the influence of qubit dephasing and oscillator heating at \old{$t=\sqrt{3\mathrm{\pi}/2}$ at which the harmful interference from the simultaneous estimation can be avoided}.
\cmmnt{The}\new{A} dip in the CFI, common to the Rabi interferometers and the prepare-and-measure setups, correspond to the parameters at $\A_k=\frac{\mathrm{\pi}}{4\sqrt{2}t}$, arising from the periodicity of the qubit detection, \old{the width of} which  gets narrower as $t$ is increased.
This dip appears due to the fragility of the estimation when one of the qubit detection probabilities is $0$ in ideal case without noise.
Beyond the values shown here the CFIs are periodic, even though the results are inconclusive \cmmnt{from}\new{due to} the periodicity of the data.
The JC interferometer has a broad dynamical range, \cmmnt{however, being}\new{but is} unable to sense the conjugate displacement.
The CFI by the direct measurement is small on this scale and is not shown.
}
 }
 \label{fig:measure-prepare}
\end{figure*}

 In Fig.~\ref{fig:measure-prepare}a, we compare the \old{CFI} of the Rabi interferometer  with \new{that of} the noise-free JC interferometer (addressed in Appendix \ref{sec:JC inter}), the noise-free direct-measurement, and prepare-and-measure strategies in the WDL  under \old{qubit} dephasing.
\cmmnt{We find that} The Rabi interferometer \cmmnt{is superior to}\new{outperforms} \old{the} ideal prepare-and-measure strategy for all dephasing levels, while they converge \old{asymptotically under complete dephasing}.
The supremacy is more pronounced over the noise-free JC interferometer.
Even at a vanishing strength, the Rabi interferometer is twice more precise, \cmmnt{and the enhancement is larger}\new{with a greater enhancement} for a larger strength. 
For $p\lesssim 0.7$ it is \cmmnt{superior to}\new{more accurate than} the direct measurement for $t\gtrsim 0.71$ and \cmmnt{to}\new{than} the JC interferometer \old{for $t\gtrsim 1$}.
When considering an input oscillator in a thermal state we attain similar comparative trends between the different schemes as for the pure state case. 
 The CFI beyond the WDL is shown in Fig.~\ref{fig:measure-prepare}c as a function of $\A_k$ for a fixed $t$  \new{which demonstrates that} the interferometer is superior to all other methods for all levels of dephasing $p$.

\textbf{Oscillator heating.---}
All mechanical oscillators (e.g. \old{trapped-ion motion} \old{\cite{LeibfriedRMP2003,Affolter2020,Gilmore2017,Wang2019,Biercuk2010}}) are inevitably coupled to a thermal bath and experience heating.
This heating \old{denoted as $\Gamma_\mathrm{h}[\rho]$ for a\old{n} \old{arbitrary initial} oscillator density matrix  $\rho$}, can be described by \old{the solution to} the Lindblad equation $\partial_{t_\mathrm{th}}\rho=\sum_{i=1,2}L_\mathrm{i} \rho L_\mathrm{i}^\dagger-\frac{1}{2}\{L_\mathrm{i}^\dagger L_\mathrm{i} \rho\}$ with Lindblad operators  $L_{1}=\sqrt{\gamma^\mathrm{th}}\sqrt{\bar{n}\old{^\mathrm{th}}+1}\hat{a}$ and $L_{2}=\sqrt{\gamma^\mathrm{th}}\sqrt{\bar{n}\old{^\mathrm{th}}}\hat{a}^\dagger$~\cite{Barnett2003}.
The effect of a thermal channel is fully characterized by \new{parameters such as} the average number of the thermal photon in the bath $\bar{n}\old{^\mathrm{th}}$, the heating rate \old{$\gamma^\mathrm{th}$}, and the heating time \old{$t_\mathrm{th}$}. 
\old{Here,} the heating parameters for scaling are chosen as in Ref. \cite{Ballance2016} where the oscillator ground state was  gaining an additional excitation of $\Delta_{\bar{n}} =1.1\times 10^{-4}$ in a single gate slot, which approximately corresponds to $\bar{n}\old{^\mathrm{th}}=1$, \old{$\gamma^\mathrm{th}=0.005\equiv\gamma_0$}, and \old{$t_\mathrm{th}=0.01$}. 
\new{Our} numerical analysis shows that the \old{CFI} \old{of an interferometer} is monotonously \old{decreasing} with \new{increasing values of} $\bar{n}\old{^\mathrm{th}}$, \old{$\gamma^\mathrm{th}$ and $t_\mathrm{th}$}.
In Fig.~\ref{fig:measure-prepare} b, we \cmmnt{plot}\new{show} CFI for the Rabi interferometer associated with different strengths for the heating effect, and we \cmmnt{see}\new{find} that \cmmnt{this interferometer is superior to}\new{it outperforms} the two other estimation strategies even under strong oscillator heating effects.

In Fig.~\ref{fig:measure-prepare}d, we \cmmnt{plot}\new{compare} the CFI beyond the WDL under various oscillator heating strengths.
\cmmnt{and we see the superiority of} The Rabi interferometer \new{remains superior to} \cmmnt{over} other methods \old{that are} assumed to be free of any decoherence.
The \cmmnt{analysis}\new{comparison} for the input thermal state under the same environments shows \old{the equivalent} \old{superiority} \cmmnt{over other benchmarks,as for vacuum/ground state oscillator} in Fig.  \ref{fig:measure-prepare}b \old{and d} despite the initial thermal noise.
\old{This robustness to \old{thermal} noise proves a clear-cut advantage of the Rabi interferometric setup over other alternatives.}
Other minor sources of noise \cmmnt{are considered}\new{is discussed} in Appendix~\ref{sec:minor}.

\cmmnt{Interestingly,} The formulas  (\ref{eq:CFIDepy}-\ref{eq:avCFI}) can be used to model the effect of an oscillator heating on CFI by replacing $p$ by a heuristic parameter $p^\mathrm{th}=1-\mathrm{e}^{-c t^2\gamma^\mathrm{th}}$ with $c=0.159$.
This universality of the CFI by dephasing formula arises because the noises or interference in the joint estimation \cmmnt{in Appendix \ref{sec:PnMFis}} linearly mixes the qubit detection probabilities which have the same mathematical description as the dephasing model.
\new{
This model predicts that the heating in the oscillator does not hinder the quadratic scaling of $2t^2$, even when subjected to significant thermal effects. 
Additionally, the impact of the thermal bath is minimal. For instance, if an uncooled oscillator (containing an average of 1 photon) is exposed to a thermal bath during the sensing stage, the resulting difference in CFI from the initial vacuum state is approximately $0.1\%$ under the considered environment.
}

\cmmnt{We briefly comment on} The similarity of the formulas for the CFI in (\ref{eq:CFIthermalDirect}), (\ref{CFIPnMnonzero}), and (\ref{eq:CFIDepy}) 
\cmmnt{This originates}\new{stems} from the qubit nature of the detection module.
For example, if the probabilities of the qubit detection outcome in the excited state is given by
$P_\mathrm{e}=\frac{1}{2}(1+\mathbb{E}\sin[2\sqrt{2}t\A_k])$ for $\mathbb{E}\in[-1,1]$ \old{after a given process}, the CFI is \new{expressed as}
$F_k[\A_k]=\frac{8 t^2 \cos ^2\left[2 \sqrt{2} t \alpha _k\right]}{\mathbb{E}^{-2}-\sin ^2\left[2 \sqrt{2} t \alpha _k\right]}$.
\cmmnt{Now} The largest value of the CFI is attained at $\mathbb{E}=\pm 1$, \cmmnt{which produces}\new{resulting in} the same scaling for CFI with $t$ as in the ideal Rabi interferometer.
This \cmmnt{proves the} optimality of the Rabi interferometer among the estimation setups using qubit ancillas
\new{is further confirmed by comparing it with other protocols for displacement, for example the estimation by squeezed thermal states and by other interferometers. 
This comparison is further discussed in more detail in Appendix \ref{sec:JC inter} and \ref{sec:minor}. 
}

\section{Discussion and outlook}

In this work, we \old{proposed} a \old{Rabi} interferometric \cmmnt{estimation} protocol for \old{the unambiguous estimation of phase space displacements} \old{in an unknown direction} of a mechanical oscillator or an electromagnetic field, e.g. represented by the motional state of a trapped ion~\cite{KienzlerPRL2016,Bruzewicz2019APRtrappedion,FluhmannHomeNature2019} or the microwave field of a superconducting circuit~\cite{Wendin2017superconducting,Touzard2019,BlaisNatPhys2020superconducting, SchoelkopfNature2020,Kwon2021Superconduting,ClerkNatPhy2020SuperCon,BlaisRMP2021CircuitQED}, \old{and in future, potentially also by light~\cite{NoriRabiRMP2019,SolanoRabiRMP2019}}.
It \old{provides} a simple and \old{robust} approach to the detection of weak forces with a sensitivity that goes beyond \cmmnt{what is possible with} other approaches that \cmmnt{are exploiting}\new{uses} similar resources.
It is \old{based on} an experimentally feasible approach that has been proven in previous experiments on Rabi interferometry~\cite{HempelNatPho2013catspectroscopy, PenasaPRA2016,GilmoreScience2021Displacement}, and \old{even for small interaction strengths,} it beats the state-of-the-art interferometer based on the standard JC coupling \old{with a bounded CFI}  in which the RWA holds.
We \cmmnt{explored}\new{analyzed} the robustness of the Rabi interferometer to various realistic noise models \new{including thermal noise and dephasing}, and \cmmnt{
notably, we} found that the thermal occupation of the probing oscillator is not critical.  
The CFI for both the interferometric and prepare-and-measure setups are scaling quadratically with the Rabi coupling strength, while the former has a 4-fold increase  \old{in contrast to} \old{the latter}. 
The \old{enhancement} is maintained for all levels of qubit dephasing and oscillator heating\cmmnt{, and thus shows an absolute superiority of the interferometric scheme}.
\cmmnt{Notably, }The dynamic range of the interferometer is limited only by the periodic nature of the qubit detector \old{that can be overcome} and is superior to those of the other methods.  
\old{In all strategies investigated here, the \old{harmful} interference that arises in the simultaneous estimation can be avoided by choosing proper Rabi strengths where the performance is equal to the individual estimation.}

These protocols can be further extended to the direct detection of the oscillators together with the qubit detectors as summarized in Appendix~\ref{sec:quaddet}, or even composite setups of Rabi interferometer and additional Rabi detector and engineered high-order Rabi interferometers as in Appendix~\ref{sec:non-linear}.
Other types of \old{nonlinear} interactions on platforms such as optics, and opto/electromechanics~\cite{AspelmeyerRMP2014cavoptomech, TeufelNature2011electromech} can be used in a similar setup, \new{ while our preliminary study in Appendix \ref{sec:JC inter} suggests that Rabi interferometer may be the optimal setup for estimating displacement}.
The non-interferometric setups are benefiting the most from the detection of the oscillator mode, and might even surpass the performance of the interferometric setup when the noise is absent.
In the future, this scheme can be further extended to the interferometric setups exploiting entangled qubit ancillas or qudit ancillas (a simple example has been analyzed \old{in} Appendix \ref{sec:entangledqubit}).
 The application of this scheme can be extended to the estimation of general Gaussian unitary or non-unitary channels~\cite{HolevoBook2019Channels}.





\section*{Acknowledgment}
KP and RF acknowledge Project 22-27431S of the Czech Science Foundation. PM and RF acknowledge 
the funding from MEYS (No. 8C22001 and 8C20002) and the European Union’s Horizon 2020 (2014-2020) research and innovation framework program under Grant Agreement No 731473 (ShoQC and SPARQL). 
Projects ShoQC and SPARQL received funding from the QuantERA ERANET Cofund in Quantum Technologies implemented within the European Union’s Horizon 2020 Program.
KP and ULA acknowledge the support from the Danish National Research Foundation (DNRF142).
We also acknowledge the European Union’s 2020 research and innovation programme (CSA-Coordination and support action, H2020-WIDESPREAD-2020-5) under grant agreement No. 951737 (NONGAUSS).
We acknowledge Jacob Hastrup for the useful discussion. 

\newpage
\tableofcontents
\newpage
\appendix
\section*{Supplementary Materials}
\addcontentsline{toc}{section}{Supplementary Materials}

\counterwithin{figure}{section}

\section{Quantum Fisher information and classical Fisher information}
\label{sec:Fisher}

To conclusively  \old{assess} \old{the} sensitivity \old{of the estimation methods by}  complex \old{setups} \old{to be considered here}, we   calculate   \old{the} Fisher information\old{s} of \old{them} which can predict the \cmmnt{unbiased} precision of estimation of  the signals\cmmnt{in a given setup}.
The QFI~\cite{Paris2009} specifies the upper bound \old{of the precision} for the \old{given} probe state \cmmnt{and all detection} \old{for the optimal estimation}   strategy, whereas CFI is the asymptotically achievable precision \cmmnt{of the infinite copies of the probe} by a specific detection. 
\old{For} a density matrix $\rho(\mathbf{\Theta})=\hat{\mathbb{U}}[\mathbf{\Theta}]\rho_\mathrm{in} \hat{\mathbb{U}}[\mathbf{\Theta}]^\dagger$ with \old{an} input state \old{density matrix} $\rho_\mathrm{in}$  \old{and }\old{the unitary evolution up to the point of the encoding of the unknown signal $\hat{\mathbb{S}}[\mathbf{\Theta}]$ with the unknown parameters $\mathbf{\Theta}$ \cmmnt{whose true value is $\mathcal{Q}$} written as $\hat{\mathbb{U}}[\mathbf{\Theta}]=\hat{\mathbb{S}}[\mathbf{\Theta}] \exp[-i t \hat{H}_\mathrm{int}]$,} \cmmnt{the unknown signal} \cmmnt{$\hat{\mathbb{U}}[\mathbf{\Theta}]$}  the QFI about target variable $\Theta_j$ \cmmnt{is expressed as $Q_{\Theta}[\rho_\mathrm{in}]=\tr[\rho(\mathbf{\Theta}) \hat{L}_{\Theta}^2]$, where the symmetric logarithmic derivative is defined as $\hat{L}_{\Theta_j}=2\partial_{\Theta_j} \rho(\mathbf{\Theta})$ \old{for pure states}}
\old{generated by a } Hamiltonian $\hat{H}_j$ \cmmnt{for a general density matrix $\rho(\mathbf{\Theta})$} \old{is expressed} as~\cite{Paris2009,Liu2016}:
\begin{align}
    Q_{j}[\rho_\mathrm{in};\Theta_j]=2\sum_{k,l}\frac{(\lambda_k-\lambda_l)^2}{\lambda_k+\lambda_l} |\bra{k}\hat{H}_j\ket{l}|^2,
    \label{eq:QFIdef}
\end{align}
where $\lambda_k$ and $\ket{k}$ are eigenvalues and eigenvectors of $\rho(\mathbf{\Theta})$ and the summation is only for $\lambda_k+\lambda_l\neq0$.

The \cmmnt{classical Fisher information}CFI about a variable $\Theta_j$  \old{in hybrid setups} \old{with a specific \old{discrete detector}} can be calculated \cmmnt{when the detector is specified} as
\begin{align}
F_{j}[\rho_\mathrm{in};\Theta_j]= \sum_n P(n;\mathbf{\Theta})^{-1} (\partial_{\Theta_j} P(n;\mathbf{\Theta}))^2.
\label{eq:CFIdef}
\end{align}
\new{Fisher information matrix (FIM) $i,j$-element is given as  $F_{\Theta_j}[\rho_\mathrm{in}]=\sum_k P_k(\mathbf{X};\mathbf{\Theta}) (\partial_{\Theta_i} P_k(\mathbf{X};\mathbf{\Theta}))(\partial_{\Theta_j} P_k(\mathbf{X};\mathbf{\Theta}))$.
This quantity explores the scenario of multiparameter estimation, with non-zero off-diagonal elements representing some correlation~\cite{FidererPRX2021QFIM}.
}
Here, a \old{discrete} detection setup consisting of  POVM elements $\{\hat{\mathbf{\mathrm{\Pi}}}[n]\}$   satisfies $\sum_{n=1}^N\hat{\mathbf{\mathrm{\Pi}}
}[n]=\hat{\mathbf{1}}$ where \old{the index} $n$ represents the \cmmnt{data}\old{measurement outcomes} from the discrete detectors. 
\old{For a continuous detection, integration substitutes for the summation. }
\old{The CFI implies common assumptions of local estimation strategy, meaning it only discerns the difference of $\mathbf{\Theta}$ and $\mathbf{\Theta}+d\mathbf{\Theta}$.
It also assumes average of infinite number of probes, where the frequency of the data can be considered as probabilities.}
\cmmnt{It}\old{The detection on many copies of the probe} generates the \old{outcome} data forming a\old{n} \old{asymptotic} probability $P(n;\mathbf{\Theta})=\tr[\mathbf{\mathrm{\Pi}}[n] \rho'(\mathbf{\Theta})]$ from a \old{final} density matrix $\rho'(\mathbf{\Theta})=\hat{\mathbb{U}}'[\mathbf{\Theta}]\rho_\mathrm{in} \hat{\mathbb{U}}'[\mathbf{\Theta}]^\dagger$ where now $\hat{\mathbb{U}}'[\mathbf{\Theta}]$ is the unitary evolution up to the point of detection, equal to $\hat{U}_\mathrm{interf}$ in the interferometer \old{in (\ref{eq:interfero})}.
The superscripts over the CFI will be used throughout the text below to specify  the setup where the estimation is performed.
A brief demonstration of the equivalence of the precision predicted by CFI and that achievable by the maximum likelihood estimation  in the asymptotic infinite copy limit applicable to the examples in the main text using a binary outcome detection \old{$N=2$} is summarized in Appendix~\ref{sec:MLE}\cmmnt{to show the accessibility of the precision predicted by CFI}.

We remark that \old{the} maximization of the CFI over all quantum POVM measurements can reach \old{the} QFI~\cite{Paris2009} in principle, but in general practices, the CFI depends on many experimental parameters, such as noises, the ancillary state, and even the true value of the target parameter.
However, as we cannot fully know the target parameter beforehand \old{of the estimation}, the choice of the setup is inevitably based on certain assumptions about it.
A brief strategy to adapt the setup based on the \old{partial} knowledge about the target parameter was described in Appendix \ref{sec:adaptive}, but it is not of the main interest.
Here, the main assumption throughout the paper  is the weakness of the force, i.e. the displacement is very weak, i.e. $|\A|\ll 1$\old{, as it is the region of the largest impact and \old{the experimental} difficulty}.

We note that the QFI in \ref{eq:QFIsim} is comparable to the QFI for the prepare-and-measure strategy as discussed   
\old{in Appendix \ref{sec:quaddet}}. 
This is to be expected due to the similarity of the \old{prepared} states \old{in the oscillator} undergoing the signal displacement \old{when they are considered as the input states for general measurements}.


\section{Estimation of oscillator phase shift and qubit rotation}
\label{sec:rotation}

\textbf{Oscillator rotation estimation and oscillator phase noise suppression}
The unknown rotation $\hat{\mathcal{R}}[\theta]$  exerted by \old{the} force where $\theta$ is the target parameter can be estimated by \cmmnt{using}\old{the} Rabi interferometer using the adjoint action \old{in place of (\ref{eq:totalaction})}:
\begin{align}
    &\exp[\mathrm{i} t \hat{\sigma}_\mathrm{x} \hat{X}]\exp[\mathrm{i} \theta \hat{n}] \exp[-\mathrm{i} t \hat{\sigma}_\mathrm{x} \hat{X}]\nonumber\\
    &=\exp[\mathrm{i}\theta(\hat{n}-t \hat{\sigma}_\mathrm{x} \hat{P}-\frac{t^2}{2})]\approx \exp[\mathrm{i} \theta \hat{n}]\exp[-\mathrm{i} t \theta \hat{\sigma}_\mathrm{x} \hat{P}].
    \label{eq:OscRot}
\end{align}
Besides the unknown signal rotation after the approximation, the second Rabi operation \old{after the final approximation} represents a momentum-dependent qubit rotation.
Therefore, using a probe state with a large momentum such as imaginary amplitude coherent state $\ket{\ii\beta}_\mathrm{C}$ can increase the qubit rotation to be estimated by \old{a} qubit detector,  similarly as in the estimation of displacement, \old{where we can replace $\hat{P}\to \sqrt{2}\beta$ in the last expression of (\ref{eq:OscRot})}, and thus all properties of the displacement estimation are equally applicable.
In the absence of dephasing, the highest precision can be expected at $\theta=0$.
The CFI by Rabi interferometer 
\old{under dephasing is given as 
\begin{align}
    F[\theta]=\frac{2 \beta ^2 t^2}{\mathbb{D}} \cmmnt{\left(2-p(1+ e^{-4 t^2})\right)^2},
\end{align}
having a scaling of $8t^2 \beta^2$ for $p=0$.}
In comparison, the optimal prepare-and-measure strategy \cmmnt{in this case}\old{in the rotation estimation by the same probe state} has CFI of $F^\mathrm{PnM,\pm,opt}[\theta]=\frac{2 \beta ^2  t^2}{\mathbb{B}_\pm}\cmmnt{ \left(1\pm e^{- t^2}-e^{-2t^2}\pm e^{-3t^2}\right)^2}$, again when the qubit ancilla is set as $\ket{\phi^\mathrm{opt}}$.
\old{From the equivalence to the estimation of displacement, the} Rabi interferometer is superior to the prepare-and-measure strategy in all cases of $\beta$ and $t$.

On the other hand, \old{if we see the rotation as the noise in the estimation of the displacement,} we can efficiently suppress \cmmnt{a}\old{this} oscillator phase noise by choosing a probe state with a small average momentum
of $\langle \hat{P}\rangle\approx 0$ such as $\ket{0}_\mathrm{C}$, where now $\hat{\mathcal{R}}[\theta]$ \cmmnt{is}\old{plays as} a noise in the estimation of displacement.
In (\ref{eq:OscRot}), the qubit rotation  then becomes effectively $0$.
This suppression implies that displacement and rotation can be independently estimated by our setups \old{by choosing  coherent states in the oscillator $\ket{\mathrm{i} \beta}_c$ with  $\beta\ll1$ or $\beta\gg1$}.

\textbf{Qubit rotation estimation and qubit rotation noise }
We can consider various scenario\old{s} where qubit rotation in an arbitrary direction is present, either as the estimation target or as noise.
For an estimation of single rotation component $\Theta_\mathrm{z}$, we again get the transformation of the qubit rotation by the Rabi interferometer:
\begin{align*}
    &\exp[-\mathrm{i} t \hat{\sigma}_\mathrm{x} \hat{X}]\exp[\mathrm{i} \Theta_\mathrm{z} \hat{\sigma}_\mathrm{z}]\exp[\mathrm{i} t \hat{\sigma}_\mathrm{x} \hat{X}]=\exp[\mathrm{i} \Theta_\mathrm{z} (\hat{\sigma}_\mathrm{z} \cos[2t \hat{X}]-\hat{\sigma}_\mathrm{y} \sin[2t\hat{X}])],\nonumber\\
    &\exp[-\mathrm{i} t \hat{\sigma}_\mathrm{x} \hat{X}]\exp[\mathrm{i} \Theta_\mathrm{y} \hat{\sigma}_\mathrm{y}]\exp[\mathrm{i} t \hat{\sigma}_\mathrm{x} \hat{X}]=\exp[\mathrm{i} \Theta_\mathrm{y} (\hat{\sigma}_\mathrm{y} \cos[2t \hat{X}]+\hat{\sigma}_\mathrm{z} \sin[2t\hat{X}])].
    \label{eq:QBRot}
\end{align*}
When these qubit rotations are estimation targets, we can calculate the Fisher informations from these equations in equivalent ways as before.
 In summary, we can say that the sensitivity in $\Theta_\mathrm{x}$ measurement is not affected, $\Theta_\mathrm{y}$ measurement deteriorated, $\Theta_\mathrm{z}$ measurement is enhanced by the Rabi interferometer.

Now we can think of the qubit rotation as noises in the estimation of displacement. 
In this case, the total operation is written using Eq.~(\ref{eq:QBRot}) as 
\begin{align}
    &\exp[-\mathrm{i} t \hat{\sigma}_\mathrm{x} \hat{X}]\exp[\mathrm{i} \Theta_\mathrm{z} \hat{\sigma}_\mathrm{z}]\exp[\mathrm{i} \Theta_\mathrm{y} \hat{\sigma}_\mathrm{y}]D[\alpha]\exp[\mathrm{i} t \hat{\sigma}_\mathrm{x} \hat{X}]\nonumber\\
    &=\exp[\mathrm{i} \Theta_\mathrm{z} (\hat{\sigma}_\mathrm{z} \cos[2t \hat{X}]-\hat{\sigma}_\mathrm{y} \sin[2t\hat{X}])]\nonumber\\
    &\times\exp[\mathrm{i} \Theta_\mathrm{y} (\hat{\sigma}_\mathrm{y} \cos[2t \hat{X}]+\hat{\sigma}_\mathrm{z} \sin[2t\hat{X}])]D[\alpha]R[\phi].
\end{align}
This relation can be again used to obtain the CFI, which is decreased from the ideal interferometer as expected.

\section{General interferometers for the estimation of a signal}
\label{sec:signal}

The target classical signal in quantum sensing is commonly composed of \old{simultaneously occurring} various \old{unitary} processes, \old{ described} as $\hat{\mathbb{S}}[\mathbf{\Theta}]=\exp[\mathrm{i}  \old{\sum_j} \Theta_j \hat{H}_j]$ where $\hat{H}_j$'s are  \old{transformation-}generating Hamiltonians and  $\Theta_j$ are the \old{unknown classical} signal  strengths.
\old{In this work, we simplify the problem and focus on a single parameter estimation $\hat{\mathbb{S}}[\Theta]=\exp[i   \Theta \hat{H}]$.}
\cmmnt{This unitary dynamics adequately describes such a unitary process below their smallest decoherence time. }
This \old{classical} signal \old{below the smallest decoherence time} transforms a known input \old{quantum} state (probe) into an unknown state, and the \old{encoded} information about this signal can be \old{drawn by} \cmmnt{using}processing \old{the} data from the available detection on many copies of probes\cmmnt{, which are detected by available detectors}.
Quantum sensing \old{aims} to optimize  the \old{estimation} setups and \old{the quantum} probes to \cmmnt{extract this information}infer \cmmnt{the}\old{such} signals efficiently. 
In many \old{experimental} systems \old{of interest,} the choice of the probes and \old{setups \cmmnt{containing}\old{including}} detectors is often pre-imposed and thus limited.
We assume \cmmnt{also} here \old{setups with probes factorized in time} where $\hat{\mathbb{S}}[\Theta]$ is occurring to \old{ensembles of} only one probe state, leaving \cmmnt{multicopy protocols}\old{the consideration of} \old{correlated probes}  for  future  investigation.

Quantum interferometers \cmmnt{exploit}\old{use} multipartite\old{, typically bipartite,} interaction Hamiltonian \old{with ancillary systems which may be of different dimensions, described by} $\hat{H}_\mathrm{int}$ \cmmnt{between many degrees of freedom } applied before and after the signal \old{$\hat{\mathbb{S}}[\mathbf{\Theta}]$} to exploit  quantum interference \cmmnt{between the modes} for a high precision \cmmnt{measurement}\old{estimation}.
 The \old{evolution in the} \old{symmetrical setup with an equal strength \old{$t$} \cmmnt{for}\old{of} the pre- and post-processing \cmmnt{of}} \cmmnt{total process} before the \old{final \cmmnt{detection}  detector }  is described by a  unitary operation composed of sequential unitary operations 
\begin{align}
\hat{U}_\mathrm{interf}=\exp[\mathrm{i} t \hat{H}_\mathrm{int}] \hat{\mathbb{S}}[\Theta] \exp[-\mathrm{i} t \hat{H}_\mathrm{int}].  \label{eq:interfero}  
\end{align}
\old{To explore \old{the} full power of given interferometry, we consider interactions \old{that are} unitary and fully controllable in \cmmnt{time}\old{strength} $t$.  }
\old{We note that for  \old{the absence of the signal} $\Theta=0$, $\hat{U}_\mathrm{interf}$ in (\ref{eq:interfero}) is reduced to the identity operation, and the following detection measures the deviation \old{from it}.}

\section{Estimation with thermal probes}
\label{sec:DirectThermal}
 \begin{figure*}[thp]
 \includegraphics[width=400px]{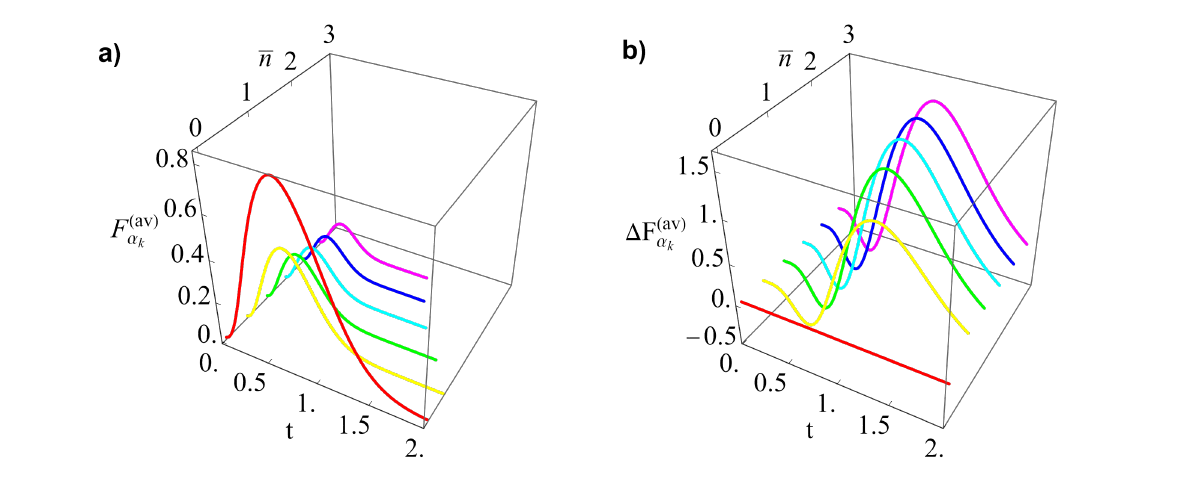} 
 \caption{\old{a) Average CFI by direct measurement using a thermal probe at various $\bar{n}$ with increment $\Delta \bar{n}=0.5$ in (\ref{eq:CFIthermalDirect}). 
 The ground state has a larger CFI than thermal probes when direct estimation is performed.
 b) The difference of CFI by a prepare-and-measure strategy in (\ref{CFIPnMnonzero}) between thermal probes with various  $\bar{n}$ from the ground state probe.
 At $t\approx 1.2$, the thermal state shows a larger estimation precision than the ground state, while approaching the latter in the asymptotic $t$.}
 } \label{fig:DirectThermal}
 \end{figure*}

In both direct measurement and prepare-and-measure strategy, using thermal probes instead of oscillator ground state probes changes the CFI.
In the direct measurement, the approximate formula for the optimal strength and the maximum average CFI at $\bar{n}\lesssim3$ is given as
\begin{align}
    (t_\mathrm{opt'}[\bar{n}],F^\mathrm{(av)}[\rho^\mathrm{th}[\bar{n}];t_\mathrm{opt'}] )\approx (0.22 +0.40*0.29^{0.71\bar{n}},0.059+0.78*3.49^{-\bar{n}^{0.69}})
\end{align} 
showing a decreasing average CFI by $\bar{n}$.
For example at $\bar{n}=1$, the maximum is given  as $F^\mathrm{(av)}[t_\mathrm{opt'}]=0.276$  at $t_\mathrm{opt'}=0.379$,  reduced by approximately half from the vacuum probe. 

\old{In the prepare-and-measure strategy, thermal probes can be beneficial to the estimation at an intermediate $t$.
As can be seen in Fig.~\ref{fig:DirectThermal} b), the CFI is increased from that of the ground state at around $t\approx 1.2$. }

\old{\section{Maximum likelihood estimation by qubit detector}
\label{sec:MLE}

Here we consider the maximum likelihood estimation (MLE) of the displacement parameter $\alpha_\mathrm{r}$ using \old{a} qubit detector, which can \old{apply} to all the protocols introduced here.
We first note that the detection outcomes from a qubit detector can be understood as a random sampling from  binary outcomes with a fixed  sampling probabilities $p$ and $1-p$.
With a fixed number of $n$ probes, if we repeat this sampling many times, the distribution of the number of one outcome, e.g. $\ket{\mathrm{g}}\bra{\mathrm{g}}$,  makes a binomial distribution by definition.
In addition, the outcomes with the same frequency of data will give the same maximum likelihood function accordingly.
We note that MLE estimation is equivalent to simply estimating the sampling probability from the frequencies.
Therefore,  the variance of MLE about $p$ is given by that of a binomial distribution $\sigma_p^2= p (1-p)/n$.
This scaling in $n$ can be described by the theory of Fisher information, where the uncertainty of an estimation method is given as $1/n F$ where $F$ is the Fisher information~\cite{LyFI2017}.
Therefore, we can identify the implied Fisher information by maximum likelihood estimation as $F[p]^\mathrm{(MLE)}=1/p(1-p)$, in a single copy limit $n=1$.
We note that this can be alternatively written also as $\sigma_p^2=(p^{-1}+
(1-p)^{-1})^{-1}$ and  $F_p^\mathrm{(MLE)}=p^{-1}+
(1-p)^{-1}$.
Now if we change our estimation target to $\alpha_k$, we can use the propagation of uncertainty as:
\begin{align}
    &\sigma_{\alpha_\mathrm{r}}^2\approx  \sigma_p^2 |\partial_{\alpha_\mathrm{r}}p|^{-2}= (p^{-1}+
(1-p)^{-1})^{-1}|\partial_{\alpha_\mathrm{r}}p|^{-2}=1/F_{k}^\mathrm{(MLE)}=1/F_{k},
\end{align}
and we can immediately see the equivalence to the CFI.
}

 \section{Non-monotonicity of the prepare-measure strategy}
 \label{sec:non-monotonocity}
 In Fig.~\ref{fig:measure-prepare}, we can see that 
 there is a non-monotonous behavior \old{vs $t$} in the CFI by the prepare-measure strategy \old{in (\ref{CFIPnMnonzero})}.
It \old{shows at $t\ll1$} \old{the scaling of $8t^2$ until it reaches} the first maximum at $t^\mathrm{max}\approx0.382$ with local maximum CFI $F_{\A_\mathrm{r}}\approx0.461$, dropping to $0$ at $t^\mathrm{min} 
\approx0.78$.  
 At this strength, the probabilities of qubit detection $P_e=P_g=1/2$ are  independent of either $\alpha_\mathrm{r}$, or the \old{initial} qubit state.
 This  behaviour \old{at $t_0$} arises because \old{the oscillator is maximally entangled to the qubit at this strength}, and the reduced  qubit state after the oscillator mode is traced out right after the second Rabi interaction before the qubit detection is given as \old{a maximally mixed qubit state}
 \begin{align}
 &\rho_\mathrm{q}=0.5 \ket{+}\bra{+}+0.5 \ket{-}\bra{-}+0.5\ii \sin\phi\sin\theta\ket{+}\bra{-}-0.5\ii \sin\phi\sin\theta\ket{-}\bra{+},
 \end{align}
 where $\theta,\phi$ are parameters of \old{initial} second ancillary qubit state \old{represented in a Bloch sphere as} $\cos[\theta/2]e^{\mathrm{i}\phi/2}\ket{\mathrm{e}}+\sin[\theta/2]\mathrm{e}^{-\mathrm{i}\phi/2}\ket{\mathrm{g}}$.
 The reduced qubit state $\rho_\mathrm{q}$ does not have any dependence on $\alpha$ and thus shows zero FI, even though the full state (two head cat entangled with qubit) has the dependence on $\A$.
 \old{For optimal initial qubit state $\ket{\pm_\mathrm{i}}$, it is given as the fully mixed state.}
 Beyond this strength again this strategy has an increased FI about $\alpha_\mathrm{r}$ vs. $t$. 

\section{Adaptive strategy for setups with varying performance}
\label{sec:adaptive}
Often, the setup with a fixed architecture works better for certain ranges of target variables than the others.
In such cases, the maximum in Fisher information can be reached \old{asymptotically by} an adaptive method.
In the  adaptive estimation protocol, the setup including the  input probe state is chosen based on all the \old{already known} partial knowledge gained about the \old{range of} target values with \old{a} subset of probe state ensembles. 
This \old{partial knowledge can be gained and accumulated by  dividing the ensembles into subensembles made of} \old{a} finite number of probes. \old{These subensembles are used for different rounds, and the gained information in all previous rounds can be used}  for the adjustment  \old{of the  setup of the current round  to increase the amount of information that can be gained per each  probe}.
Therefore, the precision can approach the maximum value \old{asymptotically as} the uncertainty in the estimation is thus decreased.

\old{For example, consider a virtual   estimation protocol of the displacement in known direction $D[\alpha_\mathrm{r}]$, whose performance may vary depending on the true value of $\alpha_\mathrm{r}$, e.g. the maximum precision is around $\alpha_\mathrm{r}^\mathrm{max}\neq 0$.
In the first round, we do not have any pre-knowledge about $\alpha_\mathrm{r}$, so we are forced to assume the true value randomly, e.g. $\alpha_\mathrm{r}^{(0)}=0$.
Now in the first round of estimation, we apply an auxiliary displacement $D[\alpha_\mathrm{r}^\mathrm{max}-\alpha_\mathrm{r}^{(0)}]$  with a finite number of the probes, and  update $\alpha_\mathrm{r}^{(1)}$ with the gained data.
Now for the second round, we apply an auxiliary displacement $D[\alpha_\mathrm{r}^\mathrm{max}-\alpha_\mathrm{r}^{(1)}]$ to make a composite displacement $D[\alpha_\mathrm{r}^\mathrm{max}-\alpha_\mathrm{r}^{(1)}+\alpha_\mathrm{r}]$, and repeat such  estimation for many rounds.
The estimated value $\alpha_\mathrm{r}^{(j)}$ can be dynamically updated so that asymptotically for $j\rightarrow \infty$ when $\alpha_\mathrm{r}^{(j)}\rightarrow \alpha_\mathrm{r}$,
the composite displacement becomes $D[\alpha_\mathrm{r}^\mathrm{max}]$ and the uncertainty can be optimally minimized against the number of probes used and reach the precision predicted by the maximum of the CFI. }


\section{Unambiguous determination of displacement parameters by the prepare and measure strategy and Rabi interferometer}
\label{sec:PnMFis}

\begin{figure}[thp]
\includegraphics[width=250px]{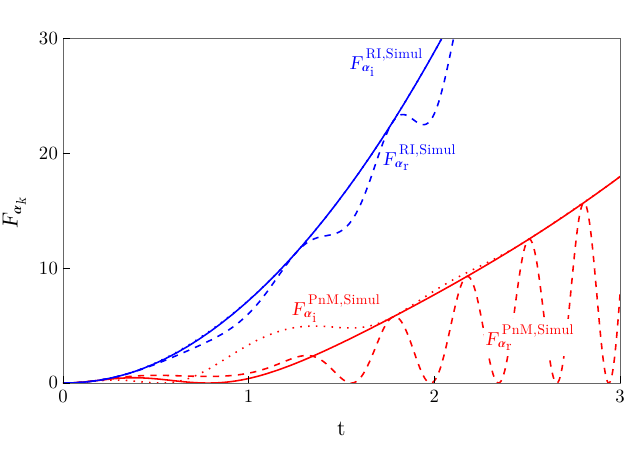}
\caption{ Comparison of CFI by prepare-and-measure strategy \old{(red) and Rabi interferometer under qubit dephasing $p=0.1$ (blue)} between an individual estimation (solid) and simultaneous estimation (dotted and dashed) at $\A_\mathrm{r}=\A_\mathrm{i}=0$.
The qubit  ancillas were set as $\ket{-_\mathrm{i}}_{A_1,A_2}$.
We note that asymptotically the CFI by simultaneous estimation in (\ref{eq:CFIPnMSimul}) is behaving similarly to that by the individual estimation in (\ref{eq:PnMCFIqubitAP}).
Here, we can see that the interference due to the simultaneous estimation can be avoided at certain strengths $t^c$.
\old{This interference can be actively cancelled using a two-qubit gate in (\ref{eq:twoqubit}).}
}
 \label{fig:simultPnM}
\end{figure}

\old{\old{In the main text}, the individual estimation of $\A_\mathrm{r}$ is described, which is equivalent to the estimation of $\A_\mathrm{i}$, when everything is substituted as $X\rightarrow P$ and $\A_\mathrm{r}\rightarrow \A_\mathrm{i}$.
}
The joint effect of the second Rabi coupling after the signal with the optimal qubit ancillary state \old{$\ket{\psi}_\mathrm{A}=c_e\ket{\mathrm{e}}_\mathrm{A}+c_g\ket{\mathrm{g}}_\mathrm{A}$} and the qubit detector afterward $\{\ket{\mathrm{e}}_\mathrm{A}\bra{\mathrm{e}}, \ket{\mathrm{g}}_\mathrm{A}\bra{\mathrm{g}}\}$ in the prepare and measure setup in Fig.~\ref{fig:RabiD} (b) is described in the operator forms as 
\begin{align}
&\hat{O}_e=\ps{\mathrm{A}}{\bra{\mathrm{e}}}\hat{R}_\mathrm{x}^\dagger(c_e\ket{\mathrm{e}}_\mathrm{A}+c_g\ket{\mathrm{g}}_\mathrm{A})=c_e\cos [t X]-\mathrm{i} c_g \sin[t X], \nonumber \\
&\hat{O}_g=\ps{\mathrm{A}}{\bra{\mathrm{g}}}\hat{R}_\mathrm{x}^\dagger(c_e\ket{\mathrm{e}}_\mathrm{A}+c_g\ket{\mathrm{g}}_\mathrm{A})=c_g\cos [t X]-\mathrm{i} c_e \sin[t X].
\end{align}
These operations conditionally act on the probe state based on the detector outcome $\ket{\mathrm{e}}_\mathrm{A}\bra{\mathrm{e}}$ or $\ket{\mathrm{g}}_\mathrm{A}\bra{\mathrm{g}}$ to give final states  $\hat{O}_{e,g}D[\alpha]\left(\ket{\mathrm{i}\frac{t}{\sqrt{2}}}_\mathrm{C}-\ket{-\mathrm{i} \frac{t}{\sqrt{2}}}_\mathrm{C}\right)$, whose norms correspond to the probability of detection given as
\begin{align}
 & P_\mathrm{e}=  \frac{\mathrm{e}^{-\frac{3 t^2}{2}} \left(\sinh \left[\frac{3 t^2}{2}\right]-\cosh \left[\frac{t^2}{2}\right]\right)
   \cos \left[\theta +2 \sqrt{2} t \alpha _\mathrm{r}\right]+1}{2}, \nonumber\\
   &P_\mathrm{g}=1-P_\mathrm{e},
\end{align}
\old{for $c_\mathrm{e}=\cos[\theta/2]\mathrm{e}^{\mathrm{i} \phi/2}$ and $c_\mathrm{g}=\sin[\theta/2]\mathrm{e}^{-\mathrm{i} \phi/2} $.}
However, the analytical form of CFI is \cmmnt{very} complex and the optimal setting depends on the target parameter $\alpha_\mathrm{r}$ and strength $t$.
When the qubit state used for the measurement  is at $\ket{-_\mathrm{i}}_{A_1}$ which is found  to be optimal for $\A_\mathrm{r}=0$,
we obtain 
\begin{widetext}
\begin{align}
   & F^{\mathrm{PnM },\ket{-_\mathrm{i}}}_{\A_\mathrm{r}}=4 \left(\mathrm{e}^{t^2}+\mathrm{e}^{2 t^2}-\mathrm{e}^{3 t^2}+1\right)^2 t^2 \cos ^2[2 \sqrt{2} t \alpha _\mathrm{r}]\nonumber\\
    &\times\frac{ \left(\mathrm{e}^{t^2}+\mathrm{e}^{2 t^2}-\mathrm{e}^{3 t^2}+1\right)^2 \cos [4 \sqrt{2} t \alpha _\mathrm{r}]-2 \mathrm{e}^{t^2}-3 \mathrm{e}^{2
   t^2}+\mathrm{e}^{4 t^2}+2 \mathrm{e}^{5 t^2}+7 \mathrm{e}^{6 t^2}-1}{\left(\left(\mathrm{e}^{t^2}+\mathrm{e}^{2 t^2}-\mathrm{e}^{3 t^2}+1\right)^2 \sin ^2[2 \sqrt{2} t \alpha _\mathrm{r}]-4 \mathrm{e}^{6 t^2}\right){}^2}.
\end{align}
\end{widetext}

When an adaptive strategy in Appendix~\ref{sec:adaptive} for optimal estimation can be used, the CFI is reduced \old{for $\alpha_\mathrm{r}\ll 1$} as 
\begin{align}
    F^\mathrm{PnM, \A_\mathrm{r}\ll 1}_{\alpha_\mathrm{r}}\old{=F^\mathrm{PnM, \A_\mathrm{i}\ll 1}_{\alpha_\mathrm{r}}}=2 t^2 e^{-6 t^2} \left(\mathrm{e}^{t^2}+\mathrm{e}^{2 t^2}-\mathrm{e}^{3 t^2}+1\right)^2 
   \label{eq:PnMCFIqubitAP}
\end{align}
in the optimal setup \old{with the second qubit input state $2^{-1/2}(\ket{+}-\ii\ket{-})$}. 
For $t\ll 1$, it scales as $8t^2$, while for $t\gg 1$, it scales as $2t^2$.

Now we summarize the simultaneous estimation precision obtained in \old{an} equivalent way in WDL:
\begin{widetext}
\begin{align}
    &F^\mathrm{PnM,Simul}_{\alpha_\mathrm{r}}= \frac{2 \mathrm{e}^{-6 t^2} t^2 \left(\mathrm{e}^{4 t^2}-\left(\mathrm{e}^{t^2}+2 \mathrm{e}^{3 t^2}-2 \mathrm{e}^{4 t^2}+\mathrm{e}^{5 t^2}\right) \cos[2 t^2]+1\right)^2}{\left(\mathrm{e}^{t^2}-\mathrm{e}^{2 t^2}+\mathrm{e}^{t^2} \cos [2 t^2]-1\right)^2}, \nonumber \\
    &F^\mathrm{PnM,Simul}_{\alpha_\mathrm{i}}=\frac{2 \mathrm{e}^{-6 t^2} t^2 \left(2 \mathrm{e}^{3 t^2}-\mathrm{e}^{4 t^2}+\mathrm{e}^{5 t^2}-2 \mathrm{e}^{4 t^2} \cos[2 t^2]+\mathrm{e}^{t^2} \cos[4 t^2]-1\right)^2}{\left(\mathrm{e}^{t^2}-\mathrm{e}^{2 t^2}+\mathrm{e}^{t^2} \cos[2 t^2]-1\right)^2}. \label{eq:CFIPnMSimul}
\end{align}
\end{widetext}
Here, the symmetry between the estimation precision of $\A_\mathrm{r}$ and $\A_\mathrm{i}$ is broken, while asymptotically it matches Eq.~(\ref{eq:PnMCFIqubitAP}) for $t>2$.
\old{The CFI of $\A_\mathrm{r}$ has an asymptotic modulation factor $\cos[2t^2]^2$ multiplied by the CFI by individual estimation.
Again, this factor is universal for all true values of $\A_\mathrm{r}$.  }
\old{We briefly note that preparation of other superposition states such as $\ket{\mathrm{i}\A}+\ket{-\mathrm{i}\A}+\ket{2\mathrm{i}\A}+\ket{-2\mathrm{i}\A}$ does not possess an enhanced CFI.}

Similarly, we can calculate the simultaneous estimation of displacement parameters by the Rabi interferometer using two ancillas as in Fig.~\ref{fig:RabiD} (c) under qubit dephasing and oscillator thermal noises.
\new{In the absence of any imperfections, the off-diagonal terms of FIM are zero, while this does not hold when imperfections such as qubit dephasing exist.}
First under the qubit dephasing,  the full expressions for CFIs are given in complex forms.
\old{Interestingly, both CFIs can be very precisely described by the CFI model in (\ref{eq:CFIDepy}-\ref{eq:avCFI}) with different dephasing levels $p_\mathrm{r}=\left(2-\cos \left[4 t^2\right]\right) p+\frac{1}{8} \left(-5+4 \cos \left[4 t^2\right]+2 \cos ^2\left[4 t^2\right]-\cos \left[8 t^2\right]\right) p^2\ge p$ and $p_\mathrm{i}=p$ in place of $p$.
This arises from the same reduction factor $\cos^2[2t^2]$ in one of the CFIs.
At certain Rabi strength $t^\mathrm{c}=\sqrt{\mathbf{c}\mathrm{\pi}/2}$ with integer $\mathbf{c}\in \mathbb{Z}$, $p_\mathrm{r}=p$ and the result is the same as the independent estimation.
Therefore, by choosing such specific strengths $t^\mathrm{c}$, we can avoid the interference occurring from the simultaneous estimation, both in Rabi interferometers, the prepare-and-measure, and the direct measurement. 
}
In WDL, they are given as
\begin{align}
&F^\mathrm{RI,Simul}_{\alpha_\mathrm{r}}=2 \mathrm{e}^{-4 t^2} t^2 \left(2-p+p \cos 4 t^2\right)^2 \left((1-p) \cosh 2 t^2+\sinh 2 t^2\right)^2,\nonumber\\
&F^\mathrm{RI,Simul}_{\alpha_\mathrm{i}}=2 \mathrm{e}^{-8 t^2} t^2 \left(\mathrm{e}^{4 t^2} (-2+p)+p \cos 4 t^2\right)^2.
\label{eq:RICFISimulAP}
\end{align}
In Fig.~\ref{fig:simultPnM}, they are compared, and we notice that asymptotically the CFI by simultaneous estimation in  (\ref{eq:CFIPnMSimul}) and (\ref{eq:RICFISimulAP}) is behaving similarly to that by the individual estimation in (\ref{eq:PnMCFIqubitAP}) and (\ref{eq:CFIDepy}) \old{at $t=t^\mathrm{c}$}.

\section{Estimation with fictitious quadrature detector and quantum Fisher information}
\label{sec:quaddet}

\begin{figure*}[thp]
\includegraphics[width=\textwidth]{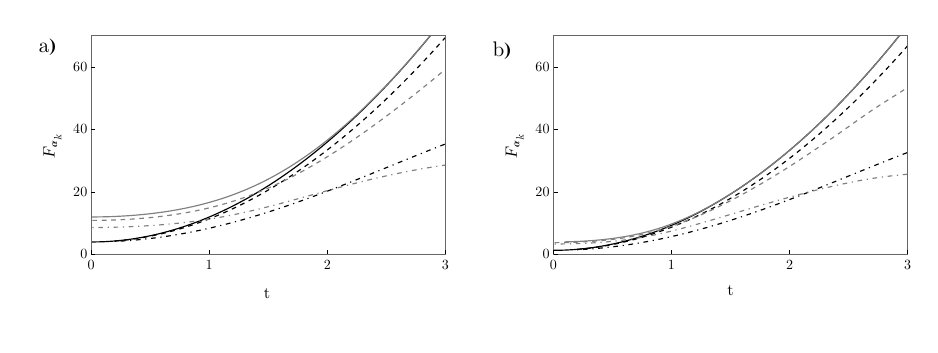}
\caption{ Comparison of CFI by Rabi interferometer and a non-interferometric setup under the same oscillator heatings $\gamma_0$, $10 \gamma_0$ and $100 \gamma_0$ as in Fig.~\ref{fig:measure-prepare} b,d) when quadrature detector is available (gray) and without it (black) for a) vacuum state and b) thermal state. 
The Rabi interferometer is superior to the prepare-and-measure strategy with an aligned quadrature detector at sufficient Rabi strengths.  
We can observe that the enhancement over the prepare-and-measure strategy with an aligned quadrature detector is even more prominent than vacuum.
}
 \label{fig:quadrature-comp}
\end{figure*}

The CFIs about $\alpha _\mathrm{r}$ and $\alpha_\mathrm{i}$ with the Rabi detector containing hypothetical field quadrature detector are  calculated to have 
$(F_{\alpha _\mathrm{r}}^\mathrm{X_{0,\mathrm{\pi}/2}}[\rho^\mathrm{th}[0]],F_{\alpha_\mathrm{i}}^\mathrm{X_{0,\mathrm{\pi}/2}})[\rho^\mathrm{th}[0]]=(4,0)$  or $(0,4)$ depending on the alignment of the phase of the quadrature detector, as this detector is sensitive to phase.
Detection of displacement in an arbitrary direction requires  an alignment of the quadrature detector to it.
 To estimate both $\alpha_\mathrm{r}$ and $\alpha_\mathrm{i}$ simultaneously, the statistical distribution of the setups into each alignment is necessary, and the CFI is halved if we equally distribute the resources between both estimations.
  For thermal states, the CFI is obtained as 
\begin{align}
    F_{\A_\mathrm{r}}^\mathrm{direct,fixed, quad}[\rho^\mathrm{th}[0];t]=\frac{4}{1+2\bar{n}},
\end{align}
reaching QFI but still below \old{the classical benchmark}.

If  a hypothetical field quadrature detector is used in the Rabi  in the prepare-and-measure setup, the CFI can be computed as
 \begin{align}
     F_{\alpha_\mathrm{r}}^\mathrm{PnM, opt, quad}[\rho^\mathrm{th}[0]]=\left(\frac{8}{\mathrm{e}^{t^2}-1}+8\right) t^2+4.
     \label{eq:PnMCFIquadrature}
 \end{align}
 \old{This is equal to QFIs about $\alpha_\mathrm{r}$ and $\alpha_\mathrm{i}$  simultaneously }  increasing monotonously by Rabi strength $t$ as $Q_{\alpha_\mathrm{r}}[\rho^\mathrm{th}[0]]=Q_{\alpha_\mathrm{i}}[\rho^\mathrm{th}[0]]=4+8 t^2+\frac{8 t^2}{\mathrm{e}^{t^2}-1}\stackrel{t\gg 1}{\rightarrow} 4+8t^2$.
This is equal to the CFI of the input cat state \old{measured by a quadrature detector}, and thus the measure-and-prepare strategy does not necessitate Rabi interaction and qubit detector.
In the  small $t$ limit, the odd cat state is reduced to a single photon state \old{which has} a larger QFI than that of the vacuum: $\lim_{t\rightarrow 0} Q_{\alpha_\old{{r,i}}}[\rho^\mathrm{th}[0]]=12$.
Again, we remark that the field mode quadrature detector is necessary to access this high sensitivity, as neither the qubit detector nor field quadrature detector alone is not sufficient.
 \old{For large $t$, it approaches scaling by a factor of $4$ \old{of the} CFI obtained without the quadrature detector, i.e. $t^2$ but with \old{a} different multiplication factor.
 It is  in  sharp contrast  to Eq.~(\ref{eq:CFIthermalDirect}) as well where  exponential decay suppresses the polynomial scaling. 
Furthermore, it shows even larger values than the CFI by the interferometric setup to be introduced below.
 }
 This can be partially explained by a high average number of excitation of the odd cat states in the oscillator, given as \old{$\langle \hat{n}\rangle=\frac{\left(\mathrm{e}^{t^2}+1\right) t^2}{2 \left(\mathrm{e}^{t^2}-1\right)}$}.
 This high average excitation number may cause it to be vulnerable to the noises on the oscillator, especially compared to the interferometric setup below.
 Again,  this \old{CFI} is accessible only by a hypothetical detector, 
and without such a detector, the CFI is reduced only back to Eq.~(\ref{CFIPnMnonzero}). 
 In addition, $\alpha_\mathrm{r}$ and $\alpha_\mathrm{i}$ cannot be measured simultaneously, because the field detector should be aligned to only one of the estimations.
 For the estimation of a displacement parameter about which the quadrature detector is not sensitive \old{(i.e. aligned to the conjugate variable estimation)}, the CFI is reduced to twice  the CFI of the previous case with only qubit detector, i.e. \old{we get} scaling of $4t^2$, with optimized qubit state at $\alpha=0$. 
 This result is interesting, as such a detector is completely insensitive to the conjugate variable displacement without the qubits.   
 Both Eq.~(\ref{eq:PnMCFIquadrature}) and (\ref{CFIPnMnonzero}) can be used as \old{a} benchmark for the estimation by \old{the} interferometric scheme in  Sec~\ref{sec:comparison}.
 
 To understand our result (\ref{eq:Freal}) better, we can compare with \old{the} maximum of CFI using virtual field quadrature detection of $\hat{X}$  after the interferometer.
The total probability density is given as 
\begin{align}
&P_\mathrm{e}(X)=\frac{\mathrm{e}^{-\mathrm{e}^{2 r} (\alpha_\mathrm{r}-X)^2}  \sin ^2\left[\sqrt{2}   \alpha_\mathrm{r} t\right]}{\sqrt{\mathrm{\pi} } \mathrm{e}^{-r}},\nonumber\\
&P_\mathrm{g}(X)=\frac{\mathrm{e}^{-\mathrm{e}^{2 r} (\alpha_\mathrm{r}-X)^2}  \cos ^2\left[\sqrt{2}   \alpha_\mathrm{r} t\right]}{\sqrt{\mathrm{\pi} } e^{-r}},
\end{align}
where $X$ is the quadrature detection outcome.
The CFI about $\alpha_\mathrm{r}$ \cmmnt{and $\alpha_\mathrm{i}$} reaches QFI $Q_{\alpha_\mathrm{r}}$
regardless of the value of $\alpha_\mathrm{r}$ and $\alpha_\mathrm{i}$.
This equality to the QFI after the first Rabi interaction  implies that this detection setup is  optimal for $\alpha_\mathrm{r}$ estimation.
We again note that for the simultaneous estimation, this detector does not enhance the CFI of the estimation of $\alpha_\mathrm{i}$.
 CFI at WDL with quadrature detector is given as
 \begin{align}
  F^\mathrm{PnM,opt,quad}_{\A_\mathrm{r}}[\rho^\mathrm{th}[\bar{n}]]=  \frac{8 \left(4 t^2 \bar{n}-\mathrm{e}^{-t^2 \left(2 \bar{n}+1\right)}+2 t^2+1\right)}{\left(2 \bar{n}+1\right) \left(2-2 \mathrm{e}^{-t^2 \left(2 \bar{n}+1\right)}\right)}.
 \end{align}
 Here again, a thermal state is not very detrimental.
 The asymptotic expression is given as $\lim_{t\rightarrow \infty}F^\mathrm{PnM,opt,quad}_{\A_\mathrm{r}}[\rho^\mathrm{th}[\bar{n}]]=8t^2$ for $t\gg 1$, regardless of $\bar{n}$. 
 In the limit of $t\ll 1$, it is given as $\lim_{t\rightarrow 0}F^\mathrm{PnM,opt,quad}_{\A_\mathrm{r}}[\rho^\mathrm{th}[\bar{n}]]=12/(1+2\bar{n})$, where the thermal probe decreases the CFI most.

If a virtual field quadrature measurement on the oscillator $\{\ket{X}_\mathrm{C}\bra{X}\}$ in eigenbasis of $\hat{X}$ is assumed in the Rabi interferometer, then we obtain  the CFI $F^{\mathrm{interf},x}_{\alpha_\mathrm{r}}[\rho^\mathrm{th}[0]]=8t^2+4$,  while the conjugate component has  $F^{\mathrm{interf},x}_{\alpha_\mathrm{i}}[\rho^\mathrm{th}[0]]=8t^2$. 
For a thermal light, it is modified as $F_{\alpha_\mathrm{r}}^{\mathrm{interf}, x}[\rho^\mathrm{th}[\bar{n}];t]=8t^2+\frac{4}{1+2\bar{n}}$ and $F_{\alpha_\mathrm{r}}^{\mathrm{interf}, x}[\rho^\mathrm{th}[\bar{n}]]=8t^2$ with quadrature detector \old{$\{\ket{X}_\mathrm{C}\bra{X}\}$}, and vice versa for the alignment  as \old{$\{\ket{P}_\mathrm{C}\bra{P}\}$}. 
The CFI  using simultaneous measurement of $\hat{X}$ and $\hat{P}$ with a POVM form $\{\mathrm{\pi}^{-1}\ket{\beta}_\mathrm{C}\bra{\beta}\}$ together with a qubit detector at the end of the setup is given as $F_{\alpha_\mathrm{r,i}}=8t^2+2$, thus still not reaching \old{but approaching} the QFIs in (\ref{eq:PnMQFI}).

\old{In Fig.~\ref{fig:simultPnM}, we compared the Rabi interferometer and prepare-and-measure strategy under noises.
As was noted above, in \old{the} ideal channel, prepare-and-measure has a larger CFI than Rabi interferometer.
Interestingly, interferometers can have a higher CFI even without \old{a} quadrature detector than \old{the} prepare-and-measure strategy with a quadrature detector with sufficient Rabi strength when the heating exist\old{s}.
This is partly due to a large number of excitation in the oscillator in \old{the} prepare-and-measure strategy, which therefore exposes a vulnerability to oscillator noises.
Remarkably, the advantage of the Rabi interferometer is more prominent in the case of \old{the} input thermal state in the oscillator.}

\section{Engineering inverse Rabi interaction using auxiliary oscillators}
\label{sec:inverseRabi}
The key element of an interferometer is the accessibility of an inverse interaction after the signal.
The inverse Rabi interaction can be engineered by \old{two methods: first  by adding a $\mathrm{\pi}$-phase shift to the oscillator to make $\hat{X}\rightarrow -\hat{X}$,
or} 
using  qubit rotations as evidenced in the following transformation equation:
\begin{align}
\exp[\mathrm{i} T \hat{\sigma}_\mathrm{y}]\hat{\sigma}_\mathrm{x}\exp[-\mathrm{i} T \hat{\sigma}_\mathrm{y}]=\hat{\sigma}_\mathrm{x} \cos[2T]+\hat{\sigma}_\mathrm{z} \sin[2T]
\label{eq:qubitrot}
\end{align}
for an arbitrary $T$, and at $T=\mathrm{\pi}/2$, it becomes $\hat{\sigma}_\mathrm{x} \rightarrow -\hat{\sigma}_\mathrm{x}$. 
By this transformation, a Rabi interaction becomes an inverse Rabi interaction $\exp[\mathrm{i} t \hat{\sigma}_\mathrm{x} \hat{X}]\rightarrow\exp[-\mathrm{i} t \hat{\sigma}_\mathrm{x}\hat{X}]$.
These qubit rotations necessary for such a conversion can in turn be engineered from other Rabi interactions with  a strong drive:
\begin{align}
\exp[\mathrm{i} k \hat{\sigma}_\mathrm{y} \hat{X}']\ket{\phi}\ket{\delta}_{C''}\approx \exp[\mathrm{i} k \hat{\sigma}_\mathrm{y} \sqrt{2}\delta]\ket{\phi}\ket{\delta}_{C''},
\end{align}
and at $k=\frac{\mathrm{\pi}/2}{\sqrt{2}\delta}$, we can achieve a flip.
Alternatively, using only JC coupling we have
\begin{align}
\exp[\mathrm{i} \tau \sigma_+ a+\mathrm{i} \tau \sigma_- a^\dagger]\ket{\phi}\ket{\mathrm{i}\delta}_{C''}\approx \exp[-\mathrm{i} \tau \sigma_\mathrm{y} \delta]\ket{\phi},
\end{align}
and at $\tau=\frac{\mathrm{\pi}}{2\delta}$ we achieve \old{an} approximate qubit flip.
If we have an access to squeezed state in this auxiliary oscillator, we can have an improved approximation by substituting $\ket{\delta}_{C''}\rightarrow D[\delta]\ket{r}_{C''}$ where $\ket{r}$ is a squeezed state.
The gate fidelity is enhanced by such a substitution as $0.962\rightarrow 0.995$ at $\mathbf{\delta}=4$ and $r=1$.

\section{Derivation of the probabilities for Rabi interferometer}
\label{sec:CFIRabi}

\old{This} operator $\old{\hat{U}_\mathrm{interf}}$ \old{in (\ref{eq:totalaction})} acts on an arbitrary input \old{pure} state  $\ket{\varphi}_\mathrm{C}$ \old{in the oscillator}  and the initial qubit  state $\ket{\phi^\mathrm{opt}}=\ket{\pm_\mathrm{i}}_{\mathrm{A}_1}$ \cmmnt{optimal at WDL under the absence of noises \old{or loss}} \cmmnt{optimal for small $\alpha$} to give the final state
\begin{align}
\ket{\Psi'}=\old{\hat{U}_\mathrm{interf}}\ket{\varphi}_\mathrm{C}\ket{\pm_\mathrm{i}}_{\mathrm{A}_1}
=\hat{D}[\alpha]\ket{\varphi}_\mathrm{C} 
\underbrace{\hat{R}_\mathrm{x}[-\sqrt{2}\alpha_\mathrm{r} t]\ket{\pm_\mathrm{i}}_{\mathrm{A}_1}}_{\equiv\ket{\phi}_{\mathrm{A}_1}}
\label{eq:finalRI}
\end{align}
This state \old{$\ket{\Psi'}$} is factorized \old{into \old{the} local oscillator and ancilla states}, \cmmnt{and}\old{where} the final qubit state $\ket{\phi}_{\mathrm{A}_1}$ is obtained independent of the initial CV state \old{$\ket{\varphi}_\mathrm{C}$}, \old{due to the geometric phase attained by the entire process}. 
\cmmnt{which} \old{This factorization} gives \cmmnt{a} many robust \old{properties} of this interferometric estimation protocol \cmmnt{which} \old{and} enables \old{it} to work on realistic noises. 
\old{We also note that the output state from a mixed state  $\rho_\mathrm{C}$ in the oscillator  is given simply as 
\begin{align}
\old{\hat{U}_\mathrm{interf}}\rho_\mathrm{C}\old{\hat{U}_\mathrm{interf}}^\dagger=\hat{D}[\A]\rho_\mathrm{C} \hat{D}[-\A]\otimes \ket{\phi}_{\mathrm{A}_1}\bra{\phi}.
\label{eq:finalRImixed}
\end{align}}

For \old{the} estimation of  $\A_\mathrm{r}$, we detect with only \old{a} qubit detector in $\hat{\sigma}_\mathrm{z}$-eigenbasis $\{\ket{\mathrm{e}}_{\mathrm{A}_1}\bra{\mathrm{e}},\ket{\mathrm{g}}_{\mathrm{A}_1}\bra{\mathrm{g}}\}$ \old{at \old{one} arm} without any detection in the oscillator \cmmnt{for realistic performance check} \cmmnt{as a realistic restriction} to get the  probabilities \old{for \old{binary} outcomes}
\begin{align}
P_\mathrm{g}(\alpha_\mathrm{r})=\cos^2[\sqrt{2}\alpha_\mathrm{r} t],~P_\mathrm{e}(\alpha_\mathrm{r})=\sin^2[\sqrt{2}\alpha_\mathrm{r} t].
\label{eq:RabiProb}
\end{align}
We can simultaneously estimate  $\alpha_\mathrm{i}$ by a natural extension of the setup \cmmnt{using more} to that with two-qubit ancillas \old{where the second} qubit ancilla interact\old{s} with the oscillator by another type of  Rabi interaction $\hat{R}'_\mathrm{P} =\exp[\ii t' \hat{\sigma}_\mathrm{x}' \hat{P}]$ and its inverse where prime represents  the second ancillary mode.
We obtain similarly the probabilities for qubit detection outcomes given as $P_\mathrm{g}=\cos^2[\sqrt{2}t'\alpha_\mathrm{i}]$ and $P_\mathrm{e}=\sin^2[\sqrt{2}t'\alpha_\mathrm{i}]$\old{, importantly, where the joint probabilities for estimation of   $\alpha_\mathrm{r}$ and $\alpha_\mathrm{i}$ factorizes}, from which the same CFI is obtained as $F^\mathrm{interf}_{\alpha_\old{{r,i}}}[\rho^\mathrm{th}[0]]=8t^2$.

\old{Under the existence of qubit dephasing noise, the probability is reduced in visibility to 
$P_\mathrm{e}=\frac{2+\left\{2-p \left(e^{-4 t^2}+1\right)\right\} \sin \left[2 \sqrt{2} t \alpha _r\right]} {4}$, and $P_\mathrm{g}=1-P_\mathrm{e}$.
This formula was used in the derivation of the CFI in (\ref{eq:CFIDepy}).
}

\section{Extended Rabi interferometer with multiple qubits}
\label{sec:entangledqubit}

\cmmnt{Moreover,} Entangled qubits such as $2^{-1/2}(\ket{\mathrm{e}}_{\mathrm{A}_1}\ket{\mathrm{e}}_{\mathrm{A}_2}+\ket{\mathrm{g}}_{\mathrm{A}_1}\ket{\mathrm{g}}_{\mathrm{A}_2})$ can make further enhancement for the Rabi interferometers over the separable qubit ancillas. 
With a straightforward extension of the calculation where the same qubit rotations $\hat{R}[-\sqrt{2}\alpha_\mathrm{r} t]$ are applied on both qubits, the qubit detectors on $\sigma_\mathrm{z}$ eigenbasis will give CFI $F_{\alpha_\mathrm{r}}=32t^2$, \old{twice} larger than $16 t^2$  from simply additive  Fisher information from separable two-qubit ancillas with $m=2$.
This result implies that the entangled state of $m$ qubits may have a larger scaling of CFI.
A simple GHZ state with $m$ qubits has a linear scaling of $16m t^2$ by the Rabi interferometer, calculated in the same way.
These states can be approximately generated using Rabi interactions before the interferometer from a vacuum oscillator in the preparation stage, as
\begin{align}
    &\exp[\mathrm{i} \frac{\mathrm{\pi}}{4t}\sum_{j=2}^m\hat{\sigma}_\mathrm{y}^{(\mathrm{A}_j)} \hat{P}]\exp[\mathrm{i} t\hat{\sigma}_\mathrm{z}\hat{X}]\ket{+}_{\mathrm{A}_1}\ket{+}_{\mathrm{A}_2}...\ket{+}_{\mathrm{A}_m}\ket{0}_\mathrm{C}\nonumber\\
    &\approx 2^{-1/2}\ket{\mathrm{e}}_{\mathrm{A}_1}\ket{\mathrm{e}}_{\mathrm{A}_2}...\ket{\mathrm{e}}_{\mathrm{A}_m}\ket{\mathrm{i}m t/\sqrt{2}}_\mathrm{C}+2^{-1/2}\ket{\mathrm{g}}_{\mathrm{A}_1}\ket{\mathrm{g}}_{\mathrm{A}_2}...\ket{\mathrm{g}}_{\mathrm{A}_m}\ket{-\mathrm{i}mt/\sqrt{2}}_\mathrm{C},
\end{align}
where the approximation works well for a large $t$.
The Rabi interactions $\exp[\mathrm{i} t \sigma_\mathrm{z} \hat{X}]$ can be obtained from $\exp[\mathrm{i} t \sigma_\mathrm{x} \hat{X}]$ by the adjoint application of qubit rotation as in Eq.~(\ref{eq:qubitrot}) setting $T=\mathrm{\pi}/4$.
This result implies that the exploration of a highly complex \old{qubit} circuit with Rabi interactions generating broader ranges of entangled states can be profitable.

\section{Asymmetric Rabi interactions}
\label{sec:AssymmetricStrength}
We can check if different strengths and types of Rabi interactions before and after the displacement may  have any advantages for the estimation by the Rabi interferometer.
With the same calculation as in the main text, we obtain a CFI
\begin{align}
F_{\alpha_\mathrm{r}}=\frac{8 \mathrm{e}^{4 t t'} t'^2 \left(-1+\mathrm{e}^{4 \mathrm{i} \sqrt{2} \alpha_\mathrm{r} t'}\right){}^2}{2 \mathrm{e}^{4 t' \left(t+\mathrm{i}
   \sqrt{2} \alpha_\mathrm{r}\right)}+\mathrm{e}^{4 t' \left(t+2 \mathrm{i} \sqrt{2} \alpha_\mathrm{r}\right)}-4 \mathrm{e}^{2 \left(t^2+t' \left(t'+2 \mathrm{i}
   \sqrt{2} \alpha_\mathrm{r}\right)\right)}+\mathrm{e}^{4 t t'}}.
   \label{eq:CFIAsym}
   \end{align}
We can see that Eq.~(\ref{eq:CFIAsym}) has \old{a} dependence on the $\alpha_\mathrm{r}$. 
The maximum value of Fisher information over $\alpha_\mathrm{r}$ reaches the FI of the value $8 \max[t,t']^2$, while for certain $\alpha_\mathrm{r}$ it drops to $0$.
A bigger difference in the strengths makes an adaptive estimation more suitable.

On the other hand, exploiting two   Rabi interferometer\old{s} utilizing  slightly different strengths $t,t'$ on two subensembles can uniquely determine $\alpha_\mathrm{r}$ in all ranges non-adaptively, as there is a unique solution of $\alpha_\mathrm{r}\in [-\infty,\infty]$ for $(P_g,P_g')=(\cos^2[\sqrt{2}\alpha_\mathrm{r} t],\cos^2[\sqrt{2}\alpha_\mathrm{r} t'])$.
If we use half of the total $n$ probes into an interferometer for duration $t$ and the remaining half into that for duration $t'$, the CFI is obtained as $8n (t^2+t'^2)/2$.
This result implies that as $t'$ approaches $t$, the whole-range estimation is attained with the same precision as Eq.~(\ref{eq:Freal}).


\section{High-order Rabi interferometers}
\label{sec:non-linear}


 A high-order Rabi gates $\hat{R}_{X^{(k)}}=\exp[\mathrm{i} t \hat{\sigma}_\mathrm{x} \hat{X}^k]$ for $k>1$ can have enhancement  over linear Rabi gate when used in the interferometers.
These gates can be engineered from linear Rabi gates by using multiple of them as in \cite{ParkNJP2018}.
In these high-order Rabi interferometers, we use the following identity to obtain the  transformation of the signal:
\begin{align}
    \exp[\mathrm{i} t \hat{\sigma}_\mathrm{x} \hat{X}^k]\hat{D}[\alpha]\exp[-\mathrm{i} t \hat{\sigma}_\mathrm{x} \hat{X}^k]=\hat{D}[\alpha]\exp[\mathrm{i} t \hat{\sigma}_\mathrm{x} \{(\hat{X}+\sqrt{2}\alpha_\mathrm{r})^k-\hat{X}^k\}]\stackrel{\alpha_\mathrm{r}\ll 1}{\approx} \hat{D}[\alpha]\exp[i t \hat{\sigma}_\mathrm{x} \sqrt{2}k \alpha_\mathrm{r} \hat{X}^{k-1}].
\end{align}
Now the  probabilities of qubit detector  outcomes are given as 
\begin{align}
    P_g=\ps{C}{\bra{\psi}}\cos[\sqrt{2} k t \hat{X}^{k-1} \alpha _\mathrm{r}]\ket{\psi}_\mathrm{C}, ~ P_e=1-P_g.
\end{align}
For vacuum/ground state in the oscillator, we obtain \old{the} following exact expressions without approximation for CFI of lowest $k$'s (superscript denotes the order of Rabi interactions):
\begin{align}
    &F_{\alpha_\mathrm{r}}^{(k=2)}=\frac{256 t^4 \alpha _\mathrm{r}^2}{\mathrm{e}^{16 t^2 \alpha _\mathrm{r}^2}-1},\nonumber\\
    &F_{\alpha_\mathrm{r}}^{(k=3)}=-\frac{9 t^2 \left(\mathrm{i} \left(\sqrt{1-6 \mathrm{i} \sqrt{2} t \alpha _\mathrm{r}}-\sqrt{1+6 \mathrm{i} \sqrt{2} t \alpha _\mathrm{r}}\right)+6
   \sqrt{2} t \alpha _\mathrm{r} \left(\sqrt{1-6 \mathrm{i} \sqrt{2} t \alpha _\mathrm{r}}+\sqrt{1+6 \mathrm{i} \sqrt{2} t \alpha
   _\mathrm{r}}\right)\right){}^2}{\left(72 t^2 \alpha _\mathrm{r}^2+1\right){}^2 \left(-144 t^2 \alpha _\mathrm{r}^2+\sqrt{72 t^2 \alpha
   _\mathrm{r}^2+1}-1\right)}.
\end{align}
In the limit of $\alpha_\mathrm{r} \rightarrow0$, they are reduced to $F_{\alpha_\mathrm{r}\ll 1}^{(k=2)}=16 t^2$, $F_{\alpha_\mathrm{r}\ll 1}^{(k=3)}=54 t^2$, and $F_{\alpha_\mathrm{r}\ll 1}^{(k=4)}=240 t^2$.
However, we note that this enhancement is present around the vicinity of a small $\alpha_\mathrm{r}$ limit, and in a larger $\alpha_\mathrm{r}$ the linear Rabi interferometer has a larger CFI.
In contrast,  a high amplitude coherent state probe $\ket{\beta}$ with $\beta\gg 1$ can be exploited in these high order Rabi interferometers in the same parameter regions for better performances, with CFI scaling as $F_{\alpha_\mathrm{r}}^{(k)}=2^{k+2} k^2 t^2 \beta ^{2 k-2}$.

\section{Enhancement of Rabi strengths by squeezing}
We briefly note that \old{online} squeezing can further enhance the estimation sensitivity simultaneously in both displacement components.
For example, we have
\begin{align}
&S[r]\exp[\mathrm{i} t \hat{\sigma}_\mathrm{x} \hat{X}]S[-r]=\exp[\mathrm{i} t e^{2r}\hat{\sigma}_\mathrm{x} \hat{X}],\nonumber\\
&S[-r]\exp[\mathrm{i} t \hat{\sigma}_\mathrm{x} \hat{P}]S[r]=\exp[\mathrm{i} t e^{2r}\hat{\sigma}_\mathrm{x} \hat{P}].
\end{align}
In principle, the squeezing transformation can \old{also} be provided by the Rabi interactions.
Eq. (\ref{eq:totalaction}) implies such an amplification, as the qubit rotation produced by the Rabi interferometer can be transferred back to the oscillator displacement by another set of interactions. 
We leave again the detailed analysis of such a strategy for future investigations.

\section{Interferometers based on other types of interactions}
\label{sec:JC inter}

\old{A} JC interaction under RWA is \old{experimentally} available in much broader systems than Rabi interactions \old{with significantly lower difficulty in the implementation}.
Overcoming an interferometer built from \old{such} JC interactions \cmmnt{\old{in Fig.~
\ref{fig:RabiD} (d)}} is an important benchmark  \cmmnt{to clarify}\old{in the clarification of} the role  of the counter-rotating term in Rabi interaction \old{and architecture built from it} for the estimation. 
We \old{first} note that \old{the} JC interferometer does not alter the ground state of joint qubit-oscillator   system
$\ket{\psi}_\mathrm{in}=U_\mathrm{JC}\ket{0}_\mathrm{C}\ket{\mathrm{g}}_\mathrm{A}=\ket{0}_\mathrm{C}\ket{\mathrm{g}}_\mathrm{A}$ \old{and is thus effectless in this case}, and \cmmnt{thus}\old{therefore} \old{an}
initial preparation of the excited  state in the qubit is required for the  estimation beyond \cmmnt{shot-noise limit}\old{the \old{classical} benchmark}. 
The action of \old{a} JC interaction on such a \old{qubit excited}  state prepares an entangled state  
$\ket{\psi}_\mathrm{in}=U_\mathrm{JC}\ket{0}_\mathrm{C}\ket{\mathrm{e}}_\mathrm{A}=\cos[\tau]\ket{0}_\mathrm{C}\ket{\mathrm{e}}_\mathrm{A}+i \sin[\tau]\ket{1}_\mathrm{C}\ket{\mathrm{g}}_\mathrm{A}$.
The QFI of this state about $\alpha_\mathrm{r}$ \old{after the displacement} is given as 
\begin{align}
Q_\mathrm{\alpha_\mathrm{r}}=8-4\cos[2\tau], \label{eq:QFIJC}
\end{align}
having the maximum value $12$ at $\tau=\frac{\mathrm{\pi}}{2}$ at which the  single quanta $\ket{1}_\mathrm{C}$ in the oscillator is prepared completely decoupled from the qubit.
This QFI \old{in (\ref{eq:QFIJC})} is observed to be always smaller than that of the state prepared by Rabi interaction of the same strength $t=\tau$.

We can now proceed to calculate the CFI of the JC interferometric setup \old{with only qubit detectors}.
After the \old{unknown signal} displacement \cmmnt{operator} and the inverse JC interaction, the state evolves to
$\ket{\psi}_\mathrm{out}=U_\mathrm{JC}^{-1}\hat{D}[\alpha]\ket{\psi}_\mathrm{in}$.
Now we can detect on the qubit \cmmnt{again} in the energy eigenbasis, \old{and} the CFI can be  calculated  from the following probability \old{of detection outcome in the qubit excited state} in Fock basis expansion:
\begin{align}
    P_e=\sum_n \frac{\mathrm{e}^{-|\alpha|^2} |\alpha|^{2n} \left\{\tau  (n-|\alpha|^2+1) \sin \tau ~ \mathrm{sinc}\left(\sqrt{n+1} \tau \right)+\cos \tau  \cos
   \left(\sqrt{n+1} \tau \right)\right\}^2}{n!}. \label{eq:JCFock}
\end{align}
This probability is  dependent only on $|\alpha|$ and thus naturally more suitable for the estimation of it, \old{but not the phase of the displacement parameter $\arg[\A]$}.
This \cmmnt{weakness}\old{limitation} in the full characterization of the displacement \old{including the phase} can be partially mitigated by exploiting a known auxiliary $D[\beta]$ \old{or equivalently a coherent state probe}, by which the phase relation between $\A$ and $\beta$ can be used.
Th\cmmnt{e}\old{is} summation \old{in (\ref{eq:JCFock})}, however, is difficult to be performed analytically, and we perform a numerical simulation.

We note the equivalence of the protocols with those achieved by the dispersive interactions available in superconducting systems~\cite{vanLoock2008HybridOptics}.
For example, the interferometer by dispersive interactions can be achieved with the total operation before the detection:
\begin{align}
  \exp[-\mathrm{i} \frac{\mathrm{\pi}}{2} \hat{\sigma}_\mathrm{z} \hat{n}_\mathrm{C}]\exp[-\mathrm{i} \frac{\mathrm{\pi}}{4} \hat{\sigma}'_\mathrm{z} \hat{n}_\mathrm{C}]  D_\mathrm{C}[\A] \exp[\mathrm{i} \frac{\mathrm{\pi}}{4} \hat{\sigma}'_\mathrm{z} \hat{n}_\mathrm{C}]\exp[\mathrm{i} \frac{\mathrm{\pi}}{2} \hat{\sigma}_\mathrm{z} \hat{n}_\mathrm{C}]\ket{-}_{\mathrm{A}_1}\ket{-}_{\mathrm{A}_1}\ket{\beta}_\mathrm{C},
\end{align}
which can be shown to have the same CFI, in a similar spirit to \cite{PenasaPRA2016}.

Furthermore, if we have an access to the squeezed state in the oscillator, we can obtain an entangled displaced squeezed state.
\new{It is well known that the most fundamental bound and optimal method in single displacement parameter estimation is given by the precision of estimation using squeezed vacuum state probes (with properly aligned homodyne detectors). 
How to perform such an estimation for the complete displacement parameter was introduced in \cite{Park2022}, but here comparison to a single component estimation bound is sufficient.}
\new{
 We consider the possibility of squeezing, but the oscillator remains in a thermal state. 
 The QFI of these squeezed thermal probes can be compared to the Rabi interferometer under the constraint of a fixed average photon number. 
 In this scenario, the QFI is given simply as $4e^{-2r}$ regardless of the average photon number in the thermal state, where r is the squeezing parameter. 
 The CFI of such a probe for the Rabi detector is given as $\frac{4e^{-1-2r}}{1+2\bar{n}}$, and that for the direct quadrature detector as $\frac{4e^{-2r}}{1+2\bar{n}}$, enhance by a factor $e$.
 The average photon number contained in the probe before and after the squeezing is given by $\bar{n}$ and $\bar{n}\cosh[r]^2+(\bar{n}+1)\sinh[r]^2$ respectively. 
 We can compare our protocols that has the average photon number after the first Rabi interaction given by $\bar{n}+t^2/2$, while the CFI is given as $8t^2$.

 On the other hand, we can generalize the analysis to general types of qubit-oscillator interaction of the form $\exp[i t \hat{H}]$, where $\hat{H}$ is  an arbitrary Hermitian matrix residing on the joint space of SU(2) and finite dimensional Fock subspaces, acting on the thermal states. 
 The random interferometers were generated by mixing random Hermitian operators $\hat{H}_\mathrm{rand}$ with the Rabi interaction operator $\hat{\sigma}_\mathrm{x} \hat{X}$ in an oscillator system with variable weights $w_1,w_2$ as $\hat{H}=w_1 \hat{H}_\mathrm{rand}+w_2 \hat{\sigma}_\mathrm{x} \hat{X}$.

\begin{figure}[thp]
\includegraphics[width=\textwidth]{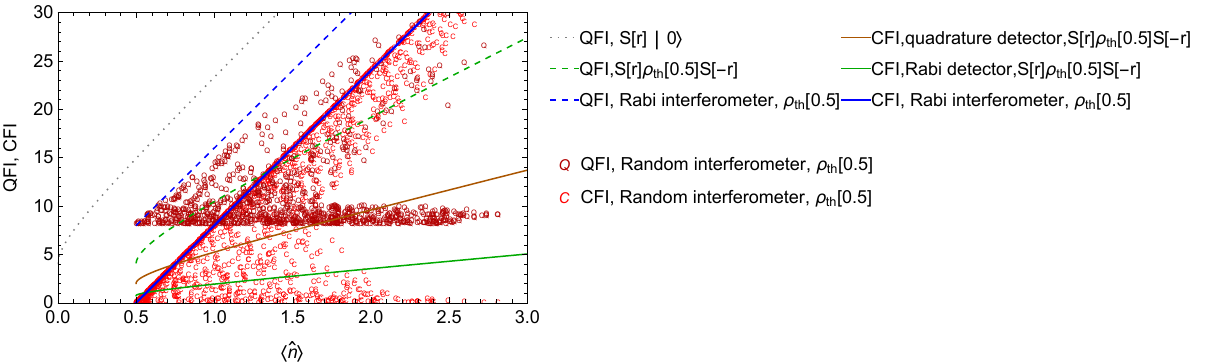}
\caption{ Comparison of the Rabi interferometer against various methods for various probe average photon numbers. 
The results show that squeezed thermal state probes (green) and thermal states interacting with randomly sampled unitary operators with qubits (red characters, Q: QFI and C: CFI) perform at most as Rabi interferometers, which are optimal in terms of both QFI and CFI by qubit detection.  
This conclusion holds for a much larger set of data, even though only a limited number of points were shown for visualization. }
 \label{fig:comprand}
\end{figure} 

Figure \ref{fig:comprand} illustrates these scenarios. 
The results indicates that the Rabi interferometer performs better than squeezed thermal probes in the large average photon number limit and is predicted to be optimal among the random interferometers studied numerically. 
Still, it leaves a space for further investigation  of other type of setups with detection on the oscillators and a larger number of qubit detection can surpass the Rabi interferometer. 
}

\section{Minor noises}
\label{sec:minor}

\textbf{Dependence of CFI on the target variable under noise} We note that \cmmnt{our scheme} \old{the} Rabi interferometer has a \old{dependence on the target parameter $\alpha_\mathrm{r}$} \old{in the precision} under noise, not well explained by CFI \cmmnt{alone}in \old{an} ideal situation.
\cmmnt{This is well shown in} \old{Let us consider} the example of \old{the} exact identity channel $\alpha_\mathrm{r}=0$, where the final qubit state \cmmnt{$\ket{\mathrm{g}}$} is \cmmnt{recovered by}\old{equal to the input qubit state after} the interferometer regardless of $t$.
This can cause an issue \old{of indeterminacy} in the calculation of CFI when \old{a} detector is detecting in that basis, as the \cmmnt{excited state detection}\old{orthogonal state} probability \cmmnt{$P_e(0)$} and  \cmmnt{the}\old{its} derivatives of the probabilities by $\alpha_\mathrm{r}$ give $0$\cmmnt{, and therefore causing an issue of indeterminacy of the CFI}.
For non-zero $\alpha_\mathrm{r}$, a similar problem happens at specific $ t_\mathrm{inst}=\mathrm{\pi}/(2\sqrt{2}\alpha_\mathrm{r})$ \old{when the noise is weak}.
Both of these instabilities can be avoided in practice, as generally the displace signal is weak but not  strictly zero, \old{and therefore $ t_\mathrm{inst}$ is much larger than experimentally available values}.
Alternatively, we can avoid this issue by adopting a different input qubit state, such as $\ket{\pm_\mathrm{i}}_\mathrm{A}$.
We notice that these instability points tend to make CFIs  prone to noises\old{, e.g. oscillator thermal noise}.
\old{This instability is impacted by any noise in the form of the fluctuation of detection probabilities, which are smeared by noise, thus sharply reducing CFI.}
\cmmnt{We note that this is a unique limitation for the interference, not existing in direct measurement scheme or prepare-and-measure scheme.}

\textbf{Loss on CV mode} The boson loss on the CV mode is a special case of the heating with the replacement $\bar{n}=0$, and can be described by a Lindblad operator with $L=\sqrt{\gamma}\hat{a}$. 
The loss parameter can be redefined as $\sqrt{\eta}=\mathrm{e}^{-\gamma \old{t_f}}$, \old{where $t_f$ is the duration in time when the loss is being applied}.
The effect of loss on the Rabi interferometer can be analytically calculated, and the probabilities are obtained as
\begin{align}
P_e=\frac{1}{2} \left(1-\mathrm{e}^{-2 t^2\left(1-\sqrt{\eta }\right) } \cos [2 \sqrt{2 \eta } t \alpha_\mathrm{r}]\right), ~P_g=1-P_e.
\end{align}
The CFI calculated from these probabilities is given as
\begin{align}
F_{\alpha_\mathrm{r}}=\frac{8 \eta t^2  \cos^2\left(2 \sqrt{2 \eta } t \alpha
   _\mathrm{r}\right)}{\mathrm{e}^{4 \left(1-\sqrt{\eta }\right) t^2}- \sin^2\left(2 \sqrt{2 \eta } t \alpha _\mathrm{r}\right)}.
\end{align}
This Fisher information can be rewritten as 
$F_{\alpha_\mathrm{r}}=8 \eta t^2  \frac{B}{A-(1-B)}$ with $A=\mathrm{e}^{4 t^2 \left(1-\sqrt{\eta }\right)}$ and $B=\cos^2\left(2 \sqrt{2 \eta } t \alpha
   _\mathrm{r}\right)$, showing monotonous behaviors in $A$ and $B$ in the range of values $A\in[1,\infty]$,$B\in [0,1]$.
   For the target values which can be most sensitively estimated satisfies $B=1$, in which case the CFI is maximized as $F_{\alpha_\mathrm{r}}=8 \eta \frac{t^2} {A} $.
   \old{The maximum CFI  can be found in WDL as $\frac{8 \eta}{f e}$ at optimized $t=f^{-1/2}$ for $A=\mathrm{e}^{f t^2}$.
   Here $f=4(1-\sqrt{\eta})$.
   }

\textbf{Gaussian distribution of the displacement signal}
We can now consider a case where the unknown displacement does not have a precise strength due to the existence of noise force but is uncertain with a Gaussian distribution.
\cmmnt{(What do we know about the noise? How is it considered in the estimation?)
The variance $\sigma_{r,i}$ of the respective parameters $\alpha_{r,i}$ are assumed to be unknown.}
This present\old{s}  one major effect, especially when an external force is weak.
The encoded state from the input state $\rho$ is then given as
\begin{align}
\rho_\mathrm{out}=(2\mathrm{\pi} \sigma_\mathrm{r} \sigma_\mathrm{i})^{-1}\int d\alpha_\mathrm{r} d\alpha_\mathrm{i} \exp[-\frac{(\alpha_\mathrm{r}-\alpha_{r0})^2}{2\old{\sigma}_\mathrm{r}^2}-\frac{(\alpha_\mathrm{i}-\alpha_{i0})^2}{2\old{\sigma}_\mathrm{i}^2}]\hat{D}[\alpha_\mathrm{r}+\ii \alpha_\mathrm{i}]\rho \hat{D}[-\alpha_\mathrm{r}-\ii \alpha_\mathrm{i}].
\end{align}
Now in the Rabi interferometer for the estimation of the real displacement parameter, the probabilities of detection are given as
\begin{align}
P_e=\frac{1}{2} \left(1-\mathrm{e}^{-4 \sigma_\mathrm{r}^2 t^2} \sin \left(2 \sqrt{2} \alpha _{r0} t\right)\right), ~P_g=1-P_e.
\end{align}
The CFI from these probabilities is given as 
\begin{align}
F_{\alpha_\mathrm{r}}=\frac{8 t^2 \cos^2\left(2 \sqrt{2} \alpha _{r0} t\right)}{\mathrm{e}^{8 \sigma_\mathrm{r}^2 t^2}-\sin^2\left(2
   \sqrt{2} \alpha _{r0} t\right)}.
   \label{eq:FIGauss}
\end{align}
We can see the correspondence with the previous result when $\sigma_\mathrm{r}\rightarrow 0$.
Interestingly, at strengths $t \alpha_{r0}=\frac{\mathrm{\pi}}{2\sqrt{2}}$, $F_{\alpha_\mathrm{r}}=0$ regardless of $\sigma_\mathrm{r}$.
Eq.~(\ref{eq:FIGauss}) can be rewritten as $F_{\alpha_\mathrm{r}}=8 t^2 \frac{B_G}{A_G-(1-B_G)}$ with $A_G=\mathrm{e}^{8 \sigma_\mathrm{r}^2 t^2}$ and $B_G=\cos^2\left(2 \sqrt{2} \alpha _{r0} t\right)$, showing a monotonous behavior in the ranges $A_G\in[1,\infty],~ B_G\in[0,1]$.
For the optimal value of $\alpha_{r0}$ accessible by an adaptive strategy, this is further reduced as $F_{\alpha_\mathrm{r}}=8t^2 A_G^{-1}$.
\old{Here, the maximal CFI is given as $\frac{8}{f_G e}$ at $t_G=f_G^{-1/2}$, with $f_G=8 \sigma_\mathrm{r}^2$.}

\textbf{Qubit heating}
Qubit heating is described by a Lindblad equation $\partial_t\rho=\sum_{i=1,2}L_\mathrm{i} \rho L_\mathrm{i}^\dagger-\frac{1}{2}\{L_\mathrm{i}^\dagger L_\mathrm{i} \rho\}$ with $L_{1,2}=\sqrt{\gamma}\hat{\sigma}_\mp$, \old{where $\gamma$ is the qubit heating rate, not necessarily equivalent to the oscillator heating rate$\gamma^\mathrm{th}$}.
The solution of this equation is given analytically as 
\begin{align}
\rho[t]=
\begin{pmatrix}
\rho_{ee}[0]\frac{1+\mathrm{e}^{-2\gamma t}}{2}+\rho_{gg}[0]\frac{1-\mathrm{e}^{-2\gamma t}}{2}  & \rho_{ge}[0]\mathrm{e}^{-\gamma t}  \\
\rho_{eg}[0]\mathrm{e}^{-\gamma t}  &\rho_{gg}[0]\frac{1+\mathrm{e}^{-2\gamma t}}{2}+\rho_{ee}[0]\frac{1-\mathrm{e}^{-2\gamma t}}{2}
\end{pmatrix}.
\end{align}
For the initial \old{qubit} density matrix $\rho[0]=(\cos[\sqrt{2}t\alpha_\mathrm{r}]\ket{\mathrm{e}}+\sin[\sqrt{2}t\alpha_\mathrm{r}]\ket{\mathrm{g}}).h.c$, we can obtain the probabilities from this solution and the Fisher information for heating time $t'$ is calculated as 
\begin{align}
F_{\alpha_\mathrm{r}}=\frac{8 t^2 \cos^2\left(2 \sqrt{2} t \alpha _\mathrm{r}\right)}{\mathrm{e}^{4 \gamma    t'}-\sin^2 \left(2 \sqrt{2} t \alpha _\mathrm{r}\right)}.
\end{align}
This Fisher information can be rewritten as $F_{\alpha_\mathrm{r}}=8t^2 \frac{B_H}{A_H-(1-B_H)}$ with $A_H=\mathrm{e}^{4 \gamma  t'}\in[1,\infty]$ and $B_H=\cos^2[2\sqrt{2}t \alpha_\mathrm{r}]\in[0,1]$, showing a monotonous bahavior of the maximal value as $8 t^2/A_H$.


\old{\textbf{Qubit relaxation} Qubit relaxation is needed for the full consideration of the qubit errors in our systems, especially in superconducting systems, even though it is a weaker error than qubit dephasing. 
We can describe qubit relaxation by spontaneous emission by solving the Lindblad master equation with Lindblad operator $\hat{\sigma}_-$.
This noise is impacting many systems of trapped ions and superconducting systems, although weaker than the dephasing or boson loss.
The solution of the Lindblad equation from any arbitrary input qubit state is given by a single parameter $q=\mathrm{e}^{-\gamma t_\mathrm{relax}/2}$, where $\gamma t_\mathrm{relax}$ is a dimensionless loss parameter as: $c_\mathrm{gg}\ket{\mathrm{g}}\bra{\mathrm{g}}+c_\mathrm{ee}\{q^2\ket{\mathrm{e}}\bra{\mathrm{e}}+(1-q^2)\ket{\mathrm{g}}\bra{\mathrm{g}}\}+q c_\mathrm{eg}\ket{\mathrm{e}}\bra{\mathrm{g}}+qc_\mathrm{ge}\ket{\mathrm{g}}\bra{\mathrm{e}}$.}\

For the same setups as before, we simply obtain the CFI as 
\begin{align}
F_{\alpha_\mathrm{r}}=\frac{8t^2 B_\mathrm{r} }{q^{-2}- (1-B_\mathrm{r})},
\end{align}
where $q=\mathrm{e}^{-\gamma t_\mathrm{relax}/2}$ is the qubit relaxation parameter, and $B_\mathrm{r}=\cos^2[2\sqrt{2}t \A_\mathrm{r}]$, again showing a monotonous behavior for the zero displacement limit $\A_\mathrm{r}=0$.
\new{We note that at $t_\mathrm{relax}\to \infty$, the CFI vanishes as $F_{\alpha_\mathrm{r}}\to 0$.}

\textbf{Qubit depolarization}
Qubit depolarizing noise channel, a more generic noise model that erases any information in the qubit if acted fully, is described by a trace preserving map $\rho \to (1-\lambda)\rho+ \frac{\lambda}{2}I$, with $\lambda\in [0,1]$. 
Following the procedures of the previous examples, we simply obtain the CFI with the same equation
\begin{align}
    F_\mathrm{depol}=\frac{8 t^2 \cos ^2\left[2 \sqrt{2} t \alpha _k\right]}{\mathbb{E}_\mathrm{depol}^{-2}-\sin ^2\left[2 \sqrt{2} t \alpha _k\right]}
\end{align}
with $\mathbb{E}_\mathrm{depol}=1-\lambda$.
As this channel completely erases the information at full depolarization $\lambda=1$, the CFI goes to zero in this limit.
The analogy as between dephasing noise and prepare-and-measure method is not established here, while this result shows a similarity to the complete qubit relaxation which also erases the entire information.



\printbibliography[sorting=none]

@article{Bruzewicz2019APRtrappedion,
    author = {Bruzewicz, Colin D. and Chiaverini, John and McConnell, Robert and Sage, Jeremy M.},
    title = "{Trapped-ion quantum computing: Progress and challenges}",
    journal = {Applied Physics Reviews},
    volume = {6},
    number = {2},
    year = {2019},
    month = {05},
    issn = {1931-9401},
    doi = {10.1063/1.5088164},
    comment = {https://doi.org/10.1063/1.5088164},
    note = {021314},
    comment = {https://pubs.aip.org/aip/apr/article-pdf/doi/10.1063/1.5088164/14577412/021314\_1\_online.pdf},
}

@article{Kjaergaard2020SuperCondQub,
author = {Kjaergaard, Morten and Schwartz, Mollie E. and Braum\"{u}ller, Jochen and Krantz, Philip and Wang, Joel I.-J. and Gustavsson, Simon and Oliver, William D.},
title = {Superconducting Qubits: Current State of Play},
journal = {Annual Review of Condensed Matter Physics},
volume = {11},
number = {1},
pages = {369-395},
year = {2020},
doi = {10.1146/annurev-conmatphys-031119-050605},
}

@article{Zurek2001Compass,
   author = {W. H. Zurek},
   doi = {10.1038/35089017},
   issn = {1476-4687},
   issue = {6848},
   journal = {Nature 2001 412:6848},
   month = {8},
   pages = {712-717},
   pmid = {11507634},
   publisher = {Nature Publishing Group},
   title = {Sub-Planck structure in phase space and its relevance for quantum decoherence},
   volume = {412},
   comment = {https://www.nature.com/articles/35089017},
   year = {2001},
}

@article{KienzlerPRL2016,
  title = {Observation of Quantum Interference between Separated Mechanical Oscillator Wave Packets},
  author = {Kienzler, D. and Fl\"uhmann, C. and Negnevitsky, V. and Lo, H.-Y. and Marinelli, M. and Nadlinger, D. and Home, J. P.},
  journal = {Phys. Rev. Lett.},
  volume = {116},
  issue = {14},
  pages = {140402},
  numpages = {5},
  year = {2016},
  month = {Apr},
  publisher = {American Physical Society},
  doi = {10.1103/PhysRevLett.116.140402},
  comment = {https://link.aps.org/doi/10.1103/PhysRevLett.116.140402}
}

@article{Hastrup2019GKP,
   author = {Jacob Hastrup and Kimin Park and Jonatan Bohr Brask and Radim Filip and Ulrik Lund Andersen},
   doi = {10.1038/s41534-020-00353-3},
   issn = {2056-6387},
   issue = {1},
   journal = {npj Quantum Information 2021 7:1},
   month = {1},
   pages = {1-8},
   publisher = {Nature Publishing Group},
   title = {Measurement-free preparation of grid states},
   volume = {7},
   comment = {https://www.nature.com/articles/s41534-020-00353-3},
   year = {2021},
}

@article{FluhmannHomeNature2019,
   author = {C. Flühmann and T. L. Nguyen and M. Marinelli and V. Negnevitsky and K. Mehta and J. P. Home},
   doi = {10.1038/s41586-019-0960-6},
   issn = {1476-4687},
   issue = {7745},
   journal = {Nature 2019 566:7745},
   month = {2},
   pages = {513-517},
   pmid = {30814715},
   publisher = {Nature Publishing Group},
   title = {Encoding a qubit in a trapped-ion mechanical oscillator},
   volume = {566},
   comment = {https://www.nature.com/articles/s41586-019-0960-6},
   year = {2019},
}

@article{SchoelkopfNature2020,
   author = {P. Campagne-Ibarcq and A. Eickbusch and S. Touzard and E. Zalys-Geller and N. E. Frattini and V. V. Sivak and P. Reinhold and S. Puri and S. Shankar and R. J. Schoelkopf and L. Frunzio and M. Mirrahimi and M. H. Devoret},
   doi = {10.1038/s41586-020-2603-3},
   issn = {1476-4687},
   issue = {7821},
   journal = {Nature 2020 584:7821},
   month = {8},
   pages = {368-372},
   pmid = {32814889},
   publisher = {Nature Publishing Group},
   title = {Quantum error correction of a qubit encoded in grid states of an oscillator},
   volume = {584},
   comment = {https://www.nature.com/articles/s41586-020-2603-3},
   year = {2020},
}

@article{SolanoRabiRMP2019,
  title = {Ultrastrong coupling regimes of light-matter interaction},
  author = {Forn-D\'{\i}az, P. and Lamata, L. and Rico, E. and Kono, J. and Solano, E.},
  journal = {Rev. Mod. Phys.},
  volume = {91},
  issue = {2},
  pages = {025005},
  numpages = {48},
  year = {2019},
  month = {Jun},
  publisher = {American Physical Society},
  doi = {10.1103/RevModPhys.91.025005},
  comment = {https://link.aps.org/doi/10.1103/RevModPhys.91.025005}
}

@article{NoriRabiRMP2019,
   author = {Anton Frisk Kockum and Adam Miranowicz and Simone De Liberato and Salvatore Savasta and Franco Nori},
   doi = {10.1038/s42254-018-0006-2},
   issn = {2522-5820},
   issue = {1},
   journal = {Nature Reviews Physics 2019 1:1},
   month = {1},
   pages = {19-40},
   publisher = {Nature Publishing Group},
   title = {Ultrastrong coupling between light and matter},
   volume = {1},
   comment = {https://www.nature.com/articles/s42254-018-0006-2},
   year = {2019},
}

@article{ParkNJP2018,
   author = {Kimin Park and Petr Marek and Radim Filip},
   doi = {10.1088/1367-2630/AABB86},
   issn = {1367-2630},
   issue = {5},
   journal = {New Journal of Physics},
   month = {5},
   pages = {053022},
   publisher = {IOP Publishing},
   title = {Deterministic nonlinear phase gates induced by a single qubit},
   volume = {20},
   comment = {https://iopscience.iop.org/article/10.1088/1367-2630/aabb86 https://iopscience.iop.org/article/10.1088/1367-2630/aabb86/meta},
   year = {2018},
}

@article{vanLoock2008HybridOptics,
  title = {Hybrid quantum computation in quantum optics},
  author = {van Loock, P. and Munro, W. J. and Nemoto, Kae and Spiller, T. P. and Ladd, T. D. and Braunstein, Samuel L. and Milburn, G. J.},
  journal = {Phys. Rev. A},
  volume = {78},
  issue = {2},
  pages = {022303},
  numpages = {5},
  year = {2008},
  month = {Aug},
  publisher = {American Physical Society},
  doi = {10.1103/PhysRevA.78.022303},
  comment = {https://link.aps.org/doi/10.1103/PhysRevA.78.022303}
}

@article{Hastrup2020squeezing,
  title = {Unconditional Preparation of Squeezed Vacuum from Rabi Interactions},
  author = {Hastrup, Jacob and Park, Kimin and Filip, Radim and Andersen, Ulrik Lund},
  journal = {Phys. Rev. Lett.},
  volume = {126},
  issue = {15},
  pages = {153602},
  numpages = {6},
  year = {2021},
  month = {Apr},
  publisher = {American Physical Society},
  doi = {10.1103/PhysRevLett.126.153602},
  comment = {https://link.aps.org/doi/10.1103/PhysRevLett.126.153602}
}

@article{SpillerNJP2006,
   author = {T. P. Spiller and Kae Nemoto and Samuel L. Braunstein and W. J. Munro and P. Van Loock and G. J. Milburn},
   doi = {10.1088/1367-2630/8/2/030},
   issn = {1367-2630},
   issue = {2},
   journal = {New Journal of Physics},
   month = {2},
   pages = {30},
   publisher = {IOP Publishing},
   title = {Quantum computation by communication},
   volume = {8},
   comment = {https://iopscience.iop.org/article/10.1088/1367-2630/8/2/030 https://iopscience.iop.org/article/10.1088/1367-2630/8/2/030/meta},
   year = {2006},
}

@article{ParkNJP2020Rabi,
   author = {Kimin Park and Julien Laurat and Radim Filip},
   doi = {10.1088/1367-2630/AB6877},
   issn = {1367-2630},
   issue = {1},
   journal = {New Journal of Physics},
   month = {1},
   pages = {013056},
   publisher = {IOP Publishing},
   title = {Hybrid Rabi interactions with traveling states of light},
   volume = {22},
   comment = {https://iopscience.iop.org/article/10.1088/1367-2630/ab6877 https://iopscience.iop.org/article/10.1088/1367-2630/ab6877/meta},
   year = {2020},
}

@article{DegenRMP2017,
   author = {C. L. Degen and F. Reinhard and P. Cappellaro},
   doi = {10.1103/REVMODPHYS.89.035002/},
   issn = {15390756},
   issue = {3},
   journal = {Reviews of Modern Physics},
   month = {7},
   pages = {035002},
   publisher = {American Physical Society},
   title = {Quantum sensing},
   volume = {89},
    year = {2017},
}

@article{PirandolaNatPhot2018PhotSens,
   author = {S. Pirandola and B. R. Bardhan and T. Gehring and C. Weedbrook and S. Lloyd},
   doi = {10.1038/s41566-018-0301-6},
   issn = {17494893},
   issue = {12},
   journal = {Nature Photonics},
   pages = {724-733},
   title = {Advances in photonic quantum sensing},
   volume = {12},
   year = {2018},
}

@article{LyFI2017,
title = {A Tutorial on Fisher information},
journal = {Journal of Mathematical Psychology},
volume = {80},
pages = {40-55},
year = {2017},
issn = {0022-2496},
doi = {https://doi.org/10.1016/j.jmp.2017.05.006},
comment = {https://www.sciencedirect.com/science/article/pii/S0022249617301396},
author = {Alexander Ly and Maarten Marsman and Josine Verhagen and Raoul P.P.P. Grasman and Eric-Jan Wagenmakers}
}

@article{IvanovPRAp2015ForceSensor,
  title = {Quantum Sensors Assisted by Spontaneous Symmetry Breaking for Detecting Very Small Forces},
  author = {Ivanov, Peter A. and Singer, Kilian and Vitanov, Nikolay V. and Porras, Diego},
  journal = {Phys. Rev. Appl.},
  volume = {4},
  issue = {5},
  pages = {054007},
  numpages = {7},
  year = {2015},
  month = {Nov},
  publisher = {American Physical Society},
  doi = {10.1103/PhysRevApplied.4.054007},
  comment = {https://link.aps.org/doi/10.1103/PhysRevApplied.4.054007}
}

@article{Ivanov2016,
author = {Ivanov, Peter A. and Vitanov, Nikolay V. and Singer, Kilian},
doi = {10.1038/srep28078},
issn = {20452322},
journal = {Scientific Reports},
number = {February},
pages = {1--8},
publisher = {Nature Publishing Group},
title = {{High-precision force sensing using a single trapped ion}},
comment = {http://dx.doi.org/10.1038/srep28078},
volume = {6},
year = {2016}
}

@article{IvanovPRA2018,
  title = {Quantum sensing of the phase-space-displacement parameters using a single trapped ion},
  author = {Ivanov, Peter A. and Vitanov, Nikolay V.},
  journal = {Phys. Rev. A},
  volume = {97},
  issue = {3},
  pages = {032308},
  numpages = {7},
  year = {2018},
  month = {Mar},
  publisher = {American Physical Society},
  doi = {10.1103/PhysRevA.97.032308},
  comment = {https://link.aps.org/doi/10.1103/PhysRevA.97.032308}
}

@article{2020FockDispersive,
 author={Wolf, Fabian and Shi, Chunyan and Heip, Jan C. and Gessner, Manuel and Pezzè, Luca and Smerzi, Augusto and Schulte, Marius and Hammerer, Klemens and Schmidt, Piet O.},title={Motional Fock states for quantum-enhanced amplitude and phase measurements with trapped ions}, volume={10}, DOI={10.1038/s41467-019-10576-4}, number={1}, journal={Nature Communications},  year={2019}}

@article{schindler2013quantum,
   author = {Philipp Schindler and Daniel Nigg and Thomas Monz and Julio T. Barreiro and Esteban Martinez and Shannon X. Wang and Stephan Quint and Matthias F. Brandl and Volckmar Nebendahl and Christian F. Roos and Michael Chwalla and Markus Hennrich and Rainer Blatt},
   doi = {10.1088/1367-2630/15/12/123012},
   issn = {1367-2630},
   issue = {12},
   journal = {New Journal of Physics},
   month = {12},
   pages = {123012},
   publisher = {IOP Publishing},
   title = {A quantum information processor with trapped ions},
   volume = {15},
   comment = {https://iopscience.iop.org/article/10.1088/1367-2630/15/12/123012 https://iopscience.iop.org/article/10.1088/1367-2630/15/12/123012/meta},
   year = {2013},
}

@article{dalvit_filho_toscano_2006, title={Quantum metrology at the Heisenberg limit with ion trap motional compass states}, volume={8}, DOI={10.1088/1367-2630/8/11/276}, number={11}, journal={New Journal of Physics}, author={Dalvit, D A and Filho, R L and Toscano, F}, year={2006}, pages={276–276}}

@article{TerhalPRA2017GKP,
  title = {Single-mode displacement sensor},
  author = {Duivenvoorden, Kasper and Terhal, Barbara M. and Weigand, Daniel},
  journal = {Phys. Rev. A},
  volume = {95},
  issue = {1},
  pages = {012305},
  numpages = {15},
  year = {2017},
  month = {Jan},
  publisher = {American Physical Society},
  doi = {10.1103/PhysRevA.95.012305},
  comment = {https://link.aps.org/doi/10.1103/PhysRevA.95.012305}
}

@article{Touzard2019,
archivePrefix = {arXiv},
author = {Touzard, S. and Kou, A. and Frattini, N. E. and Sivak, V. V. and Puri, S. and Grimm, A. and Frunzio, L. and Shankar, S. and Devoret, M. H.},
doi = {10.1103/PhysRevLett.122.080502},
issn = {10797114},
journal = {Physical Review Letters},
month = {feb},
number = {8},
pages = {080502},
pmid = {30932609},
publisher = {American Physical Society},
title = {{Gated Conditional Displacement Readout of Superconducting Qubits}},
volume = {122},
year = {2019}
}

@article{Kwon2021Superconduting,
author = {Kwon, Sangil and Tomonaga, Akiyoshi and {Lakshmi Bhai}, Gopika and Devitt, Simon J. and Tsai, Jaw Shen},
doi = {10.1063/5.0029735},
issn = {10897550},
journal = {Journal of Applied Physics},
number = {4},
publisher = {arxiv.org},
title = {{Gate-based superconducting quantum computing}},
type = {PDF},
comment = {https://arxiv.org/pdf/2009.08021},
volume = {129},
year = {2021}
}

@article{ChenPRL2015,
author = {Chen, Bing and Qiu, Cheng and Chen, Shuying and Guo, Jinxian and Chen, L. Q. and Ou, Z. Y. and Zhang, Weiping},
doi = {10.1103/PhysRevLett.115.043602},
issn = {10797114},
journal = {Physical Review Letters},
month = {jul},
number = {4},
pages = {043602},
publisher = {American Physical Society},
title = {{Atom-Light Hybrid Interferometer}},
comment = {https://journals.aps.org/prl/abstract/10.1103/PhysRevLett.115.043602},
volume = {115},
year = {2015}
}

@book{Barnett2003,
    author = {Barnett, Stephen M. and Radmore, Paul M.},
title = {Methods in Theoretical Quantum Optics},
isbn = {978-0198563617},
    publisher = {Oxford University Press},
    year = {2002},
    month = {11},
    doi = {10.1093/acprof:oso/9780198563617.001.0001},
    comment = {https://doi.org/10.1093/acprof:oso/9780198563617.001.0001},
}

@article{Lang2013microwaveHOM,
author = {Lang, C. and Eichler, C. and Steffen, L. and Fink, J. M. and Woolley, M. J. and Blais, A. and Wallraff, A.},
doi = {10.1038/nphys2612},
issn = {17452481},
journal = {Nature Physics},
number = {6},
pages = {345--348},
title = {{Correlations, indistinguishability and entanglement in Hong-Ou-Mandel experiments at microwave frequencies}},
volume = {9},
year = {2013}
}

@article{Gao2018MwMemInterf,
archivePrefix = {arXiv},
comment = {1802.08510},
author = {Gao, Yvonne Y. and Lester, Brian J. and Zhang, Yaxing and Wang, Chen and Rosenblum, Serge and Frunzio, Luigi and Jiang, Liang and Girvin, S. M. and Schoelkopf, Robert J.},
doi = {10.1103/PhysRevX.8.021073},
comment = {1802.08510},
issn = {21603308},
journal = {Physical Review X},
number = {2},
title = {{Programmable Interference between Two Microwave Quantum Memories}},
volume = {8},
year = {2018}
}

@article{Ramsey1949,
author = {Ramsey, Norman F.},
doi = {10.1103/PhysRev.76.996},
issn = {0031899X},
journal = {Physical Review},
number = {7},
pages = {996},
title = {{A new molecular beam resonance method}},
volume = {76},
year = {1949}
}

@article{Leibfried2004HeisenbergMultipartite,
author = {Leibfried, D. and Barrett, M. D. and Schaetz, T. and Britton, J. and Chiaverini, J. and Itano, W. M. and Jost, J. D. and Langer, C. and Wineland, D. J.},
doi = {10.1126/science.1097576},
issn = {00368075},
journal = {Science},
number = {5676},
pages = {1476--1478},
title = {{Toward Heisenberg-limited spectroscopy with multiparticle entangled states}},
volume = {304},
year = {2004}
}

@article{Wang2019HeisenSupercond,
author = {Wang, W. and Wu, Y. and Ma, Y. and Cai, W. and Hu, L. and Mu, X. and Xu, Y. and Chen, Zi Jie and Wang, H. and Song, Y. P. and Yuan, H. and Zou, C. L. and Duan, L. M. and Sun, L.},
doi = {10.1038/s41467-019-12290-7},
issn = {20411723},
journal = {Nature Communications},
number = {1},
pmid = {31558721},
title = {{Heisenberg-limited single-mode quantum metrology in a superconducting circuit}},
volume = {10},
year = {2019}
}

@article{Cai2021,
archivePrefix = {arXiv},
comment = {2102.05409},
author = {Cai, M. -L. L. and Liu, Z. D. -D. and Zhao, W. -D. D. and Wu, Y. K. -K. and Mei, Q. X. -X. and Jiang, Y. and He, L. and Zhang, X. and Zhou, Z. -C. C. and Duan, L. M. -M.},
doi = {10.1038/s41467-021-21425-8},
comment = {2102.05409},
issn = {2041-1723},
journal = {Nature Communications},
month = {feb},
number = {1},
pages = {1126},
pmid = {33602942},
publisher = {Nature Research},
title = {{Observation of a quantum phase transition in the quantum Rabi model with a single trapped ion}},
comment = {http://www.nature.com/articles/s41467-021-21425-8 http://arxiv.org/abs/2102.05409 http://dx.doi.org/10.1038/s41467-021-21425-8 https://doi.org/10.1038/s41467-021-21425-8},
volume = {12},
year = {2021}
}

@article{McCormick2019,
author = {McCormick, Katherine C. and Keller, Jonas and Burd, Shaun C. and Wineland, David J. and Wilson, Andrew C. and Leibfried, Dietrich},
doi = {10.1038/s41586-019-1421-y},
issn = {1476-4687},
journal = {Nature},
month = {aug},
number = {7767},
pages = {86--90},
pmid = {31332388},
publisher = {Nature Publishing Group},
title = {{Quantum-enhanced sensing of a single-ion mechanical oscillator.}},
comment = {http://www.nature.com/articles/s41586-019-1421-y http://www.ncbi.nlm.nih.gov/pubmed/31332388 http://www.pubmedcentral.nih.gov/articlerender.fcgi?artid=PMC6986265 https://doi.org/10.1038/s41586-019-1421-y},
volume = {572},
year = {2019}
}

@article{Premaratne2017,
author = {Premaratne, Shavindra P. and Wellstood, F. C. and Palmer, B. S.},
doi = {10.1038/ncomms14148},
issn = {20411723},
journal = {Nature Communications},
pmid = {28128205},
title = {{Microwave photon Fock state generation by stimulated Raman adiabatic passage}},
volume = {8},
year = {2017}
}

@article{Wang2017,
archivePrefix = {arXiv},
comment = {1703.03316},
author = {Wang, W. and Hu, L. and Xu, Y. and Liu, K. and Ma, Y. and Zheng, Shi Biao and Vijay, R. and Song, Y. P. and Duan, L. M. and Sun, L.},
doi = {10.1103/PhysRevLett.118.223604},
comment = {1703.03316},
issn = {10797114},
journal = {Physical Review Letters},
number = {22},
pmid = {28621980},
title = {{Converting Quasiclassical States into Arbitrary Fock State Superpositions in a Superconducting Circuit}},
volume = {118},
year = {2017}
}

@article{Pfaff2017NatureConversion,
author = {Pfaff, Wolfgang and Axline, Christopher J. and Burkhart, Luke D. and Vool, Uri and Reinhold, Philip and Frunzio, Luigi and Jiang, Liang and Devoret, Michel H. and Schoelkopf, Robert J.},
doi = {10.1038/nphys4143},
issn = {17452481},
journal = {Nature Physics},
number = {9},
pages = {882--887},
title = {{Controlled release of multiphoton quantum states from a microwave cavity memory}},
volume = {13},
year = {2017}
}

@article{Gely2019,
archivePrefix = {arXiv},
comment = {1901.07267},
author = {Gely, Mario F. and Kounalakis, Marios and Dickel, Christian and Dalle, Jacob and Vatr{\'{e}}, R{\'{e}}my and Baker, Brian and Jenkins, Mark D. and Steele, Gary A.},
doi = {10.1126/science.aaw3101},
comment = {1901.07267},
issn = {10959203},
journal = {Science},
number = {6431},
pages = {1072--1075},
pmid = {30846596},
title = {{Observation and stabilization of photonic Fock states in a hot radio-frequency resonator}},
volume = {363},
year = {2019}
}

@article{Chu2018,
archivePrefix = {arXiv},
comment = {1804.07426},
author = {Chu, Yiwen and Kharel, Prashanta and Yoon, Taekwan and Frunzio, Luigi and Rakich, Peter T. and Schoelkopf, Robert J.},
doi = {10.1038/s41586-018-0717-7},
comment = {1804.07426},
issn = {14764687},
journal = {Nature},
number = {7733},
pages = {666--670},
pmid = {30464340},
title = {{Creation and control of multi-phonon Fock states in a bulk acoustic-wave resonator}},
volume = {563},
year = {2018}
}

@article{Kristen2019,
   author = {M. Kristen and A. Schneider and A. Stehli and T. Wolz and S. Danilin and H. S. Ku and J. Long and X. Wu and R. Lake and D. P. Pappas and A. V. Ustinov and M. Weides},
   doi = {10.1038/s41534-020-00287-w},
   isbn = {4153402000287},
   issn = {2056-6387},
   issue = {1},
   journal = {npj Quantum Information 2020 6:1},
   month = {6},
   pages = {1-5},
   publisher = {Nature Publishing Group},
   title = {Amplitude and frequency sensing of microwave fields with a superconducting transmon qudit},
   volume = {6},
   comment = {https://www.nature.com/articles/s41534-020-00287-w},
   year = {2020},
}

@article{Dixit2020,
  title = {Searching for Dark Matter with a Superconducting Qubit},
  author = {Dixit, Akash V. and Chakram, Srivatsan and He, Kevin and Agrawal, Ankur and Naik, Ravi K. and Schuster, David I. and Chou, Aaron},
  journal = {Phys. Rev. Lett.},
  volume = {126},
  issue = {14},
  pages = {141302},
  numpages = {7},
  year = {2021},
  month = {Apr},
  publisher = {American Physical Society},
  doi = {10.1103/PhysRevLett.126.141302},
}

@article{Lachance-Quirion2020,
archivePrefix = {arXiv},
comment = {1910.09096},
author = {Lachance-Quirion, Dany and Wolski, Samuel Piotr and Tabuchi, Yutaka and Kono, Shingo and Usami, Koji and Nakamura, Yasunobu},
doi = {10.1126/science.aaz9236},
comment = {1910.09096},
issn = {10959203},
journal = {Science},
number = {6476},
pages = {425--428},
pmid = {31974250},
title = {{Entanglement-based single-shot detection of a single magnon with a superconducting qubit}},
volume = {367},
year = {2020}
}

@article{Wolski2020,
  title = {Dissipation-Based Quantum Sensing of Magnons with a Superconducting Qubit},
  author = {Wolski, S. P. and Lachance-Quirion, D. and Tabuchi, Y. and Kono, S. and Noguchi, A. and Usami, K. and Nakamura, Y.},
  journal = {Phys. Rev. Lett.},
  volume = {125},
  issue = {11},
  pages = {117701},
  numpages = {6},
  year = {2020},
  month = {Sep},
  publisher = {American Physical Society},
  doi = {10.1103/PhysRevLett.125.117701},
}

@article{Lachance-Quirion2017,
author = {Lachance-Quirion, Dany and Tabuchi, Yutaka and Ishino, Seiichiro and Noguchi, Atsushi and Ishikawa, Toyofumi and Yamazaki, Rekishu and Nakamura, Yasunobu},
doi = {10.1126/sciadv.1603150},
issn = {23752548},
journal = {Science Advances},
number = {7},
pmid = {28695204},
title = {{Resolving quanta of collective spin excitations in a millimeter-sized ferromagnet}},
volume = {3},
year = {2017}
}

@article{Wang2019,
  title = {Quantum Microwave Radiometry with a Superconducting Qubit},
  author = {Wang, Zhixin and Xu, Mingrui and Han, Xu and Fu, Wei and Puri, Shruti and Girvin, S. M. and Tang, Hong X. and Shankar, S. and Devoret, M. H.},
  journal = {Phys. Rev. Lett.},
  volume = {126},
  issue = {18},
  pages = {180501},
  numpages = {7},
  year = {2021},
  month = {May},
  publisher = {American Physical Society},
  doi = {10.1103/PhysRevLett.126.180501},
}

@article{Wang2021,
   author = {W. Wang and Z. J. Chen and X. Liu and W. Cai and Y. Ma and X. Mu and X. Pan and Z. Hua and L. Hu and Y. Xu and H. Wang and Y. P. Song and X. B. Zou and C. L. Zou and L. Sun},
   doi = {10.1038/s41467-022-30410-8},
   issn = {2041-1723},
   issue = {1},
   journal = {Nature Communications 2022 13:1},
   month = {6},
   pages = {1-8},
   pmid = {35680786},
   publisher = {Nature Publishing Group},
   title = {Quantum-enhanced radiometry via approximate quantum error correction},
   volume = {13},
   comment = {https://www.nature.com/articles/s41467-022-30410-8},
   year = {2022},
}

@article{pop00001,
author = {Li, Meixiu and Chen, Tao and Gooding, J. Justin and Liu, Jingquan},
doi = {10.1021/acssensors.9b00514},
issn = {23793694},
journal = {ACS Sensors},
number = {7},
pages = {1732--1748},
pmid = {31267734},
publisher = {ACS Publications},
title = {{Review of carbon and graphene quantum dots for sensing}},
comment = {https://pubs.acs.org/doi/abs/10.1021/acssensors.9b00514},
volume = {4},
year = {2019}
}

@article{pop00003,
   author = {Xueshi Guo and Casper R. Breum and Johannes Borregaard and Shuro Izumi and Mikkel V. Larsen and Tobias Gehring and Matthias Christandl and Jonas S. Neergaard-Nielsen and Ulrik L. Andersen},
   doi = {10.1038/s41567-019-0743-x},
   issn = {1745-2481},
   issue = {3},
   journal = {Nature Physics 2019 16:3},
   month = {12},
   pages = {281-284},
   publisher = {Nature Publishing Group},
   title = {Distributed quantum sensing in a continuous-variable entangled network},
   volume = {16},
     year = {2019},
}

@article{GiovannettiNatPho2011Metrology,
archivePrefix = {arXiv},
comment = {1102.2318},
author = {Giovannetti, Vittorio and Lloyd, Seth and MacCone, Lorenzo},
doi = {10.1038/nphoton.2011.35},
comment = {1102.2318},
issn = {17494885},
journal = {Nature Photonics},
number = {4},
pages = {222--229},
publisher = {Nature Publishing Group},
title = {{Advances in quantum metrology}},
volume = {5},
year = {2011}
}

@article{pop00009,
author = {Lawrie, B. J. and Lett, P. D. and Marino, A. M. and Pooser, R. C.},
doi = {10.1021/acsphotonics.9b00250},
issn = {23304022},
journal = {ACS Photonics},
number = {6},
pages = {1307--1318},
publisher = {ACS Publications},
title = {{Quantum Sensing with Squeezed Light}},
comment = {https://pubs.acs.org/doi/abs/10.1021/acsphotonics.9b00250},
volume = {6},
year = {2019}
}

@article{Gilmore2017,
archivePrefix = {arXiv},
comment = {1703.05369},
author = {Gilmore, K. A. and Bohnet, J. G. and Sawyer, B. C. and Britton, J. W. and Bollinger, J. J.},
doi = {10.1103/PhysRevLett.118.263602},
comment = {1703.05369},
issn = {10797114},
journal = {Physical Review Letters},
number = {26},
pages = {1--5},
pmid = {28707910},
title = {{Amplitude Sensing below the Zero-Point Fluctuations with a Two-Dimensional Trapped-Ion Mechanical Oscillator}},
volume = {118},
year = {2017}
}

@article{Biercuk2010,
author = {Biercuk, Michael J and Uys, Hermann and Britton, Joe W and Vandevender, Aaron P and Bollinger, John J},
doi = {10.1038/nnano.2010.165},
issn = {17483395},
journal = {Nature Nanotechnology},
number = {9},
pages = {646--650},
title = {{Ultrasensitive detection of force and displacement using trapped ions}},
comment = {www.nature.com/naturenanotechnology},
volume = {5},
year = {2010}
}

@article{Schirhagl2014,
author = {Schirhagl, Romana and Chang, Kevin and Loretz, Michael and Degen, Christian L.},
doi = {10.1146/annurev-physchem-040513-103659},
issn = {0066426X},
journal = {Annual Review of Physical Chemistry},
month = {apr},
pages = {83--105},
pmid = {24274702},
publisher = {Annual Reviews Inc.},
title = {{Nitrogen-vacancy centers in diamond: Nanoscale sensors for physics and biology}},
comment = {https://www.annualreviews.org/doi/abs/10.1146/annurev-physchem-040513-103659},
volume = {65},
year = {2014}
}

@article{LigoPRL2016,
archivePrefix = {arXiv},
comment = {1602.03837},
author={LIGO Scientific Collaboration and Virgo Collaboration},
doi = {10.1103/PhysRevLett.116.061102},
comment = {1602.03837},
issn = {10797114},
journal = {Physical Review Letters},
month = {feb},
number = {6},
pages = {061102},
pmid = {26918975},
publisher = {American Physical Society},
title = {{Observation of gravitational waves from a binary black hole merger}},
comment = {https://journals.aps.org/prl/abstract/10.1103/PhysRevLett.116.061102},
volume = {116},
year = {2016}
}

@article{Bongs2019atominterferometer,
author = {Bongs, Kai and Holynski, Michael and Vovrosh, Jamie and Bouyer, Philippe and Condon, Gabriel and Rasel, Ernst and Schubert, Christian and Schleich, Wolfgang P. and Roura, Albert},
journal = {Nature Reviews Physics},
doi = {10.1038/s42254-019-0117-4},
issn = {25225820},
month = {dec},
number = {12},
pages = {731--739},
publisher = {Springer Nature},
title = {{Taking atom interferometric quantum sensors from the laboratory to real-world applications}},
volume = {1},
year = {2019}
}

@article{Cronin2009atominterferometer,
author = {Cronin, Alexander D. and Schmiedmayer, J{\"{o}}rg and Pritchard, David E.},
doi = {10.1103/RevModPhys.81.1051},
issn = {00346861},
journal = {Reviews of Modern Physics},
month = {aug},
number = {3},
pages = {1051--1129},
title = {{Optics and interferometry with atoms and molecules}},
volume = {81},
year = {2009}
}

@article{ShoreKnight1993JC,
author = { Bruce W.   Shore  and  Peter L.   Knight },
title = {The Jaynes-Cummings Model},
journal = {Journal of Modern Optics},
volume = {40},
number = {7},
pages = {1195-1238},
year  = {1993},
publisher = {Taylor & Francis},
doi = {10.1080/09500349314551321},
}

@article{FinkNature2008,
archivePrefix = {arXiv},
comment = {0902.1827},
author = {Fink, J. M. and G{\"{o}}ppl, M. and Baur, M. and Bianchetti, R. and Leek, P. J. and Blais, A. and Wallraff, A.},
doi = {10.1038/nature07112},
comment = {0902.1827},
issn = {14764687},
journal = {Nature},
number = {7202},
pages = {315--318},
pmid = {18633413},
publisher = {nature.com},
title = {{Climbing the Jaynes-Cummings ladder and observing its $\sqrt{n}$ nonlinearity in a cavity QED system}},
comment = {https://www.nature.com/articles/nature07112},
volume = {454},
year = {2008}
}

@article{Casanova2010JC,
archivePrefix = {arXiv},
comment = {1008.1240},
author = {Casanova, J. and Romero, G. and Lizuain, I. and Garc{\'{i}}a-Ripoll, J. J. and Solano, E.},
doi = {10.1103/PhysRevLett.105.263603},
comment = {1008.1240},
issn = {00319007},
journal = {Physical Review Letters},
number = {26},
publisher = {APS},
title = {{Deep strong coupling regime of the Jaynes-Cummings model}},
comment = {https://journals.aps.org/prl/abstract/10.1103/PhysRevLett.105.263603},
volume = {105},
year = {2010}
}

@article{RiehlePRL1991Ramsey,
author = {Riehle, F. and Kisters, Th and Witte, A. and Helmcke, J. and Bord{\'{e}}, Ch J.},
doi = {10.1103/PhysRevLett.67.177},
issn = {00319007},
journal = {Physical Review Letters},
number = {2},
pages = {177--180},
title = {{Optical Ramsey spectroscopy in a rotating frame: Sagnac effect in a matter-wave interferometer}},
volume = {67},
year = {1991}
}

@article{CadoretPRL2008Ramsey,
author = {Cadoret, Malo and {De Mirandes}, Estefania and Clad{\'{e}}, Pierre and Guellati-Kh{\'{e}}lifa, Sa{\"{i}}da and Schwob, Catherine and Nez, Fran{\c{c}}ois and Julien, Lucile and Biraben, Fran{\c{c}}ois},
doi = {10.1103/PhysRevLett.101.230801},
issn = {00319007},
journal = {Physical Review Letters},
number = {23},
publisher = {APS},
title = {{Combination of bloch oscillations with a Ramsey-Bord{\'{e}} interferometer: New determination of the fine structure constant}},
comment = {https://journals.aps.org/prl/abstract/10.1103/PhysRevLett.101.230801},
volume = {101},
year = {2008}
}

@article{AriasPRL2018Ramsey,
  title = {Realization of a Rydberg-Dressed Ramsey Interferometer and Electrometer},
  author = {Arias, A. and Lochead, G. and Wintermantel, T. M. and Helmrich, S. and Whitlock, S.},
  journal = {Phys. Rev. Lett.},
  volume = {122},
  issue = {5},
  pages = {053601},
  numpages = {6},
  year = {2019},
  month = {Feb},
  publisher = {American Physical Society},
  doi = {10.1103/PhysRevLett.122.053601},
}

@article{Liu2016,
archivePrefix = {arXiv},
comment = {1501.04290},
author = {Liu, Jing and Chen, Jie and Jing, Xiao Xing and Wang, Xiaoguang},
doi = {10.1088/1751-8113/49/27/275302},
comment = {1501.04290},
issn = {17518121},
journal = {Journal of Physics A: Mathematical and Theoretical},
number = {27},
title = {{Quantum Fisher information and symmetric logarithmic derivative via anti-commutators}},
volume = {49},
year = {2016}
}

@article{Paris2009,
archivePrefix = {arXiv},
comment = {0804.2981},
author = {Paris, Matteo G.A.},
doi = {10.1142/S0219749909004839},
comment = {0804.2981},
issn = {02197499},
journal = {International Journal of Quantum Information},
number = {SUPPL.},
pages = {125--137},
title = {{Quantum estimation for quantum technology}},
volume = {7},
year = {2009}
}

@article{Bonato2016,
author = {Bonato, C. and Blok, M. S. and Dinani, H. T. and Berry, D. W. and Markham, M. L. and Twitchen, D. J. and Hanson, R.},
doi = {10.1038/nnano.2015.261},
issn = {17483395},
journal = {Nature Nanotechnology},
month = {mar},
number = {3},
pages = {247--252},
publisher = {Nature Publishing Group},
title = {{Optimized quantum sensing with a single electron spin using real-time adaptive measurements}},
comment = {www.nature.com/naturenanotechnology},
volume = {11},
year = {2016}
}

@article{hwang15,
  title = {Quantum Phase Transition and Universal Dynamics in the Rabi Model},
  author = {Hwang, Myung-Joong and Puebla, Ricardo and Plenio, Martin B.},
  journal = {Phys. Rev. Lett.},
  volume = {115},
  issue = {18},
  pages = {180404},
  numpages = {5},
  year = {2015},
  month = {Oct},
  publisher = {American Physical Society},
  doi = {10.1103/PhysRevLett.115.180404},
  comment = {https://link.aps.org/doi/10.1103/PhysRevLett.115.180404}
}

@article{Ballance2016,
archivePrefix = {arXiv},
comment = {1512.04600},
author = {Ballance, C J and Harty, T P and Linke, N M and Sepiol, M A and Lucas, D M},
doi = {10.1103/PhysRevLett.117.060504},
comment = {1512.04600},
issn = {10797114},
journal = {Physical Review Letters},
number = {6},
title = {{High-Fidelity Quantum Logic Gates Using Trapped-Ion Hyperfine Qubits}},
volume = {117},
year = {2016}
}

@article{yin13,
  title = {Large quantum superpositions of a levitated nanodiamond through spin-optomechanical coupling},
  author = {Yin, Zhang-qi and Li, Tongcang and Zhang, Xiang and Duan, L. M.},
  journal = {Phys. Rev. A},
  volume = {88},
  issue = {3},
  pages = {033614},
  numpages = {6},
  year = {2013},
  month = {Sep},
  publisher = {American Physical Society},
  doi = {10.1103/PhysRevA.88.033614},
  comment = {https://link.aps.org/doi/10.1103/PhysRevA.88.033614}
}

@article{xiang12,
  title = {Hybrid quantum circuits: Superconducting circuits interacting with other quantum systems},
  author = {Xiang, Ze-Liang and Ashhab, Sahel and You, J. Q. and Nori, Franco},
  journal = {Rev. Mod. Phys.},
  volume = {85},
  issue = {2},
  pages = {623--653},
  numpages = {0},
  year = {2013},
  month = {Apr},
  publisher = {American Physical Society},
  doi = {10.1103/RevModPhys.85.623},
  comment = {https://link.aps.org/doi/10.1103/RevModPhys.85.623}
}

@article{kotler16,
  title = {Hybrid quantum systems with trapped charged particles},
  author = {Kotler, Shlomi and Simmonds, Raymond W. and Leibfried, Dietrich and Wineland, David J.},
  journal = {Phys. Rev. A},
  volume = {95},
  issue = {2},
  pages = {022327},
  numpages = {29},
  year = {2017},
  month = {Feb},
  publisher = {American Physical Society},
  doi = {10.1103/PhysRevA.95.022327},
  comment = {https://link.aps.org/doi/10.1103/PhysRevA.95.022327}
}

@article{MonroeRMP2021,
  title = {Programmable quantum simulations of spin systems with trapped ions},
  author = {Monroe, C. and Campbell, W. C. and Duan, L.-M. and Gong, Z.-X. and Gorshkov, A. V. and Hess, P. W. and Islam, R. and Kim, K. and Linke, N. M. and Pagano, G. and Richerme, P. and Senko, C. and Yao, N. Y.},
  journal = {Rev. Mod. Phys.},
  volume = {93},
  issue = {2},
  pages = {025001},
  numpages = {57},
  year = {2021},
  month = {Apr},
  publisher = {American Physical Society},
  doi = {10.1103/RevModPhys.93.025001},
  comment = {https://link.aps.org/doi/10.1103/RevModPhys.93.025001}
}

@article {Kurizki2015,
	author = {Kurizki, Gershon and Bertet, Patrice and Kubo, Yuimaru and M{\o}lmer, Klaus and Petrosyan, David and Rabl, Peter and Schmiedmayer, J{\"o}rg},
	title = {Quantum technologies with hybrid systems},
	volume = {112},
	number = {13},
	pages = {3866--3873},
	year = {2015},
	doi = {10.1073/pnas.1419326112},
	publisher = {National Academy of Sciences},
	issn = {0027-8424},
	comment = {https://www.pnas.org/content/112/13/3866},
	journal = {Proceedings of the National Academy of Sciences}
}

@article{Brownnutt2015,
archivePrefix = {arXiv},
comment = {1409.6572},
author = {Brownnutt, M. and Kumph, M. and Rabl, P. and Blatt, R.},
doi = {10.1103/RevModPhys.87.1419},
comment = {1409.6572},
issn = {15390756},
journal = {Reviews of Modern Physics},
month = {dec},
number = {4},
pages = {1419},
publisher = {American Physical Society},
title = {{Ion-trap measurements of electric-field noise near surfaces}},
comment = {https://journals.aps.org/rmp/abstract/10.1103/RevModPhys.87.1419},
volume = {87},
year = {2015}
}

@article{LeibfriedRMP2003,
  title = {Quantum dynamics of single trapped ions},
  author = {Leibfried, D. and Blatt, R. and Monroe, C. and Wineland, D.},
  journal = {Rev. Mod. Phys.},
  volume = {75},
  issue = {1},
  pages = {281--324},
  numpages = {0},
  year = {2003},
  month = {Mar},
  publisher = {American Physical Society},
  doi = {10.1103/RevModPhys.75.281},
  comment = {https://link.aps.org/doi/10.1103/RevModPhys.75.281}
}

@article{Affolter2020,
archivePrefix = {arXiv},
comment = {2008.00035},
author = {Affolter, M. and Gilmore, K. A. and Jordan, J. E. and Bollinger, J. J.},
doi = {10.1103/PhysRevA.102.052609},
comment = {2008.00035},
issn = {24699934},
journal = {Physical Review A},
month = {jul},
number = {5},
pages = {052609},
publisher = {American Physical Society},
title = {{Phase-coherent sensing of the center-of-mass motion of trapped-ion crystals}},
comment = {https://journals.aps.org/pra/abstract/10.1103/PhysRevA.102.052609 http://arxiv.org/abs/2008.00035 http://dx.doi.org/10.1103/PhysRevA.102.052609},
volume = {102},
year = {2020}
}

@article{ritschRMP2012,
  title = {Cold atoms in cavity-generated dynamical optical potentials},
  author = {Ritsch, Helmut and Domokos, Peter and Brennecke, Ferdinand and Esslinger, Tilman},
  journal = {Rev. Mod. Phys.},
  volume = {85},
  issue = {2},
  pages = {553--601},
  numpages = {0},
  year = {2013},
  month = {Apr},
  publisher = {American Physical Society},
  doi = {10.1103/RevModPhys.85.553},
  comment = {https://link.aps.org/doi/10.1103/RevModPhys.85.553}
}

@article{ZurekNature2001,
   author = {W. H. Zurek},
   doi = {10.1038/35089017},
   issn = {1476-4687},
   issue = {6848},
   journal = {Nature 2001 412:6848},
   month = {8},
   pages = {712-717},
   pmid = {11507634},
   publisher = {Nature Publishing Group},
   title = {Sub-Planck structure in phase space and its relevance for quantum decoherence},
   volume = {412},
   comment = {https://www.nature.com/articles/35089017},
   year = {2001},
}

@article{sensinghighenergyphysics2018,
      title={Quantum Sensing for High Energy Physics}, 
      author={Zeeshan Ahmed and Yuri Alexeev and Giorgio Apollinari and Asimina Arvanitaki and David Awschalom and Karl K. Berggren and Karl Van Bibber and Przemyslaw Bienias and Geoffrey Bodwin and Malcolm Boshier and Daniel Bowring and Davide Braga and Karen Byrum and Gustavo Cancelo and Gianpaolo Carosi and Tom Cecil and Clarence Chang and Mattia Checchin and Sergei Chekanov and Aaron Chou and Aashish Clerk and Ian Cloet and Michael Crisler and Marcel Demarteau and Ranjan Dharmapalan and Matthew Dietrich and Junjia Ding and Zelimir Djurcic and John Doyle and James Fast and Michael Fazio and Peter Fierlinger and Hal Finkel and Patrick Fox and Gerald Gabrielse and Andrei Gaponenko and Maurice Garcia-Sciveres and Andrew Geraci and Jeffrey Guest and Supratik Guha and Salman Habib and Ron Harnik and Amr Helmy and Yuekun Heng and Jason Henning and Joseph Heremans and Phay Ho and Jason Hogan and Johannes Hubmayr and David Hume and Kent Irwin and Cynthia Jenks and Nick Karonis and Raj Kettimuthu and Derek Kimball and Jonathan King and Eve Kovacs and Richard Kriske and Donna Kubik and Akito Kusaka and Benjamin Lawrie and Konrad Lehnert and Paul Lett and Jonathan Lewis and Pavel Lougovski and Larry Lurio and Xuedan Ma and Edward May and Petra Merkel and Jessica Metcalfe and Antonino Miceli and Misun Min and Sandeep Miryala and John Mitchell and Vesna Mitrovic and Holger Mueller and Sae Woo Nam and Hogan Nguyen and Howard Nicholson and Andrei Nomerotski and Michael Norman and Kevin O'Brien and Roger O'Brient and Umeshkumar Patel and Bjoern Penning and Sergey Perverzev and Nicholas Peters and Raphael Pooser and Chrystian Posada and James Proudfoot and Tenzin Rabga and Tijana Rajh and Sergio Rescia and Alexander Romanenko and Roger Rusack and Monika Schleier-Smith and Keith Schwab and Julie Segal and Ian Shipsey and Erik Shirokoff and Andrew Sonnenschein and Valerie Taylor and Robert Tschirhart and Chris Tully and David Underwood and Vladan Vuletic and Robert Wagner and Gensheng Wang and Harry Weerts and Nathan Woollett and Junqi Xie and Volodymyr Yefremenko and John Zasadzinski and Jinlong Zhang and Xufeng Zhang and Vishnu Zutshi},
      year={2018},
      eprint={1803.11306},
      archivePrefix={arXiv},
      primaryClass={hep-ex}
}

@article{HackerCatNatPho2019,
archivePrefix = {arXiv},
comment = {1812.09604},
author = {Hacker, Bastian and Welte, Stephan and Daiss, Severin and Shaukat, Armin and Ritter, Stephan and Li, Lin and Rempe, Gerhard},
doi = {10.1038/s41566-018-0339-5},
comment = {1812.09604},
isbn = {4156601803},
issn = {17494893},
journal = {Nature Photonics},
number = {2},
pages = {110--115},
publisher = {Springer US},
title = {{Deterministic creation of entangled atom–light Schr{\"{o}}dinger-cat states}},
comment = {http://dx.doi.org/10.1038/s41566-018-0339-5 https://arxiv.org/pdf/1812.09604.pdf},
volume = {13},
year = {2019}
}

@article{Bemani2021,
  title = {Force Sensing in an Optomechanical System with Feedback-Controlled In-Loop Light},
  author = {Bemani, F. and \ifmmode \check{C}\else \v{C}\fi{}ernot\'{\i}k, O. and Ruppert, L. and Vitali, D. and Filip, R.},
  journal = {Phys. Rev. Appl.},
  volume = {17},
  issue = {3},
  pages = {034020},
  numpages = {14},
  year = {2022},
  month = {Mar},
  publisher = {American Physical Society},
  doi = {10.1103/PhysRevApplied.17.034020},
  comment = {https://link.aps.org/doi/10.1103/PhysRevApplied.17.034020}
}

@article{tsangPRL2010,
   title = {Coherent Quantum-Noise Cancellation for Optomechanical Sensors},
  author = {Tsang, Mankei and Caves, Carlton M.},
  journal = {Phys. Rev. Lett.},
  volume = {105},
  issue = {12},
  pages = {123601},
  numpages = {4},
  year = {2010},
  month = {Sep},
  publisher = {American Physical Society},
  doi = {10.1103/PhysRevLett.105.123601},
  comment = {https://link.aps.org/doi/10.1103/PhysRevLett.105.123601}
}

@article{motazedifard2020,
   author = {Ali Motazedifard and A. Dalafi and M. H. Naderi},
   doi = {10.1116/5.0035952/997321},
   issn = {26390213},
   issue = {2},
   journal = {AVS Quantum Science},
   month = {6},
   pages = {24701},
   publisher = {American Institute of Physics Inc.},
   title = {Ultraprecision quantum sensing and measurement based on nonlinear hybrid optomechanical systems containing ultracold atoms or atomic Bose-Einstein condensate},
   volume = {3},
   comment = {/avs/aqs/article/3/2/024701/997321/Ultraprecision-quantum-sensing-and-measurement},
   year = {2021},
}

@article{Wendin2017superconducting,
doi = {10.1088/1361-6633/aa7e1a},
comment = {https://dx.doi.org/10.1088/1361-6633/aa7e1a},
year = {2017},
month = {sep},
publisher = {IOP Publishing},
volume = {80},
number = {10},
pages = {106001},
author = {G Wendin},
title = {Quantum information processing with superconducting circuits: a review},
journal = {Reports on Progress in Physics},
}

@article{BlaisNatPhys2020superconducting,
author = {Blais, Alexandre and Girvin, Steven M. and Oliver, William D.},
doi = {10.1038/s41567-020-0806-z},
issn = {17452481},
journal = {Nature Physics},
number = {3},
pages = {247--256},
publisher = {Springer US},
title = {{Quantum information processing and quantum optics with circuit quantum electrodynamics}},
comment = {http://dx.doi.org/10.1038/s41567-020-0806-z},
volume = {16},
year = {2020}
}

@article{munroPRA2001cat,
  title = {Weak-force detection with superposed coherent states},
  author = {Munro, W. J. and Nemoto, K. and Milburn, G. J. and Braunstein, S. L.},
  journal = {Phys. Rev. A},
  volume = {66},
  issue = {2},
  pages = {023819},
  numpages = {6},
  year = {2002},
  month = {Aug},
  publisher = {American Physical Society},
  doi = {10.1103/PhysRevA.66.023819},
  comment = {https://link.aps.org/doi/10.1103/PhysRevA.66.023819}
}

@article{GilmoreScience2021Displacement,
archivePrefix = {arXiv},
comment = {2103.08690v1},
author = {Gilmore, Kevin A. and Affolter, Matthew and Lewis-Swan, Robert J. and Barberena, Diego and Jordan, Elena and Rey, Ana Maria and Bollinger, John J.},
doi = {10.1126/science.abi5226},
comment = {2103.08690v1},
issn = {10959203},
journal = {Science},
number = {6555},
pages = {673--678},
pmid = {34353950},
title = {Quantum-enhanced sensing of displacements and electric fields with two-dimensional trapped-ion crystals},
volume = {373},
year = {2021}
}

@article{HempelNatPho2013catspectroscopy,
archivePrefix = {arXiv},
comment = {1304.3270},
author = {Hempel, C. and Lanyon, B. P. and Jurcevic, P. and Gerritsma, R. and Blatt, R. and Roos, C. F.},
doi = {10.1038/nphoton.2013.172},
comment = {1304.3270},
issn = {1749-4885},
journal = {Nature Photonics},
month = {apr},
number = {8},
pages = {630--633},
title = {{Entanglement-enhanced detection of single-photon scattering events}},
comment = {www.nature.com/naturephotonics http://www.nature.com/articles/nphoton.2013.172 http://arxiv.org/abs/1304.3270 http://dx.doi.org/10.1038/nphoton.2013.172},
volume = {7},
year = {2013}
}

@article{BurdScience2018SqzAmpDisp,
archivePrefix = {arXiv},
comment = {1812.01812},
author = {Burd, S C and Srinivas, R and Bollinger, J J and Wilson, A C and Wineland, D J and Leibfried, D and Slichter, D H and Allcock, D. T.C.},
doi = {10.1126/science.aaw2884},
comment = {1812.01812},
issn = {10959203},
journal = {Science},
number = {6446},
pages = {1163--1165},
pmid = {31221854},
title = {{Quantum amplification of mechanical oscillator motion}},
volume = {364},
year = {2019}
}

@article{PenasaPRA2016,
archivePrefix = {arXiv},
comment = {1605.09568},
author = {Penasa, M. and Gerlich, S. and Rybarczyk, T. and M{\'{e}}tillon, V. and Brune, M. and Raimond, J. M. and Haroche, S. and Davidovich, L. and Dotsenko, I.},
doi = {10.1103/PhysRevA.94.022313},
comment = {1605.09568},
issn = {24699934},
journal = {Physical Review A},
number = {2},
pages = {1--7},
title = {{Measurement of a microwave field amplitude beyond the standard quantum limit}},
volume = {94},
year = {2016}
}

@article{LIGO2020,
	author = {BP Abbott and R Abbott and TD Abbott and S Abraham et al.s},
	type = {HTML},
	title = {Prospects for observing and localizing gravitational-wave transients with Advanced LIGO, Advanced Virgo and KAGRA},
	journal = {Living Rev Relativ},
	publisher = {Springer},
	doi = {10.1007/s41114-020-00026-9},
	comment = {https://link.springer.com/article/10.1007/s41114-020-00026-9},
	fulltext = {https://link.springer.com/article/10.1007/s41114-020-00026-9},
		year = {2020},
	}

@article{BraunRMP2018Sensing,
author = {Braun, Daniel and Adesso, Gerardo and Benatti, Fabio and Floreanini, Roberto and Marzolino, Ugo and Mitchell, Morgan W. and Pirandola, Stefano},
doi = {10.1103/RevModPhys.90.035006},
issn = {15390756},
journal = {Reviews of Modern Physics},
number = {3},
pages = {1--52},
title = {{Quantum-enhanced measurements without entanglement}},
volume = {90},
year = {2018}
}

@article{Sidhu2020,
archivePrefix = {arXiv},
comment = {1907.06628},
author = {Sidhu, Jasminder S and Kok, Pieter},
doi = {10.1116/1.5119961},
comment = {1907.06628},
issn = {2639-0213},
journal = {AVS Quantum Science},
number = {1},
pages = {014701},
title = {{Geometric perspective on quantum parameter estimation}},
comment = {https://doi.org/10.1116/1.5119961},
volume = {2},
year = {2020}
}

@article{Polino2020,
author = {Polino, Emanuele and Valeri, Mauro and Spagnolo, Nicol{\`{o}} and Sciarrino, Fabio},
doi = {10.1116/5.0007577},
comment = {2003.05821},
issn = {2639-0213},
journal = {AVS Quantum Science},
number = {2},
pages = {024703},
title = {{Photonic quantum metrology}},
comment = {https://doi.org/10.1116/5.0007577},
volume = {2},
year = {2020}
}

@article{AspelmeyerRMP2014cavoptomech,
	author = {M Aspelmeyer and TJ Kippenberg and F Marquardt},
	title = {Cavity optomechanics},
	journal = {Reviews of Modern Physics},
	publisher = {APS},
	doi = {10.1103/RevModPhys.86.1391},
	comment = {https://journals.aps.org/rmp/abstract/10.1103/RevModPhys.86.1391},
		year = {2014},
}

@article{TeufelNature2011electromech,
   author = {J. D. Teufel and Dale Li and M. S. Allman and K. Cicak and A. J. Sirois and J. D. Whittaker and R. W. Simmonds},
   doi = {10.1038/nature09898},
   issn = {1476-4687},
   issue = {7337},
   journal = {Nature 2011 471:7337},
   month = {3},
   pages = {204-208},
   publisher = {Nature Publishing Group},
   title = {Circuit cavity electromechanics in the strong-coupling regime},
   volume = {471},
   comment = {https://www.nature.com/articles/nature09898},
   year = {2011},
}

@book{HolevoBook2019Channels,
	author = {AS Holevo},
	type = {BOOK},
	title = {Quantum systems, channels, information},
	publisher = {degruyter.com},
	doi = {10.1515/9783110642490},
	year = {2019},
		}

@article{IvanovPhysRevA2020TwoDisplac,
  title = {Enhanced two-parameter phase-space-displacement estimation close to a dissipative phase transition},
  author = {Ivanov, Peter A.},
  journal = {Phys. Rev. A},
  volume = {102},
  issue = {5},
  pages = {052611},
  numpages = {10},
  year = {2020},
  month = {Nov},
  publisher = {American Physical Society},
  doi = {10.1103/PhysRevA.102.052611},
  comment = {https://link.aps.org/doi/10.1103/PhysRevA.102.052611}
}

@article{HerbschlebNatComm2021DynamicRange,
author = {Herbschleb, E. D. and Kato, H. and Makino, T. and Yamasaki, S. and Mizuochi, N.},
doi = {10.1038/s41467-020-20561-x},
issn = {2041-1723},
journal = {Nature Communications 2021 12:1},
month = {jan},
number = {1},
pages = {1--8},
pmid = {33436617},
publisher = {Nature Publishing Group},
title = {{Ultra-high dynamic range quantum measurement retaining its sensitivity}},
comment = {https://www.nature.com/articles/s41467-020-20561-x},
volume = {12},
year = {2021}
}

@article{AlbarelliPRA2018,
archivePrefix = {arXiv},
comment = {1804.05763},
author = {Albarelli, Francesco and Genoni, Marco G. and Paris, Matteo G.A. A and Ferraro, Alessandro},
doi = {10.1103/PhysRevA.98.052350},
comment = {1804.05763},
issn = {24699934},
journal = {Physical Review A},
month = {nov},
number = {5},
pages = {52350},
publisher = {American Physical Society},
title = {{Resource theory of quantum non-Gaussianity and Wigner negativity}},
volume = {98},
year = {2018}
}

@book{DAlessandroBook2021control,
author = {D'Alessandro, Domenico},
booktitle = {Introduction to Quantum Control and Dynamics},
doi = {10.1201/9781003051268},
isbn = {9781003051268},
month = {jul},
publisher = {Chapman and Hall/CRC},
title = {{Introduction to Quantum Control and Dynamics}},
comment = {https://www.taylorfrancis.com/books/mono/10.1201/9781003051268/introduction-quantum-control-dynamics-domenico-alessandro},
year = {2021}
}

@article{PezzeRMP2018AtomEnsemble,
archivePrefix = {arXiv},
comment = {1609.01609},
author = {Pezz{\`{e}}, Luca and Smerzi, Augusto and Oberthaler, Markus K. and Schmied, Roman and Treutlein, Philipp},
doi = {10.1103/RevModPhys.90.035005},
comment = {1609.01609},
issn = {15390756},
journal = {Reviews of Modern Physics},
number = {3},
publisher = {APS},
title = {{Quantum metrology with nonclassical states of atomic ensembles}},
comment = {https://journals.aps.org/rmp/abstract/10.1103/RevModPhys.90.035005},
volume = {90},
year = {2018}
}

@article{GuPR2017Superconducting,
title = {Microwave photonics with superconducting quantum circuits},
journal = {Physics Reports},
volume = {718-719},
pages = {1-102},
year = {2017},
note = {Microwave photonics with superconducting quantum circuits},
issn = {0370-1573},
doi = {https://doi.org/10.1016/j.physrep.2017.10.002},
comment = {https://www.sciencedirect.com/science/article/pii/S0370157317303290},
author = {Xiu Gu and Anton Frisk Kockum and Adam Miranowicz and Yu-xi Liu and Franco Nori},
}

@article{BlaisRMP2021CircuitQED,
author = {Blais, Alexandre and Grimsmo, Arne L and Girvin, S M and Wallraff, Andreas},
doi = {10.1103/RevModPhys.93.025005},
journal = {Reviews of Modern Physics},
title = {{Circuit quantum electrodynamics}},
volume = {93},
year = {2021}
}

@incollection{DEMKOWICZDOBRZANSKI2015345,
title = {Chapter Four - Quantum Limits in Optical Interferometry},
editor = {E. Wolf},
series = {Progress in Optics},
publisher = {Elsevier},
volume = {60},
pages = {345-435},
year = {2015},
issn = {0079-6638},
doi = {https://doi.org/10.1016/bs.po.2015.02.003},
comment = {https://www.sciencedirect.com/science/article/pii/S0079663815000049},
author = {Rafal Demkowicz-Dobrzański and Marcin Jarzyna and Jan Kołodyński},
}

@article{ClerkNatPhy2020SuperCon,
   author = {A. A. Clerk and K. W. Lehnert and P. Bertet and J. R. Petta and Y. Nakamura},
   doi = {10.1038/s41567-020-0797-9},
   issn = {1745-2481},
   issue = {3},
   journal = {Nature Physics 2020 16:3},
   month = {3},
   pages = {257-267},
   publisher = {Nature Publishing Group},
   title = {Hybrid quantum systems with circuit quantum electrodynamics},
   volume = {16},
   comment = {https://www.nature.com/articles/s41567-020-0797-9},
   year = {2020},
}

@article{Martinez-Garaot2018,
archivePrefix = {arXiv},
comment = {1807.07535},
author = {Mart{\'{i}}nez-Garaot, S. and Rodriguez-Prieto, A. and Muga, J. G.},
doi = {10.1103/PhysRevA.98.043622},
comment = {1807.07535},
issn = {24699934},
journal = {Physical Review A},
month = {jul},
number = {4},
publisher = {American Physical Society},
title = {{Interferometer with a driven trapped ion}},
comment = {http://arxiv.org/abs/1807.07535 http://dx.doi.org/10.1103/PhysRevA.98.043622},
volume = {98},
year = {2018}
}

@article{McCormick2018,
archivePrefix = {arXiv},
comment = {1811.00668},
author = {McCormick, Katherine C. and Keller, Jonas and Wineland, David J. and Wilson, Andrew C. and Leibfried, Dietrich},
doi = {10.1088/2058-9565/ab0513},
comment = {1811.00668},
issn = {20589565},
journal = {Quantum Science and Technology},
month = {nov},
number = {2},
publisher = {Institute of Physics Publishing},
title = {{Coherently displaced oscillator quantum states of a single trapped atom}},
comment = {http://arxiv.org/abs/1811.00668 http://dx.doi.org/10.1088/2058-9565/ab0513},
volume = {4},
year = {2018}
}

@article{Park2022,
  title = {Optimal Estimation of Conjugate Shifts in Position and Momentum by Classically Correlated Probes and Measurements},
  author = {Park, Kimin and Oh, Changhun and Filip, Radim and Marek, Petr},
  journal = {Phys. Rev. Appl.},
  volume = {18},
  issue = {1},
  pages = {014060},
  numpages = {8},
  year = {2022},
  month = {Jul},
  publisher = {American Physical Society},
  doi = {10.1103/PhysRevApplied.18.014060},
  comment = {https://link.aps.org/doi/10.1103/PhysRevApplied.18.014060}
}

@article{Park2021,
   author = {Kimin Park and Jacob Hastrup and Jonas Schou Neergaard-Nielsen and Jonatan Bohr Brask and Radim Filip and Ulrik L. Andersen},
   doi = {10.1038/s41534-022-00577-5},
   issn = {2056-6387},
   issue = {1},
   journal = {npj Quantum Information 2022 8:1},
   month = {6},
   pages = {1-8},
   publisher = {Nature Publishing Group},
   title = {Slowing quantum decoherence of oscillators by hybrid processing},
   volume = {8},
   comment = {https://www.nature.com/articles/s41534-022-00577-5},
   year = {2022},
}

@article{HastrupPRL2022,
archivePrefix = {arXiv},
comment = {2106.12272},
author = {Hastrup, Jacob and Park, Kimin and Brask, Jonatan Bohr and Filip, Radim and Andersen, Ulrik Lund},
doi = {10.1103/PhysRevLett.128.110503},
comment = {2106.12272},
issn = {0031-9007},
journal = {Physical Review Letters},
month = {mar},
number = {11},
pages = {110503},
title = {{Universal Unitary Transfer of Continuous-Variable Quantum States into a Few Qubits}},
comment = {http://arxiv.org/abs/2106.12272 https://link.aps.org/doi/10.1103/PhysRevLett.128.110503},
volume = {128},
year = {2022}
}

@article{GarbePRL2020CriticalRabiSensing,
archivePrefix = {arXiv},
comment = {1910.00604},
author = {Garbe, Louis and Bina, Matteo and Keller, Arne and Paris, Matteo G.A. and Felicetti, Simone},
doi = {10.1103/PhysRevLett.124.120504},
comment = {1910.00604},
issn = {10797114},
journal = {Physical Review Letters},
number = {12},
pages = {120504},
pmid = {32281838},
publisher = {American Physical Society},
title = {{Critical Quantum Metrology with a Finite-Component Quantum Phase Transition}},
comment = {https://doi.org/10.1103/PhysRevLett.124.120504},
volume = {124},
year = {2020}
}

@article{DiCandia2021KerrCritialSensor,
   author = {R. Di Candia and F. Minganti and K. V. Petrovnin and G. S. Paraoanu and S. Felicetti},
   doi = {10.1038/s41534-023-00690-z},
   issn = {2056-6387},
   issue = {1},
   journal = {npj Quantum Information 2023 9:1},
   month = {3},
   pages = {1-9},
   publisher = {Nature Publishing Group},
   title = {Critical parametric quantum sensing},
   volume = {9},
   comment = {https://www.nature.com/articles/s41534-023-00690-z},
   year = {2023},
}

@article{ChuPRL2021,
archivePrefix = {arXiv},
comment = {2008.11381},
author = {Chu, Yaoming and Zhang, Shaoliang and Yu, Baiyi and Cai, Jianming},
doi = {10.1103/PhysRevLett.126.010502},
comment = {2008.11381},
issn = {10797114},
journal = {Physical Review Letters},
number = {1},
pages = {10502},
pmid = {33480770},
publisher = {American Physical Society},
title = {{Dynamic Framework for Criticality-Enhanced Quantum Sensing}},
comment = {https://doi.org/10.1103/PhysRevLett.126.010502},
volume = {126},
year = {2021}
}

@article{GoreckiPRL2019,
   author = {Wojciech Gorecki and Rafal Demkowicz-Dobrzanski and Howard M. Wiseman and Dominic W. Berry},
   doi = {10.1103/PhysRevLett.124.030501},
   issn = {10797114},
   issue = {3},
   journal = {Physical Review Letters},
   month = {7},
   pmid = {32031843},
   title = {$\pi$-Corrected Heisenberg Limit},
   volume = {124},
   comment = {http://arxiv.org/abs/1907.05428 http://dx.doi.org/10.1103/PhysRevLett.124.030501},
   year = {2019},
}

@article{FidererPRX2021QFIM,
   author = {Lukas J. Fiderer and Tommaso Tufarelli and Samanta Piano and Gerardo Adesso},
   doi = {10.1103/PRXQUANTUM.2.020308},
   issn = {26913399},
   issue = {2},
   journal = {PRX Quantum},
   month = {4},
   pages = {020308},
   publisher = {American Physical Society},
   title = {General Expressions for the Quantum Fisher Information Matrix with Applications to Discrete Quantum Imaging},
   volume = {2},
   comment = {https://journals.aps.org/prxquantum/abstract/10.1103/PRXQuantum.2.020308},
   year = {2021},
}

\end{document}